\documentclass{jfm}
\usepackage{tikz}
\usepackage{graphicx}
\usetikzlibrary{positioning}
\usepackage{xcolor}
\usepackage{epstopdf, epsfig}
\usepackage[T1]{fontenc}
\usepackage{upquote}
\usepackage{amsmath}

\shorttitle{Magnetoconvection in a horizontal duct flow}
\shortauthor{R. Akhmedagaev, O. Zikanov, and Y. Listratov}

\title{Magnetoconvection in a horizontal duct flow at very high Hartmann and Grashof numbers}

\author{R. Akhmedagaev\aff{1}, O. Zikanov\aff{1} \corresp{\email{zikanov@umich.edu}},
    \and Y. Listratov\aff{2,3}}

\affiliation{\aff{1}University of Michigan - Dearborn, 4901 Evergreen Road, Dearborn, MI 48128-1491, USA
\aff{2}Moscow Power Engineering Institute, 14 Krasnokazarmennaya Street, Moscow, 111250, Russia
\aff{3}Joint Institute for High Temperatures Russian Academy of Science, Izhorskaya Street 13, Building 2, Moscow, 125412, Russia}

\begin{document}

\maketitle

\begin{abstract}
Direct numerical simulations and linear stability analysis are carried out to study mixed convection in a horizontal duct with constant-rate heating applied at the bottom and imposed transverse horizontal magnetic field. A two-dimensional approximation corresponding to the asymptotic limit of very strong magnetic field effect is validated and applied, together with full three-dimensional analysis, to investigate the flow's behaviour in the previously unexplored range of control parameters corresponding to typical conditions of a liquid metal blanket of a nuclear fusion reactor (Hartmann numbers up to $10^4$ and Grashof numbers up to $10^{10}$). It is found that the instability to quasi-two-dimensional rolls parallel to the magnetic field discovered at smaller Hartmann and Grashof numbers in earlier studies also occurs in this parameter range. Transport of the rolls by the mean flow leads to magnetoconvective temperature fluctuations of exceptionally high amplitudes. It is also demonstrated that quasi-two-dimensional structure of flows at very high Hartmann numbers does not guarantee accuracy of the classical two-dimensional approximation. The accuracy deteriorates at the highest Grashof numbers considered in the study.

\end{abstract}


\section{\textbf{Introduction}}

Combined convection and magnetohydrodynamic (MHD) effects dramatically change the nature of flows of electrically conducting fluids. The combination appears in many technological applications such as metallurgy, liquid metal batteries, and growth of semiconductor crystals \citep{Ozoe05, Davidson16}. Another prominent example is the liquid metal blankets of nuclear fusion reactors where an electrically conducting fluid (e.g., a PbLi alloy) serves as a coolant, radiation shield and tritium breeder \citep{Abdou15}. A distinctive feature of this system is that the convection and magnetic filed effects are both exceptionally strong.

Many aspects of the transformation of flows of electrically conducting fluid under the influence of a strong magnetic field, such as suppression of turbulent fluctuations, anisotropic or quasi-two-dimensional (Q2D) states with zero or weak velocity gradients along the field lines, formation of MHD boundary layers, and delay of laminar-turbulent transition, are relatively well understood \citep[see, e.g.,][]{Branover78,  Davidson16, Sommeria82, Zikanov14}. This paper addresses a recently discovered and still poorly understood phenomenon - the high amplitude fluctuations in flows in ducts and pipes \citep[see, e.g.,][]{Genin11Pamir,Vetcha13,Zikanov13,Belyaev21}. The term magneto-convective fluctuations (MCFs) proposed for the phenomenon by \citet{Belyaev21} will be used in this paper. As discussed in detail in the review of \citet{Zikanov21} and references therein, the fluctuations have been detected in experimental and computational studies of a large variety of systems: pipes and ducts of various orientations with respect to gravity, various heating arrangements, and various configurations of the magnetic field. 

The fluctuations were called anomalous in some earlier works, e.g., by \citet{Zikanov13} and \citet{Zhang14}. This term now appears imprecise and somewhat misleading since it has been understood that the fluctuations are rather common. They occur in a wide variety of magnetoconvection flows. It must also be mentioned that, in a broader context, the magnetoconvective fluctuations are a part of the general phenomenon of large-amplitude fluctuations commonly found in flows, where turbulence is suppressed by a strong magnetic field and flow fields are strongly anisotropic or Q2D \citep[see, e.g.,][for discussion and references]{Smolentsev21, Zikanov21}.

The nature of the magnetoconvective fluctuations can be briefly described as follows. They appear in the conditions of a very strong magnetic field effect, i.e. in the range of Hartmann numbers, where turbulence is fully suppressed by magnetic damping. In experiments, the MCFs are manifested by oscillations of temperature with very high amplitude (up to $50 K$ in some cases) and typical frequencies much lower than the frequencies of turbulence-induced fluctuations. Specific properties of the MCFs vary with the flow's configuration and values of the control parameters \citep{Zikanov21}. The effect has potentially serious consequences for design and operation of liquid metal blankets of future fusion reactors. Should the fluctuations appear in an actual blanket, they may lead to strong and unsteady thermal stresses in the walls \citep[see, e.g.,][]{Belyaev18} possibly under the condition of significantly reduced strength of the wall material \citep{Kolmakov16}.  Due to their possibly very large amplitude, the stresses will threaten the structural integrity of a fusion reactor system. Significant effects on heat transfer, transport of tritium, and wall corrosion are also anticipated. As we discuss later in this section, it is yet impossible to say how realistic these expectations are, since no experiments or computations at very high $Ha$ and $Gr$ typical for reactor conditions have been conducted so far. 

\begin{figure}
	\centering 

\raisebox{11em}	
	
\begin{tikzpicture}

\node (img1) {\includegraphics[scale=0.20]{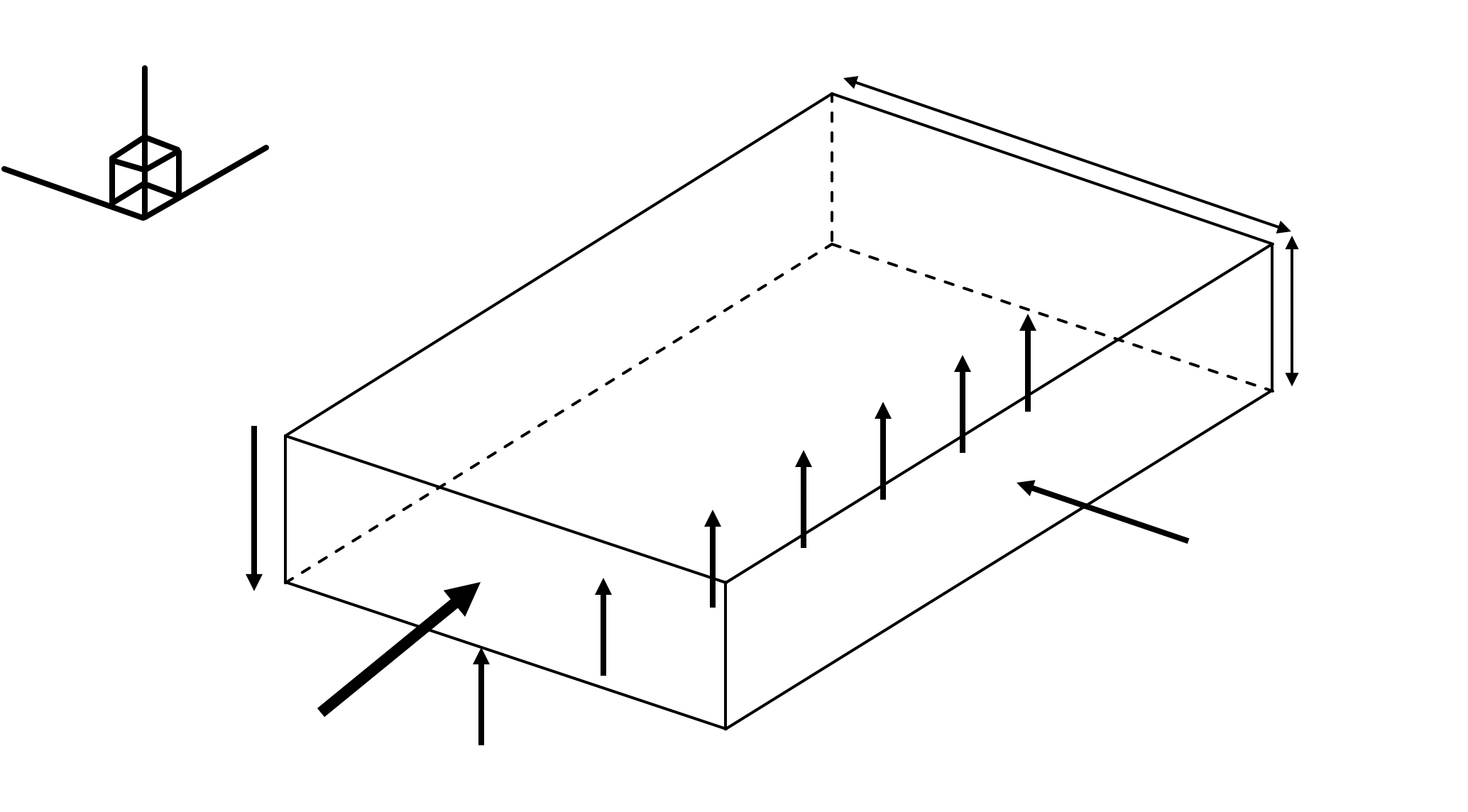}};

	\node[left=of img1, xshift=2.1cm ,yshift=1.90cm,scale=1.00, font=\color{black}] {$Z$};
	\node[left=of img1, xshift=1.25cm ,yshift=1.25cm,scale=1.00, font=\color{black}] {$Y$};
	\node[left=of img1, xshift=2.95cm ,yshift=1.35cm,scale=1.00, font=\color{black}] {$X$};
	
	\node[left=of img1, xshift=2.85cm ,yshift=-1.25cm,scale=1.00, font=\color{black}] {${\boldsymbol{U}}$};
	\node[left=of img1, xshift=3.45cm ,yshift=-1.70cm,scale=1.00, font=\color{black}] {$q$};
	\node[left=of img1, xshift=7.10cm ,yshift=-0.40cm,scale=1.00, font=\color{black}] {$\boldsymbol{B}$};
	\node[left=of img1, xshift=2.30cm ,yshift=-0.55cm,scale=1.00, font=\color{black}] {$\boldsymbol{g}$};
	
	\node[left=of img1, xshift=7.90cm ,yshift=0.5cm,scale=1.00, font=\color{black}] {$h$};
	\node[left=of img1, xshift=6.70cm ,yshift=1.50cm,scale=1.00, font=\color{black}] {$2d$};

\end{tikzpicture}

  \caption{ Flow geometry and coordinate system. The arrows marked by letters $\boldsymbol{g}$, $\boldsymbol{B}$ and $q$ denote, respectively, the orientations of the gravity acceleration, magnetic field and wall heating.}
  
\label{fig1}
\end{figure}
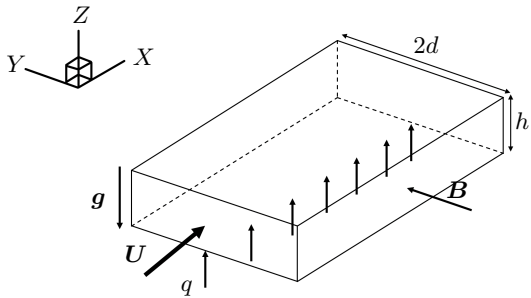

Flows in a rectangular duct with heating applied at the bottom and imposed transverse horizontal magnetic field (see figure \ref{fig1}) are considered in this paper. The configuration is not found in currently developed specific designs of liquid metal blankets of fusion reactors, although it may occur in future designs of an upper divertor and top blanket modules \citep{Kirillov97}. It is also important as an archetypal system, in which the MCFs were first identified (in \citet{Genin11Pamir} and \citet{Zikanov13}, where they were named anomalous fluctuations) and explained.


Similar systems for either ducts or round pipes have been studied experimentally \citep{Genin11Pamir, Belyaev15, Sahu20} and numerically \citep{Zikanov13, Zhang14, Vo17, Listratov18}. The flow is controlled by four dimensionless parameters: the Reynolds, Prandtl, Grashof and Hartmann numbers,
\begin{equation} \label{eq0}
    {\it Re} = \frac{Ud}{\nu}, \ \ \ {\it Pr} = \frac{\nu}{\chi},  \ \ \	{\it Gr} = \frac{g\beta q d^4}{\nu^2\kappa}, \ \ \ {\it Ha} = Bd \sqrt{{\frac{\sigma}{\rho\nu}}}, 
\end{equation} with the duct half-width $d$, the mean streamwise velocity $U$, the kinematic viscosity $\nu$, the temperature diffusivity $\chi$, the acceleration due to gravity $g$, the coefficient of thermal expansion $\beta$, the heat flux of constant rate $q$, the thermal conductivity $\kappa$, the electrical conductivity $\sigma$ and the mass density $\rho$. Rectangular duct geometry adds the aspect ratio $\Gamma = {2d}/{h}$ as a parameter, where $h$ is the height of the duct.

Interesting results were obtained in the linear stability analysis of the Poiseulle-Rayleigh-B\'{e}rnard duct flow with a transverse magnetic field performed by \citet{Vo17}. Two-dimensional approximation valid in the limit of strong magnetic field presented later in this paper was used. One important result of \citet{Vo17} is relevant to our work even though different boundary conditions were used. It was demonstrated that the convection instability occurs at moderate and high Grashof number (approximately above $10^6$) at the Hartmann numbers ($\sim 10^4$) typical for reactor blanket conditions. 

The presence of MCFs in a horizontal round pipe with a lower half of the wall heated was detected in experiments \citep{Genin11Pamir, Belyaev15} and explained in the linear stability analysis and direct numerical simulations (DNS) by \citet{Zikanov13}. Flows of mercury with ${\it Pr} \approx 0.022$,  ${\it Re}$ up to $10^5$, ${\it Gr}$ up to $10^8$, and ${\it Ha}$ up to $500$ were investigated. It was shown that at a strong magnetic field the suppression of flow structures having large gradients along the field lines resulted in the most unstable modes in the form of convection rolls with axes aligned with the field. The instability led to development of convection structures in the form of Q2D rolls. Transport of the rolls by the mean flow generated the MCFs. 

The analysis was extended in the numerical simulations of \citet{Zhang14}. Flows in a horizontal duct of aspect ratio $\Gamma = 1$ with bottom heating and transverse magnetic field at ${\it Pr} = 0.0321$,  ${\it Re} = 5000$, $50 \le {\it Ha} \le 800$ and $10^5 \le {\it Gr} \le 10^9$ were investigated. The instability leading to the formation of Q2D rolls similar to those found in the pipe flow was detected at sufficiently high $Gr$ and $Ha$. 

Investigations of \citet{Zhang14} conducted in the broader range of parameters than for the pipe flow demonstrated existence of two distinct secondary flow regimes. The realization of the regimes depended on the relative strength of the convection and MHD effects. The low-${\it Gr}$ type characterized by Q2D distributions of velocity and temperature dominated by spanwise rolls appeared at $Gr$  below a certain $Gr^{\ast}(Ha)$. At higher ${\it Gr}$, stronger convection resulted in three-dimensional (3D) flow states combining the spanwise rolls with streamwise ones (the geometrically preferred convection structure in pipes and ducts with bottom heating).

Flows of liquid metals in fusion reactor blankets and divertors are subject to very strong effects of convection ($Gr$ $\sim 10^{10} - 10^{12}$) and magnetic fields ($Ha \sim 10^{4}$) \citep[see, e.g.,][]{Smolentsev08,  Smolentsev10}. Such extreme parameters present serious obstacles to analysis, because neither laboratory experiments nor 3D simulations of unsteady flow regimes in realistic blanket or divertor geometries can, at this moment, achieve such values. 

In an attempt to reach the typical blanket flow conditions, the data on two types of the secondary flow regime in a horizontal duct were extrapolated to high $Gr$ and $Ha$ by \citet{Zhang14}. The extrapolation predicted existence of MCFs at the typical blanket parameters. It also predicted that the flow would likely be of the low-${\it Gr}$ type at ${\it Gr} \le 10^{10}$ and of the high-${\it Gr}$ type at higher ${\it Gr}$. The experiments in the pipe flow \citep[see the review of recent results in][]{Zikanov21}, on the contrary, indicate that MCFs may disappear at high ${\it Ha}$, so the extrapolation can be wrong. The nature of the convection flow at the parameters corresponding to ducts in blankets and divertors of an operating fusion reactor remains unknown, setting up the motivation for the present study. 

The focus of our investigation is on the magnetoconvection in the range of very high $Gr$ and $Ha$ including the values typical for a reactor blanket and divertor.  To the best of our knowledge, this study is the first to analyze the MCF effect in this range. Linear stability analysis and DNS of flows in a horizontal duct with ${\it Pr} = 0.025$, $Re = 5000$, $10^8 \le {\it Gr} \le 10^{10}$ and $10^3 \le {\it Ha} \le 10^4$ are performed. The study follows the work of \citet{Zhang14} but differs by much larger values of $Gr$ and $Ha$ and the aspect ratio $\Gamma = 3.5$ selected to match the new experimental facility \citep[see, e.g.,][]{Belyaev17}, on which the same configuration is to be explored at ${\it Ha} \lesssim 10^3$ and ${\it Gr} \lesssim 10^8$ in the near future. Another essential difference between our work and the work by \citet{Zhang14} is that we carry out an in-depth analysis of the accuracy of the two-dimensional approximation applied to Q2D flows at such high $Ha$.


\section{\textbf{Presentation of the problem}}

The flow of an incompressible, Newtonian, viscous, electrically conducting fluid (a liquid metal) with constant physical properties is considered. The fluid moves through a horizontal duct of aspect ratio $\Gamma = 3.5$ (see figure \ref{fig1}). Spatially uniform and time-independent magnetic field $\boldsymbol{B}=B\boldsymbol{e}_y$ is imposed in the horizontal transverse direction. All walls are perfectly electrically insulated. The top and side walls are perfectly thermally insulated. The bottom wall is subject to uniform heating with the heat flux of constant rate $q$. The no-slip boundary conditions for velocity are applied at the walls.


\subsection{Physical model}  \label{physmod1}

The Boussinesq and quasi-static approximations are applied. The quasi-static approximation is valid at small Reynolds and Prandtl numbers and usually utilized in numerical and theoretical studies of MHD flows of liquid metals \citep{Davidson16}. The approximation implies that the imposed magnetic field $\boldsymbol{B}$ is much stronger than the perturbations of the magnetic field $\boldsymbol{b}$ induced by the electric currents caused by the fluid motion. The induced magnetic field can be neglected in the expressions of the Lorentz force and Ohm's law. Furthermore, the induced field is assumed to adjust instantaneously to changes of velocity field. 

The governing equations are rendered non-dimensional using the duct half-width in the magnetic field direction $d$ as the length scale, mean streamwise velocity $U$ as the velocity scale, wall heating-based group $qd/\kappa$ as the temperature scale, $B$ as the scale of the magnetic field strength, and $dUB$ as the scale of electric potential. The equations can be written as:
\begin{equation} \label{eq1}
	\frac{\partial \boldsymbol{u}}{\partial t} + (\boldsymbol{u}\bcdot \bnabla)\boldsymbol{u} = 
	-\bnabla p -\bnabla \hat{p} -\bnabla \tilde{p} + \frac{1}{Re}\bnabla^2 \boldsymbol{u} + {\it \boldsymbol{F}_b} + {\it \boldsymbol{F}_L},
\end{equation}
\begin{equation} \label{eq2}
	\bnabla\bcdot\boldsymbol{u} = 0,
\end{equation}
\begin{equation} \label{eq3}
	\frac{\partial \theta}{\partial t} + \boldsymbol{u}\bcdot\bnabla \theta = {\frac{1}{{\it Re}{\it Pr}}} \bnabla^2 \theta - u_x \frac{\text{d}{\it T_m}}{\text{d}x},
\end{equation} where $\boldsymbol{u}$ is the velocity field. The decompositions of the temperature and pressure fields commonly used in studies of mixed convection in ducts and pipes \citep[see, e.g.,][]{Alboussiere93, Lyubimova09, Zikanov13, Zhang14,Zikanov21} are applied. The decompositions are convenient, since they allow one to recast the problem in terms of the fluctuation fields, which are statistically uniform in the streamwise direction and, thus, study the flow in a relatively short segment of the channel with periodic inlet-exit conditions. The temperature field is written as a sum
\begin{equation} \label{eq6}
	T(\boldsymbol{x},t) = \it {T_m}(x) + \theta(\boldsymbol{x},t)
\end{equation} of fluctuations $\theta$ and the mean-mixed temperature 
\begin{equation} \label{eq6_}
	 \it {T_m}(x) = \frac{\int_{A} u_xT \text{d}A}{\int_{A} u_x \text{d}A}= A^{-1} \int_{A} u_xT \text{d}A , 
\end{equation} where $A = 2 h/d$ is the cross-section area of the duct. One can also use the decomposition into fluctuations and simple mean temperature $\bar{T}(x)=A^{-1}\int_{A} T \text{d}A$. Applying the energy balance between the wall heating and the streamwise convection heat transfer, we find that $\it {T_m}(x)$ and $\bar{T}(x)$ are linear functions with the same derivative:
\begin{equation} \label{eq_Tm_lin}
	\frac{\text{d}{\it T_m}}{\text{d}x} = \frac{\text{d}{\it \bar{T}}}{\text{d}x} = \frac{\Pi}{A RePr} = \frac{\Gamma}{2RePr} , 
\end{equation} where $\Pi = 2$ is the perimeter of the heated portion of the wall. 

The total pressure $P$ is presented in (\ref{eq1}) as
\begin{equation} \label{eq4}
	P = \hat{p}(x) + \tilde{p}(x,z) + p(\boldsymbol{x},t),
\end{equation} where $p(\boldsymbol{x},t)$ is the field of pressure fluctuations statistically homogeneous in the streamwise direction, and $\hat{p}$ is a linear function of $x$ corresponding to the spatially uniform streamwise gradient ${\text{d}{\hat{p}}}/{\text{d}x}$ applied as a flow-driving mechanism. In the simulations discussed in this paper, the gradient is adjusted at every time step to maintain constant mean velocity. 

The second term of the decomposition becomes necessary in numerical models of mixed convection in non-vertical channels with periodic inlet–exit conditions. The component
\begin{equation} \label{eq5}
	\tilde{p}(x,z) = \frac{\text{d}{\it T_m}}{\text{d}x}\frac{Gr}{Re^2} xz = \frac{\Pi}{A{\it Re}{\it Pr}}\frac{Gr}{Re^2} xz,
\end{equation} arises due to the buoyancy force caused by the mean–mixed temperature ${\it T_m}$:
\begin{equation} \label{ch3_buoyancy_Tm}
	\boldsymbol{F}_{b, m} = Gr Re^{-2} \boldsymbol{e}_z T_m,
\end{equation} where $\boldsymbol{e}_z$ is the unit vector opposite to the direction of gravity (see figure \ref{fig1}). The force has a non-zero curl and, therefore, modifies the velocity field. Its action on the flow can be described by introducing the pressure field $\tilde{p}$, such that its vertical gradient balances $\boldsymbol{F}_{b, m}$. The pressure field is a two-dimensional function increasing with the streamwise coordinate $x$ and vertical coordinate $z$. Its $z$-dependent $x$-gradient, which appears in the respective momentum equation, generates a flow in the positive $x$-direction in the lower part of the channel and in the negative $x$-direction in the upper part. The result is a top-bottom asymmetry of the streamwise velocity profile and of the associated convection heat flux, which can dramatically change the structure of the flow at high $Gr$ and $Ha$ \citep[see ][]{Zikanov13,Zhang14,Zhang17,Zikanov21}.

The buoyancy force in (\ref{eq1}) is 

\begin{equation} \label{ch3_buoyancy}
	\boldsymbol{F }_b = Gr Re^{-2} \boldsymbol{e}_z T.
\end{equation}

The Lorentz force is computed as
\begin{equation} \label{ch3_Lorentz}
	\boldsymbol{F}_L  = Ha^2 Re^{-1} \boldsymbol{j}\times \boldsymbol{e}_y,
\end{equation}
where $\boldsymbol{e}_y$ is the unit vector along the imposed magnetic field (see figure \ref{fig1}). The electric current $\boldsymbol{j}$ is determined by the Ohm's law
\begin{equation} \label{eq7}
	\boldsymbol{j} = -\nabla \phi + (\boldsymbol{u} \times \boldsymbol{e}_y),
\end{equation} where the electric potential $\phi$ is a solution of the Poisson equation expressing the instantaneous electric neutrality of the fluid:
\begin{equation} \label{eq8}
	\nabla^2\phi = \nabla \cdot (\boldsymbol{u} \times \boldsymbol{e}_y).
\end{equation} 

The inlet–exit conditions are those of periodicity of the velocity $\boldsymbol{u}$, temperature fluctuations $\theta$, pressure fluctuations $p$, and potential $\phi$.


\subsection{Two-dimensional approximation} \label{physmod2}

Flows with a very strong imposed magnetic field are considered, so the Hartmann number and the Stuart number satisfy ${Ha} \gg 1$ and $N \equiv Ha^2/Re\gg 1$, respectively. The flows are anticipated to have Q2D form with nearly zero gradients along the magnetic field lines except in the thin Hartmann layers at the walls perpendicular to the field. The 2D approximation proposed by \citet{Sommeria82} can be applied in this asymptotic limit. The problem can be expressed in terms of the variables integrated wall-to-wall along the direction of the magnetic field, leading to 2D dynamics for y-averaged quantities. The approximation has been verified and examined by \citet{Potherat00,Potherat05}, and utilized in numerical studies of liquid metal flows in rectangular ducts \citep[see, e.g.,][]{Potherat07, Smolentsev12, Vetcha13, Vo17, Zhang18}.

The often applied abbreviation SM82 will be used for the model in the following. The $y-$independent solutions obtained in the framework of the model will be referred to as 2D solutions, while the full solutions obtained numerically without resorting to the model will be designated as 3D.

The SM82 model is derived for flows with ${Ha} \gg 1$ and $N \gg 1$, in domains with electrically insulating walls and constant wall-to-wall distance in the field direction. It utilizes the fact that the Lorentz force becomes nearly zero in the bulk region of Q2D flows in such geometries, and that the effect of the magnetic field on the flow is largely reduced to thin Hartmann layers and can be accurately modelled by the linear friction term in the momentum equation. 

It must be noted that the original SM82 model was developed for isothermal flows. Its extension to flows with heat transfer and temperature variations was, to our best knowledge, first proposed by \citet{Smolentsev08}. As demonstrated in this and in the following studies \citep[see, e.g.,][]{Gelfgat11,Vetcha13, Zhang18}, the model can be extended to 2D approximation of temperature if the imposed heat flux is perpendicular to the magnetic field.

The SM82 version of (\ref{eq1})$-$(\ref{eq3}) is
\begin{equation} \label{eqsm1}
	\frac{\partial \boldsymbol{u}}{\partial t} + (\boldsymbol{u}\bcdot \bnabla)\boldsymbol{u} = 
	-\bnabla p -\bnabla \hat{p} -\bnabla \tilde{p} + \frac{1}{Re}\bnabla^2 \boldsymbol{u} - \frac{Ha}{Re}\boldsymbol{u} + \frac{Gr}{Re^2}T\hat{\boldsymbol{e}}_z,
\end{equation}
\begin{equation} \label{eqsm2}
	\bnabla\bcdot\boldsymbol{u} = 0,
\end{equation}
\begin{equation} \label{eqsm3}
	\frac{\partial \theta}{\partial t} + \boldsymbol{u}\bcdot\bnabla \theta = {\frac{1}{{\it Re}{\it Pr}}} \bnabla^2 \theta - u_x \frac{\text{d}{\it T_m}}{\text{d}x},
\end{equation} where all the flow variables are now 2D fields obtained by wall-to-wall averaging. The term (${Ha}/{Re}$)$\boldsymbol{u}$ represents the effect of friction in the Hartmann layers. The same notation as in (\ref{eq1})$-$(\ref{eq3}) is used. The boundary conditions on velocity and temperature on the remaining two wall are the same as in the 3D model. 


\subsection{Numerical method} \label{physmod3}

The governing equations (\ref{eq1})$-$(\ref{eq3}) and (\ref{eqsm1})$-$(\ref{eqsm3}) are solved numerically using the finite difference scheme introduced by \citet{Krasnov11}, and later developed, tested and applied to high-${Ha}$ flows and flows with thermal convection in numerous works including those by  \citet{Krasnov12}, \citet{Zhao12}, \citet{Zikanov13}, \citet{Zhang14}, and \citet{Gelfgat18}. The spatial discretization is of the second order and nearly fully conservative with regards to the mass, momentum, electric charge, kinetic energy and thermal energy conservation principles \citep{Ni07, Krasnov11}. The computational grid is clustered towards the walls according to the coordinate transformation in the horizontal direction $y = \tanh(A_y \eta)/\tanh(A_y) $ and in the vertical direction $ z = \tanh(A_z \xi)/\tanh(A_z)$. Here $\eta$ and $\xi$ are the transformed coordinates, in which the grid is uniform, and $A_y$ and $A_z$ are the coefficients determining the degrees of clustering. The time discretization is implicit for the conduction and viscosity terms and based on the Adams–Bashforth/backward-differentiation method of the second order and the standard projection algorithm \citep[see, e.g.,][]{Zikanov19}. The nonlinear convection and body force terms are treated explicitly. The elliptic equations for potential, pressure, temperature and velocity components are solved using the Fourier decomposition in the streamwise coordinate and the direct cyclic reduction solution of the 2D equations for Fourier components conducted on the transformed grid \citep[see][]{Krasnov11}.

The algorithm is parallelized using the hybrid MPI-OpenMP approach. The MPI memory distribution is along the $y-$coordinate in the physical space and along the streamwise wavenumber in the Fourier space.


\subsection{Approach to linear stability analysis} \label{approach}

The base flow needs to be selected before conducting the linear stability analysis. We note that an archetypal structure of a laminar flow with convection in a horizontal channel heated from below is a superposition of the streamwise flow $u_x(y,z)$ and one or several streamwise-uniform convection rolls $(u_y(y,z),u_z(y,z))$. At high $Ha$, the structure can be modified by the magnetic field and replaced by a 3D structure with the rolls aligned with the magnetic field and, thus, $x$-dependent velocity and temperature at high $Ha$. Following \citet{Zikanov13} and \citet{Zhang14}, we treat the problem as that of the instability of the laminar steady-state streamwise-uniform base flow $\boldsymbol{U}(y,z)$, $\Theta(y,z)$, $P(y,z)$ to $x$-dependent perturbations.

The base flow is calculated by artificially imposing uniformity in the streamwise direction, i.e. by applying $x$-averaging after every time step. In order to assure that a fully developed state of the base flow is reached, each solution is computed for sufficiently long time. Long evolution, in some regimes up to $1000$ time units, is typically required in order to arrive at this state. No unsteady base flow solutions have been detected in the studied range of parameters. The steady-state solutions are discussed in section \ref{results1}.
  
  The linear stability analysis is conducted using a modified version of the numerical model described in section \ref{physmod3}. We follow evolution of perturbations - solutions of the equations linearized around the base flow $\boldsymbol{U}(y,z)$, $\Theta(y,z)$, $P(y,z)$. Individual Fourier modes determined by their streamwise wavelength $\lambda$ are computed. This is practically achieved by setting the length of the computational domain to $\lambda$ and filtering out all the Fourier modes except the zero mode corresponding to the base flow and the first mode corresponding to the perturbations of wavelength $\lambda$. All simulations start with random noise distributions of velocity and temperature. 
  
  The linear instability is identified by the exponential growth of the perturbations with the growth rate determined as
  
\begin{equation} \label{eq_gamma}
    \gamma = \frac{1}{2E'}\frac{dE'}{dt},
\end{equation} where $E'=\langle f^2\rangle$, $\langle \ldots \rangle$ stands for volume averaging, and  $f$ stands for perturbations of a velocity component or temperature. The growth rate coefficient is recorded after its values computed for all three velocity components and temperature coincide with each other and remain constant within the third digit after the decimal point for at least $100$ time units. The results of the linear stability analysis are presented in sections \ref{results3} and \ref{results4}.


\subsection{Grid sensitivity study} \label{gridstudy}

The grid sensitivity study has been conducted for the base flow. A detailed description of the various flow regimes is provided in section \ref{results1}. For the present discussion, it is sufficient to say that accurate resolution of the internal flow structure, along with two boundary layers: the Hartmann layers of thickness $\delta_{Ha} \sim {\it Ha}^{-1}$ at the vertical walls and the Shercliff layers of thickness $\delta_{Sh} \sim {\it Ha}^{-1/2}$ at the top and bottom walls, is critically important for accurate representation of the flow behavior. 

As an example, the results obtained at $Ha = 1200$, $Gr = 10^{8}$ are presented in table \ref{table1}. In a fully developed steady-state flow, the integrated Lorentz and buoyancy forces are zero. The wall friction must be balanced by the driving pressure gradient according to
\begin{equation} \label{wallfric1}
    \frac{d\hat{p}}{dx} = A^{-1}\left(\tau_{Ha}+\tau_{Sh}\right),
\end{equation} where $\tau_{Ha}$ and $\tau_{Sh}$ are the computed values of the integrated friction forces at the
Hartmann and Shercliff walls of the duct, respectively, expressed as
\begin{equation} \label{wallfric2}
\tau_{Ha}=\tau_y  =  \frac{1}{Re}\sum_{y=\pm 1}\int_{-1/\Gamma}^{1/\Gamma}\frac{\partial U_x}{\partial y}dz, \ \ \ \  
\tau_{Sh}=\tau_z  =  \frac{1}{Re}\sum_{z=\pm 1/\Gamma}\int_{-1}^1\frac{\partial U_x}{\partial z}dy.
\end{equation}

Values of $\tau_{Ha}$, $\tau_{Sh}$ and the error $\epsilon$, with which the computed solution satisfies (\ref{wallfric1}), found on various grids are compared in table \ref{table1}. On the basis of these data, we conclude that the grid with $N_y \times N_z = 192 \times 96$, $A_y = 4.0$ and $A_z = 2.0$ is sufficient. The maximum and minimum grid steps of such a grid are $\Delta y_{min}\approx 0.0001$, $\Delta y_{max}\approx 0.042$, $\Delta z_{min} \approx 0.0009$, $\Delta z_{max} \approx 0.012$. The Hartmann and Shercliff layers are resolved by, respectively, $9$ and $32$ grid points.


\begin{table}
  \begin{center}
  \begin{tabular}{cccccccccc}

    $N_y$ & $N_z$ & $A_y$ & $A_z$ & $-\tau_{Ha}$ & $-\tau_{Sh}$ & $\epsilon$ & $N_{Ha}$ & $N_{Sh}$ \\ [3pt]
  \hline \\ [3pt]
    
    128 &  32   & 4.0 & 2.0 & 0.23425 & 0.02914 & 0.00074 &  6 & 10 \\
    128 &  64   & 4.0 & 2.0 & 0.23460 & 0.26417 & 0.00070 &  6 & 21 \\
    128 &  64   & 4.3 & 2.0 & 0.23477 & 0.26387 & 0.00025 &  8 & 21 \\
    128 &  96   & 4.0 & 2.0 & 0.23468 & 0.26407 & 0.00069 &  6 & 32 \\
    128 &  96   & 4.3 & 2.0 & 0.23484 & 0.26377 & 0.00024 &  8 & 32 \\
    192 &  96   & 4.0 & 2.0 & 0.23459 & 0.26356 & 0.00030 &  9 & 32 \\
    192 &128   & 4.0 & 2.0 & 0.23462 & 0.26347 & 0.00029 &  9 & 43 \\

  \end{tabular}
  \end{center}
  \caption{Grid sensitivity study conducted for $Ha = 1200$ and $Gr = 10^{8}$: $\tau_{Ha}$ and $\tau_{Sh}$ are the wall friction forces, $\epsilon$ is the absolute error of the balance (\ref{wallfric1}). The number of grid points inside the Hartmann and Shercliff boundary layers are $N_{Ha}$ and $N_{Sh}$, respectively.}
  \label{table1}
\end{table}

The parameters of the grids in the entire studied parameter range of $Ha$ and $Gr$ have been determined in the same way. It has been found that the value of $Gr$ does not affect the selection at fixed $Ha$. This effect can be explained by the presence of very strong magnetic fields which fully suppress transverse circulation (see section \ref{results1} for a discussion). The summary of the grids used in the simulations is presented in table \ref{table2}.

A grid sensitivity study has also been conducted to determine the minimum number of grid points $N_x$ required in the linear stability analysis. It has been found that the growth of linear unstable modes is accurately reproduced at $N_x = 32$ for modes with $\lambda \lesssim 2$, which needs to be increased to $N_x = 64$ for greater $\lambda$. 

Computational domains of length $L_x = 4\pi$ or $L_x = 2\pi$ are used in DNS. As we will see below, these lengths are substantially larger than the streamwise wavelength of the fastest growing instability modes. The flow structures have been accurately resolved with, respectively, $384$ or $192$ grid points in the $x-$direction.

The time steps adjusted to secure numerical instability and, thus, varying with $Ha$ and $Gr$, but never exceeding $2.5 \times 10^{-3}$, are used in linear stability and DNS simulations.


\begin{table}
  \begin{center}
  \begin{tabular}{cccccccccc}

    $Ha$ & $N_y$ & $N_z$ & $A_y$ & $A_z$  & $N_{Ha}$ & $N_{Sh}$ \\ [3pt]
  \hline \\ [3pt]

    1000    & 192 & 96 & 4.0 & 2.0 &  9 & 32  \\
    2000    & 256 & 96 & 4.0 & 2.0 &  8 & 28  \\
    3000    & 384 & 96 & 4.0 & 2.0 &  9 & 24 \\
    4000    & 384 & 96 & 4.0 & 2.0 &  7 & 22 \\
    5000    & 512 & 96 & 4.0 & 2.0 &  8 & 20  \\
    6000    & 512 & 96 & 4.0 & 2.0 &  7 & 19 \\
    7000    & 512 & 96 & 4.3 & 2.0 &  9 & 18 \\
    8000    & 512 & 96 & 4.3 & 2.0 &  8 & 17 \\
    9000    & 512 & 96 & 4.3 & 2.0 &  7 & 16 \\
    10000  & 512 & 96 & 4.3 & 2.0 &  6 & 15  \\

  \end{tabular}
  \end{center}
  \caption{Parameters of the computational grids used in the simulations for $Gr = 10^8 - 10^{10}$. The number of grid points inside the Hartmann and Shercliff boundary layers are $N_{Ha}$ and $N_{Sh}$, respectively.}
  \label{table2}
\end{table}


 \section{\textbf{Results}} \label{results}
 
 
  \subsection{Base flow} \label{results1}
  
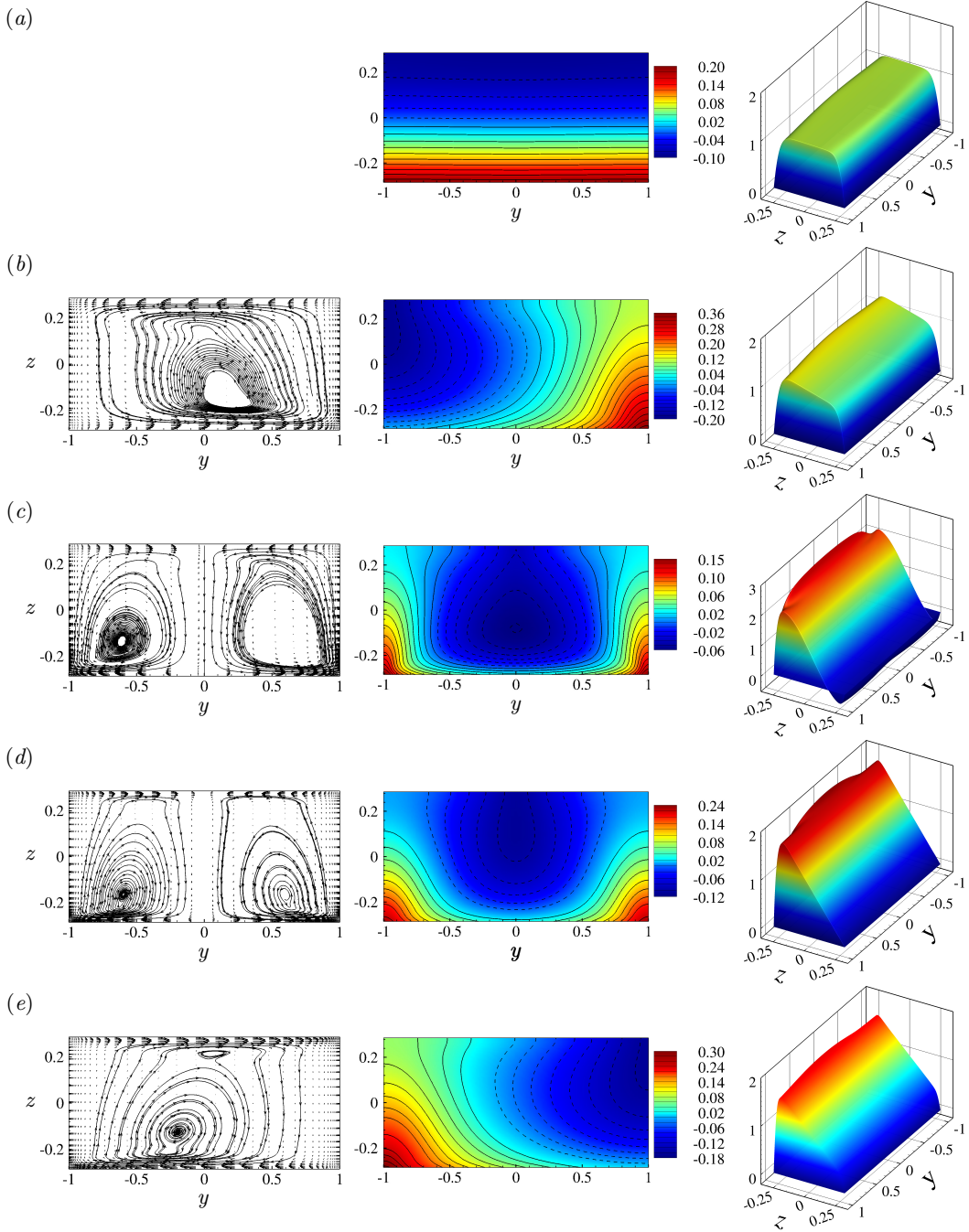
\begin{figure}
	\centering 
	
\begin{tikzpicture}


\node (img1a) {\includegraphics[scale=0.225]{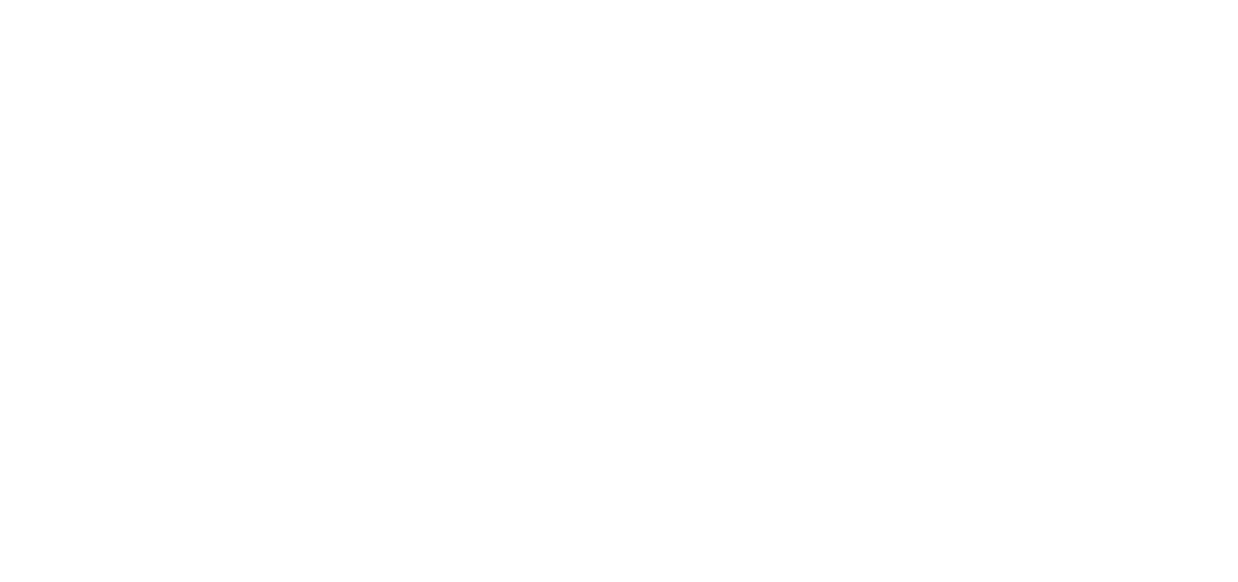}}; 
  	\node[left=of img1a, xshift=1.15cm ,yshift=1.5cm,rotate=0,font=\color{black}] {({\it a})};
	
\node [right=of img1a, xshift=-1.40cm, yshift=0.00cm]  (img2a)  {\includegraphics[scale=0.260]{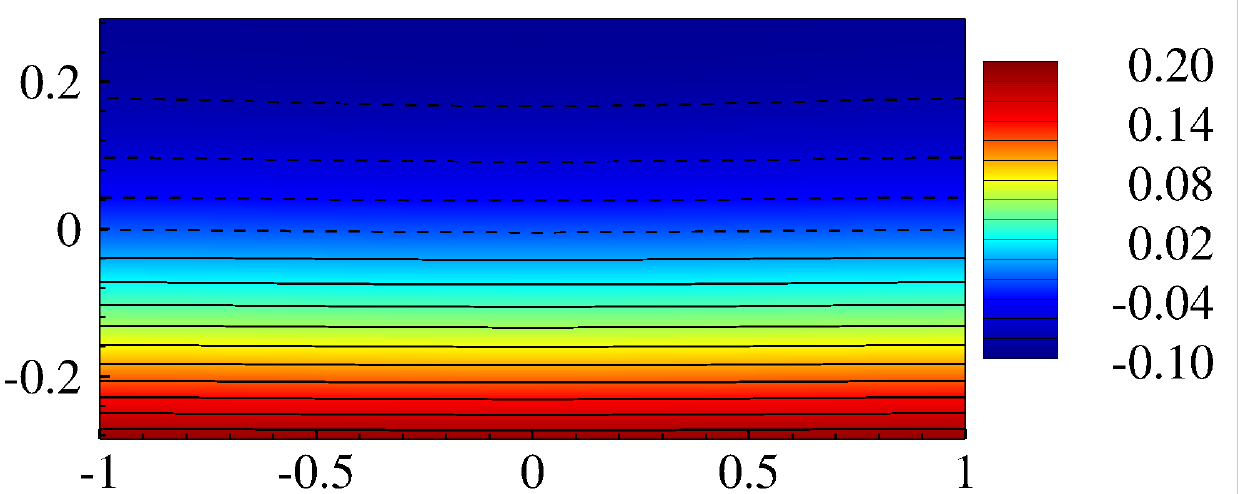}};
	\node[below=of img2a, xshift=-0.40cm ,yshift=1.1cm,rotate=0,font=\color{black}] {{\it y}};
	
\node [right=of img2a, xshift=-1.30cm, yshift=0.00cm] (img3a)  {\includegraphics[scale=0.170]{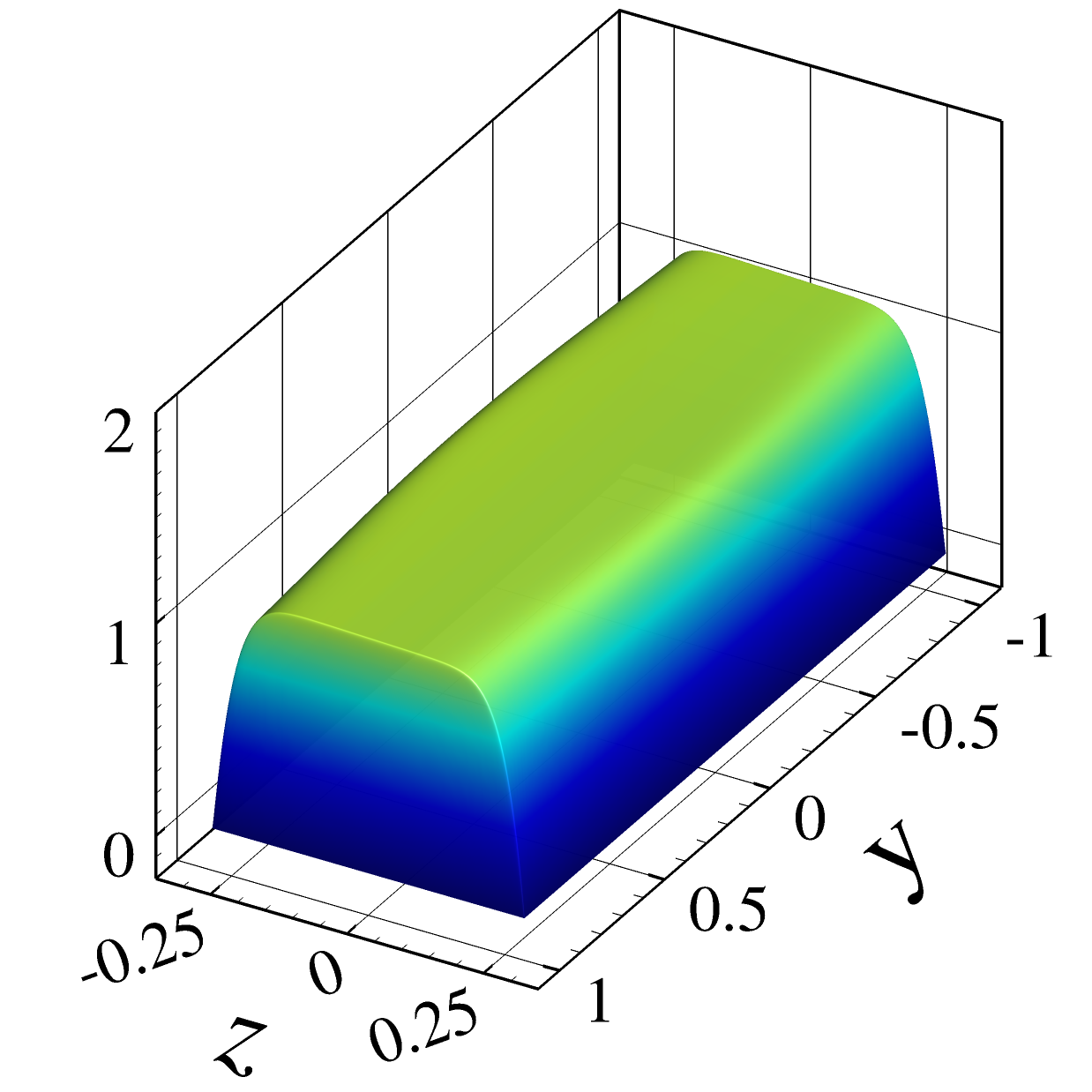}};


\node [below=of img1a, yshift=-0.1cm] (img1b)  {\includegraphics[scale=0.225] {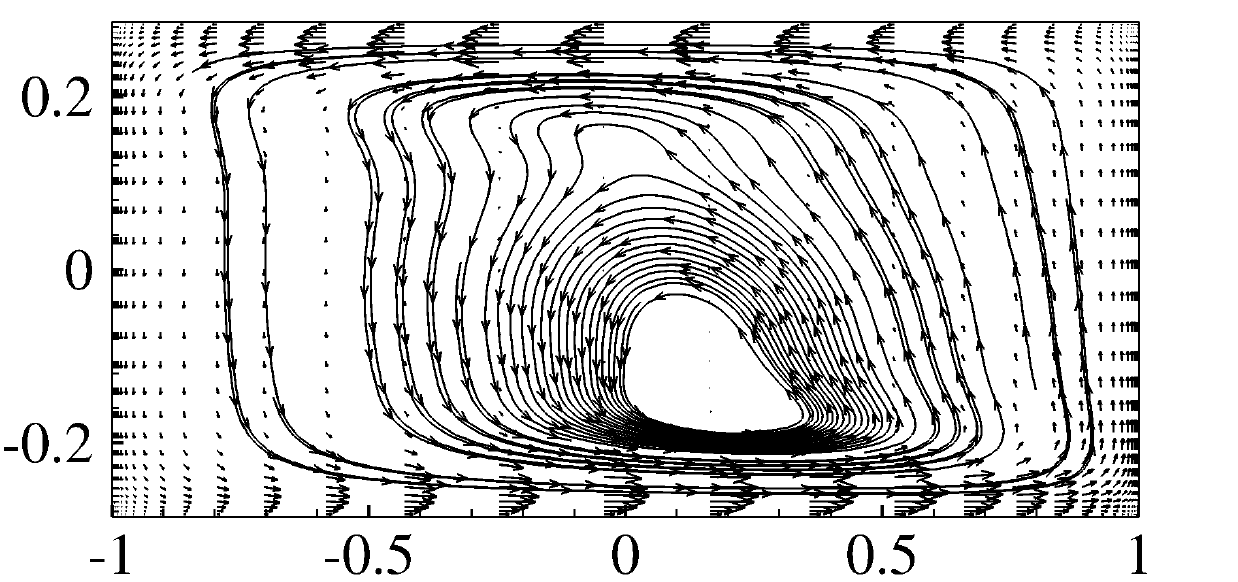}};
  	\node[left=of img1b, xshift=1.15cm ,yshift=1.5cm,rotate=0,font=\color{black}] {({\it b})};
	\node[left=of img1b, xshift=1.15cm ,yshift=0.1cm,rotate=0,font=\color{black}] {{\it z}};
	\node[below=of img1b, xshift=0.00cm ,yshift=1.1cm,rotate=0,font=\color{black}] {{\it y}};
		
\node [right=of img1b, xshift=-1.40cm, yshift=0.00cm]  (img2b)  {\includegraphics[scale=0.260]{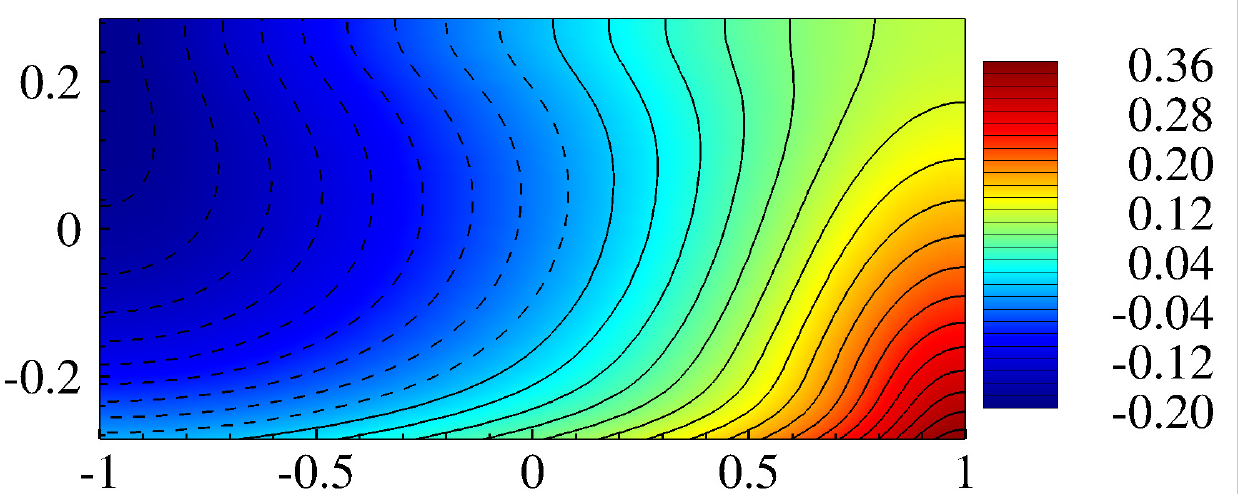}};
	\node[below=of img2b, xshift=-0.40cm ,yshift=1.1cm,rotate=0,font=\color{black}] {{\it y}};
	
\node [right=of img2b, xshift=-1.30cm, yshift=0.00cm] (img3b)  {\includegraphics[scale=0.170]{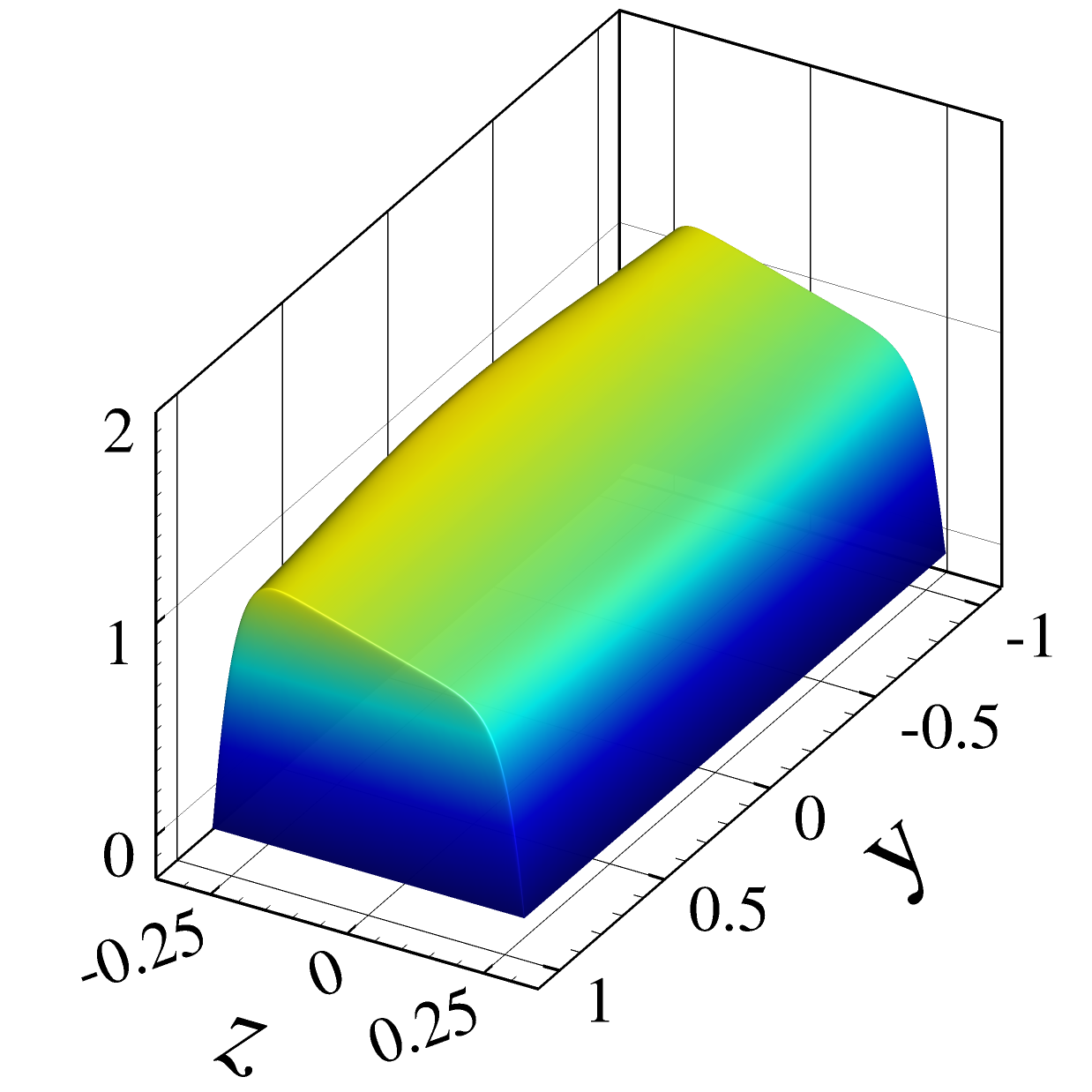}}; 


\node [below=of img1b, yshift=-0.1cm] (img1c)  {\includegraphics[scale=0.225] {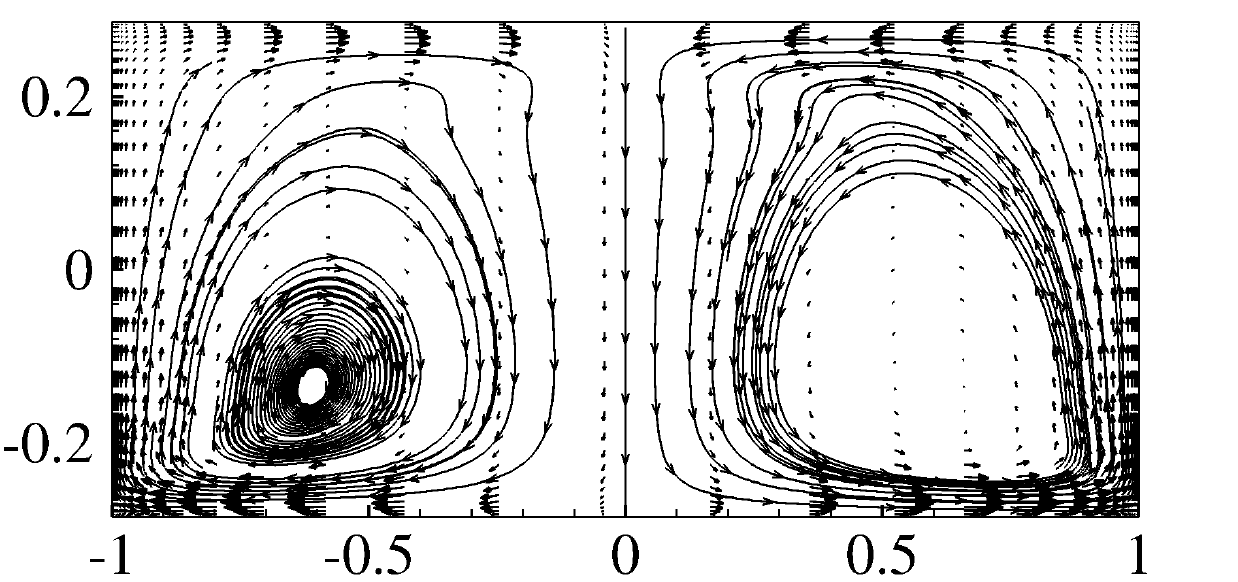}};
  	\node[left=of img1c, xshift=1.15cm ,yshift=1.5cm,rotate=0,font=\color{black}] {({\it c})};
	\node[left=of img1c, xshift=1.15cm ,yshift=0.1cm,rotate=0,font=\color{black}] {{\it z}};
	\node[below=of img1c, xshift=0.00cm ,yshift=1.1cm,rotate=0,font=\color{black}] {{\it y}};
	
\node [right=of img1c, xshift=-1.40cm, yshift=0.00cm]  (img2c)  {\includegraphics[scale=0.260]{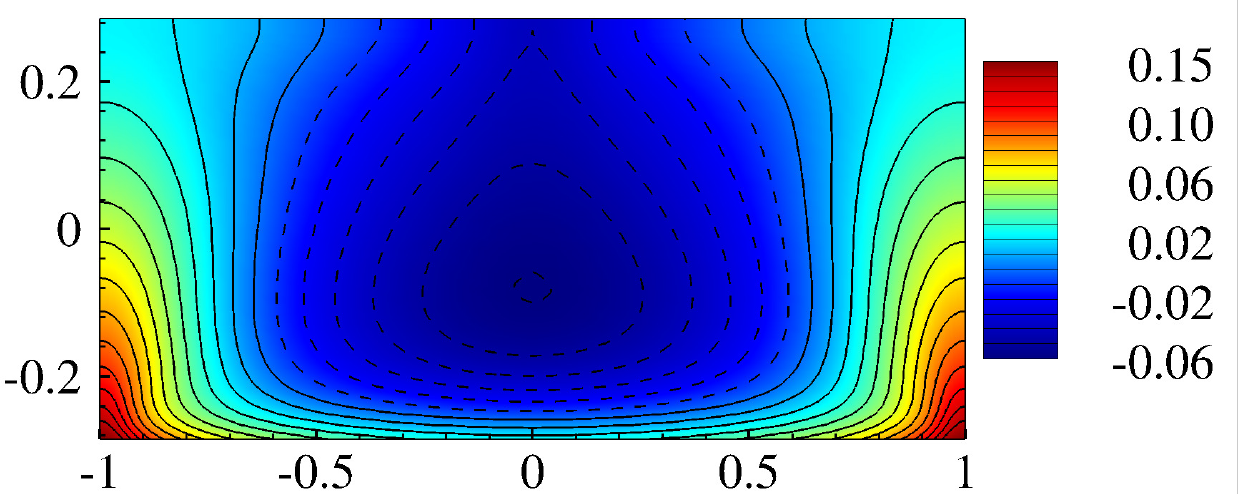}};
	\node[below=of img2c, xshift=-0.40cm ,yshift=1.1cm,rotate=0,font=\color{black}] {{\it y}};

\node [right=of img2c, xshift=-1.30cm, yshift=0.00cm] (img3c)  {\includegraphics[scale=0.170]{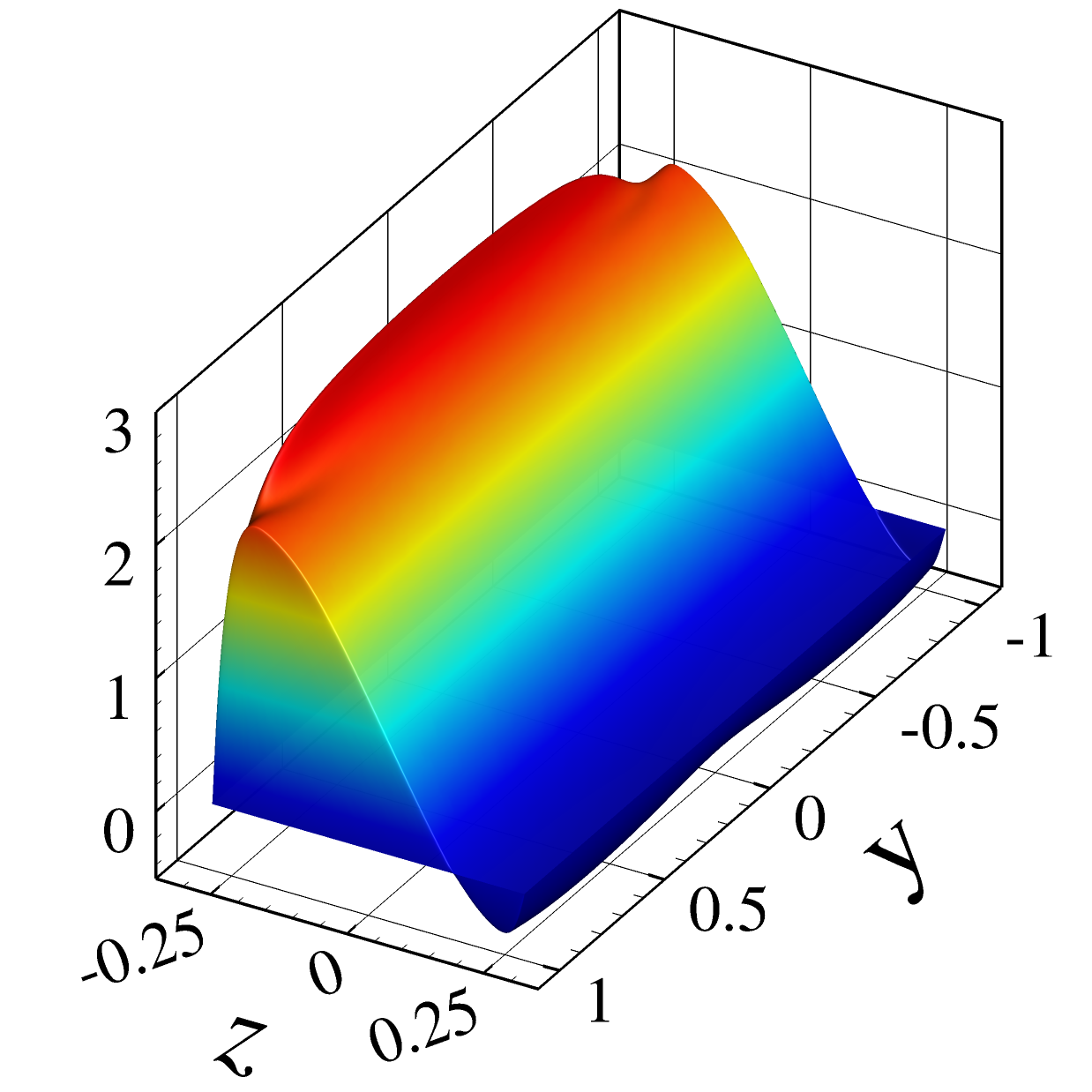}}; 


\node [below=of img1c, yshift=-0.1cm] (img1d)  {\includegraphics[scale=0.225] {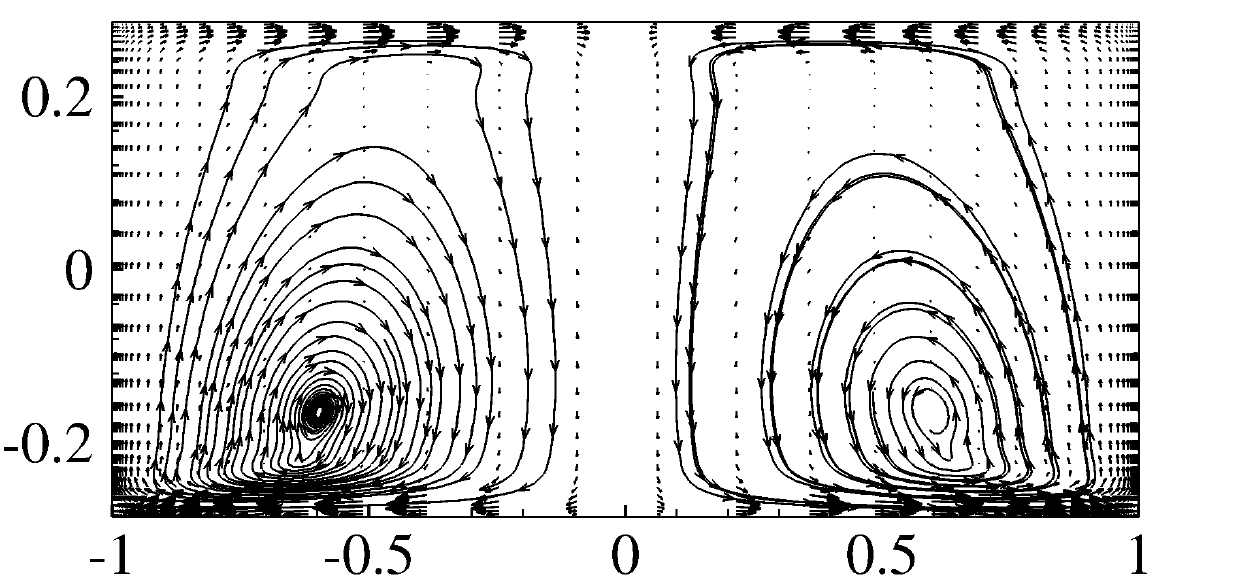}};
  	\node[left=of img1d, xshift=1.15cm ,yshift=1.5cm,rotate=0,font=\color{black}] {({\it d})};
	\node[left=of img1d, xshift=1.15cm ,yshift=0.1cm,rotate=0,font=\color{black}] {{\it z}};
	\node[below=of img1d, xshift=0.00cm ,yshift=1.1cm,rotate=0,font=\color{black}] {{\it y}};
	
\node [right=of img1d, xshift=-1.40cm, yshift=0.00cm]  (img2d)  {\includegraphics[scale=0.260]{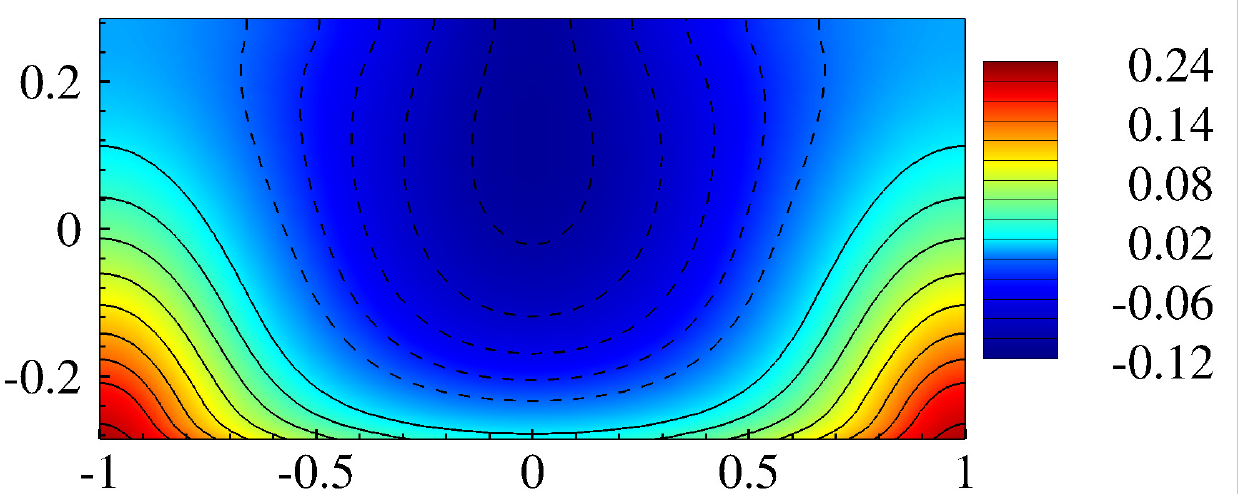}};
	\node[below=of img2d, xshift=-0.40cm ,yshift=1.1cm,rotate=0,font=\color{black}] {{\it y}};
	
\node [right=of img2d, xshift=-1.30cm, yshift=0.00cm] (img3d)  {\includegraphics[scale=0.170]{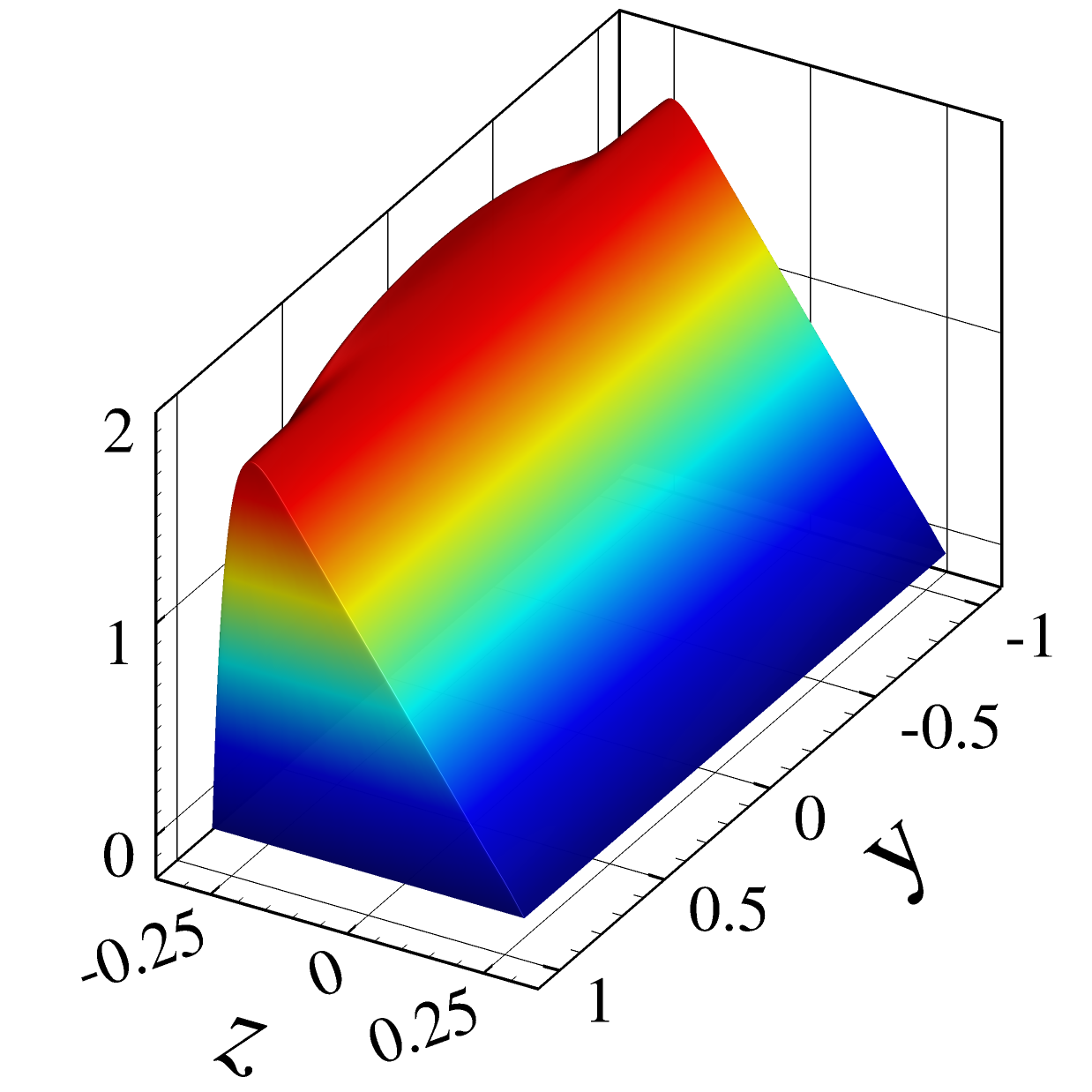}}; 


\node [below=of img1d, yshift=-0.1cm] (img1e)  {\includegraphics[scale=0.225] {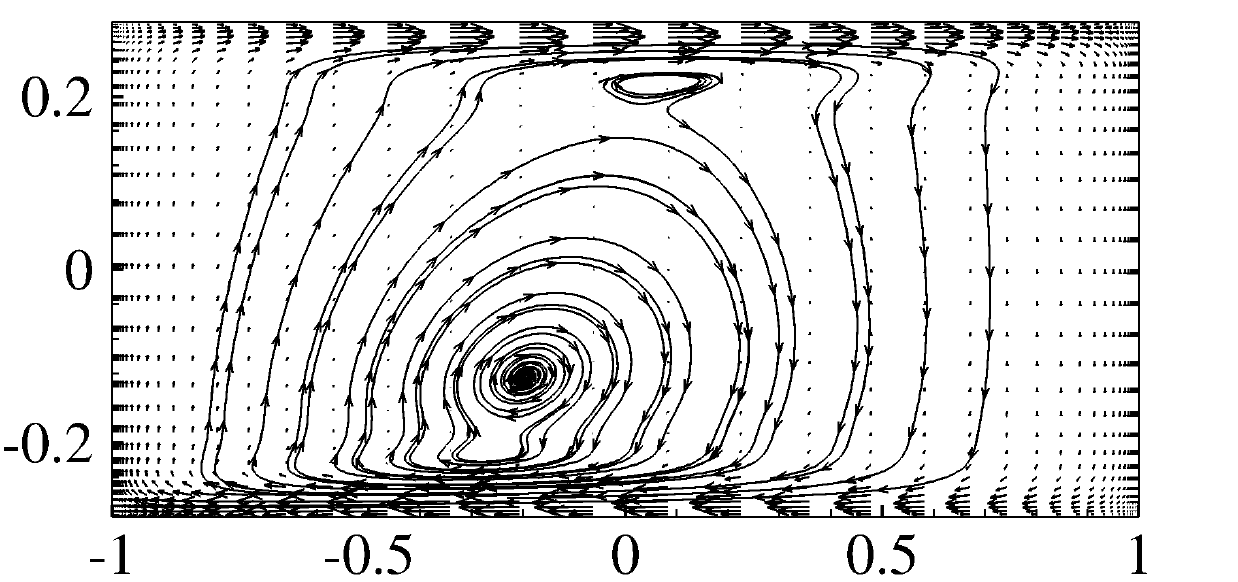}};
  	\node[left=of img1e, xshift=1.15cm ,yshift=1.5cm,rotate=0,font=\color{black}] {({\it e})};
	\node[left=of img1e, xshift=1.15cm ,yshift=0.1cm,rotate=0,font=\color{black}] {{\it z}};
	\node[below=of img1e, xshift=0.00cm ,yshift=1.1cm,rotate=0,font=\color{black}] {{\it y}};
	
\node [right=of img1e, xshift=-1.40cm, yshift=0.00cm]  (img2e)  {\includegraphics[scale=0.260]{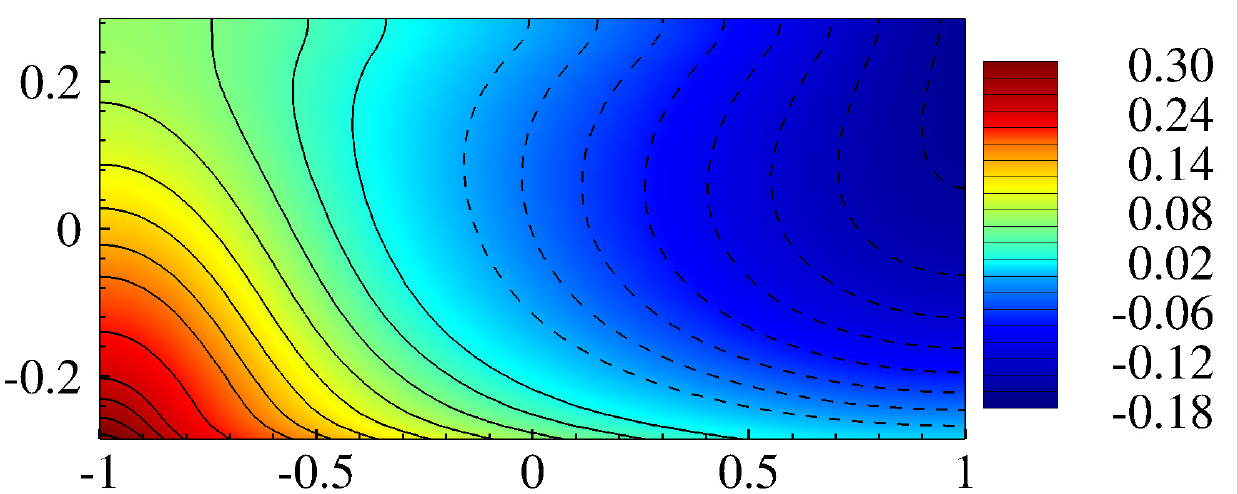}};
	\node[below=of img2d, xshift=-0.40cm ,yshift=1.1cm,rotate=0,font=\color{black}] {{\it y}};
	
\node [right=of img2e, xshift=-1.30cm, yshift=0.00cm] (img3e)  {\includegraphics[scale=0.170]{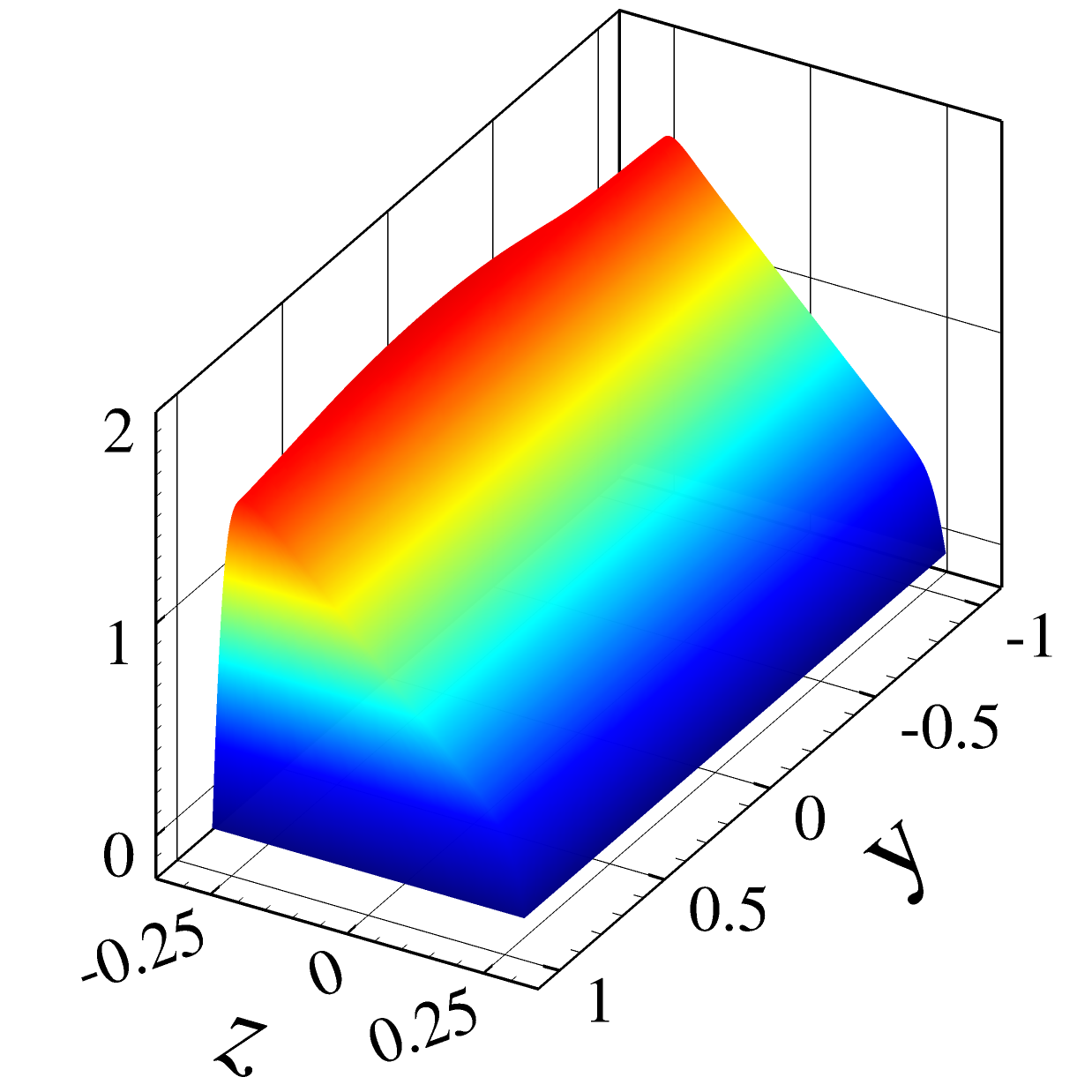}};

\end{tikzpicture}

  \caption{Base flow at $Ha = 1000$, $Gr = 10^8$ ($a$), $Ha = 1000$, $Gr = 10^9$ ($b$), $Ha = 1000$, $Gr = 10^{10}$ ($c$), $Ha = 2000$, $Gr = 10^{10}$ ($d$) and $Ha = 3000$, $Gr = 10^{10}$ ($e$). Vector fields and streamlines of transverse circulation ($u_y$, $u_z$) are shown in the left column (not in ($a$), since the velocity’s amplitude is virtually zero in this case). The middle and right columns show distributions of temperature $\Theta$ and streamwise velocity $U_x$, respectively. Solid and dashed isolines in the middle column indicate positive and negative values, respectively. The wall heating is at $z = -0.2857$, and the magnetic field is in the $y$-direction.}

\label{fig2}
\end{figure}

\begin{figure}
	\centering 

\raisebox{11em}	
	
\begin{tikzpicture}


\node (img2a) {\includegraphics[scale=0.165]{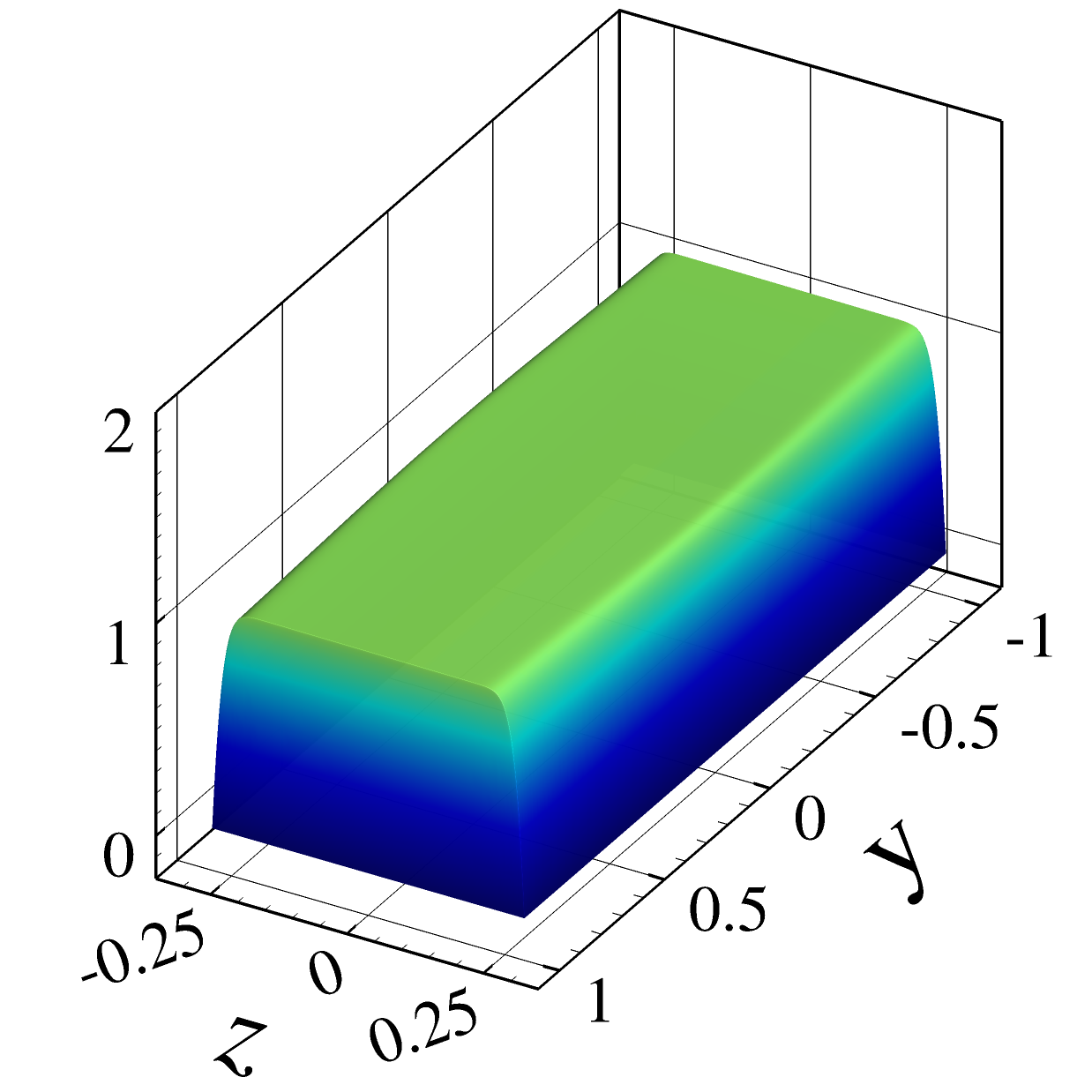}}; 
  	\node[left=of img2a, xshift=2.00cm ,yshift=1.65cm,rotate=0,font=\color{black}] {({\it a})};
	
	
\node [right=of img2a, xshift=-1.36cm, yshift=0.00cm]  (img2b)  {\includegraphics[scale=0.165]{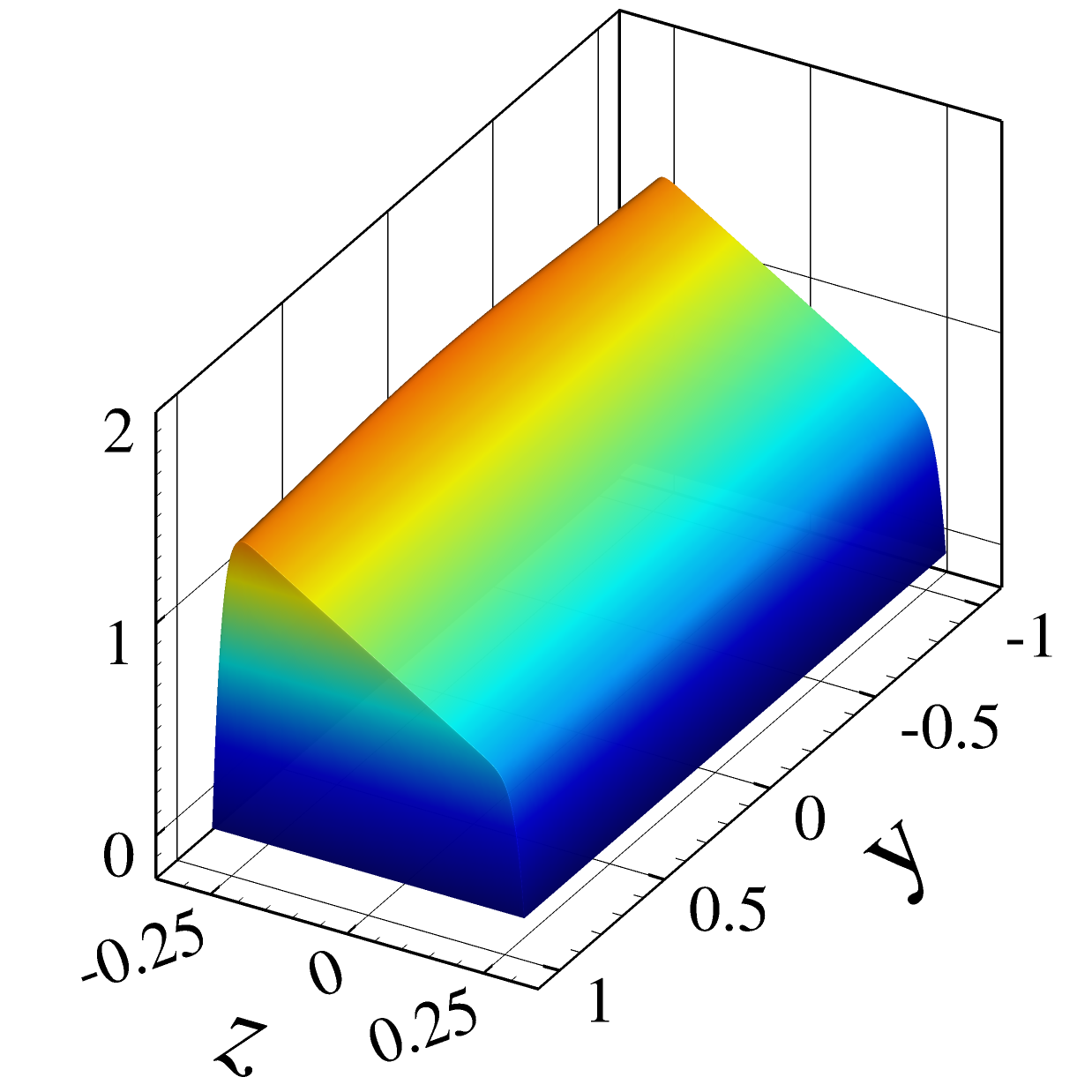}};
  	\node[left=of img2b, xshift=2.00cm ,yshift=1.65cm,rotate=0,font=\color{black}]{({\it b})};
	
	
\node [right=of img2b, xshift=-1.36cm, yshift=-0.05cm] (img2c)  {\includegraphics[scale=0.165]{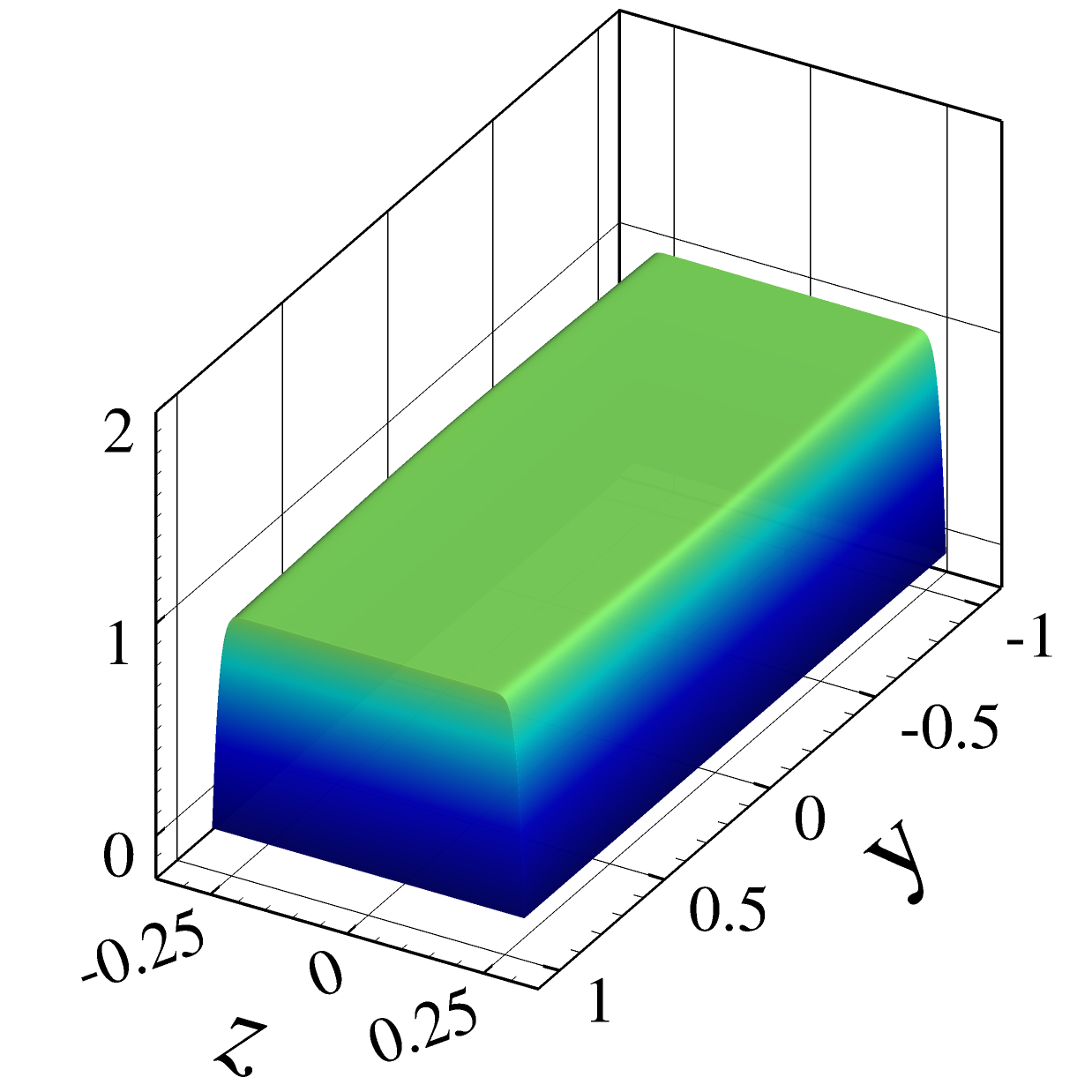}};
  	\node[left=of img2c, xshift=2.00cm ,yshift=1.70cm,rotate=0,font=\color{black}]{({\it c})};
	
	
\node [right=of img2c, xshift=-1.36cm, yshift=-0.05cm] (img2d)  {\includegraphics[scale=0.165]{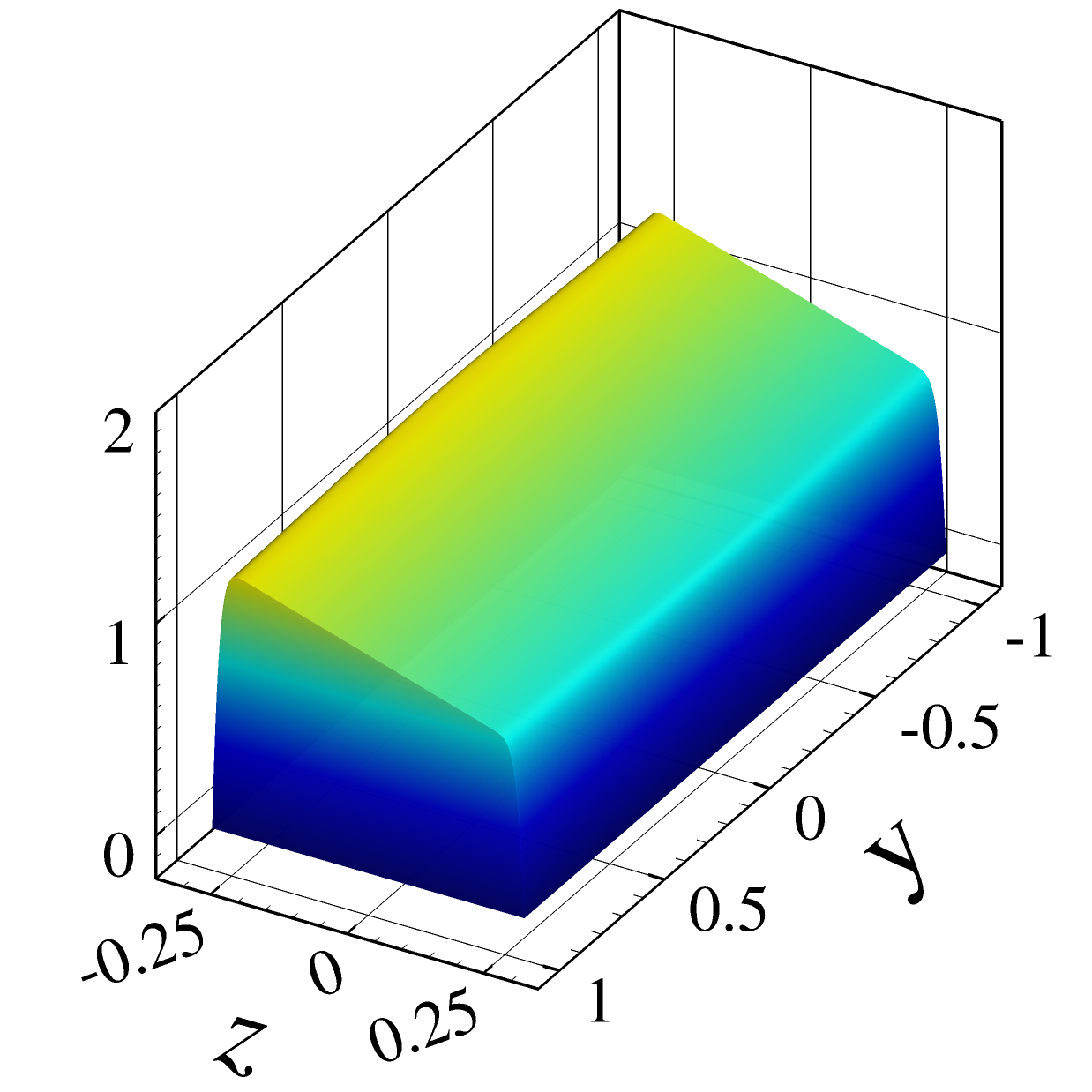}};
  	\node[left=of img2d, xshift=2.00cm ,yshift=1.72cm,rotate=0,font=\color{black}]{({\it d})};

\end{tikzpicture}

  \caption{Q2D base flow at $Ha = 5000$, $Gr = 10^9$ ($a$), $Ha = 5000$, $Gr = 10^{10}$ ($b$), $Ha = 10000$, $Gr = 10^9$ ($c$) and $Ha = 10000$, $Gr = 10^{10}$ ($d$). Distribution of streamwise velocity $u_x$ is shown. The wall heating is at $z = -0.2857$, and the magnetic field is in the $y$-direction.}
  
\label{fig3}
\end{figure}
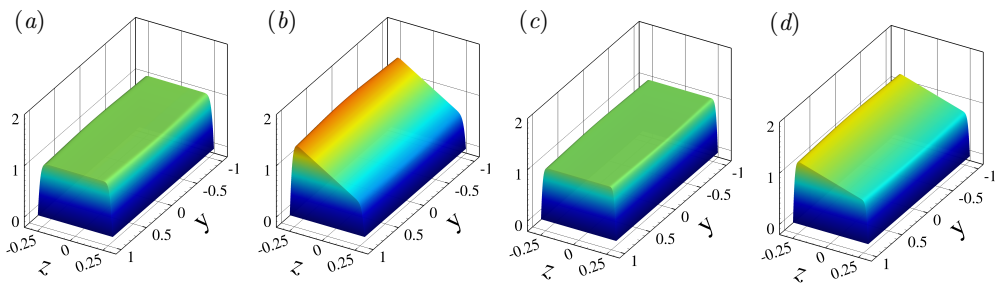

  	The structure of the base flow, as it is defined in section \ref{approach}, for several typical cases is illustrated in figures \ref{fig2} and \ref{fig3}. The results for all the completed simulations are summarized in table \ref{tablebase1}. The table shows the type of the flow for a particular regime (Q2D or 3D flows to be discussed shortly), the maximum and minimum values of $U_x$, integral quantities, such as the wall friction force $d\hat{p}/dx$, the volume-averaged kinetic energies of streamwise and transverse velocities
\begin{equation} \label{ke_eq1}
E_x  =  A^{-1}\int_{-1/\Gamma}^{1/\Gamma}\int_{-1}^1 U_x^2dydz, \  \ E_t  =  A^{-1}\int_{-1/\Gamma}^{1/\Gamma}\int_{-1}^1 \left(U_y^2+U_z^2\right)dydz,
\end{equation} and the mean square of temperature perturbations

\begin{equation} \label{ke_eq1}    
E_{\theta}  =  A^{-1}\int_{-1/\Gamma}^{1/\Gamma}\int_{-1}^1 \Theta^2 dydz.
\end{equation}

Similar data for flows with lower values of $Gr$ and $Ha$ in a duct with $\Gamma = 1$ can be found in \citep{Zhang14}.

The flow structure is predominantly determined by the effect of magnetoconvection. Similarly to the findings of \citet{Zhang14}, we observe two regimes of the base flow depending on whether $Gr$ is smaller or larger than a certain threshold $Gr^{\ast}(Ha)$. The Q2D regime observed at $Gr < Gr^{\ast}$ is characterized by the transverse convection-induced circulation entirely suppressed by the strong magnetic field. The distributions of the temperature and streamwise velocity are nearly one-dimensional outside of the Hartmann boundary layers ($\Theta \approx \Theta(z)$ and $U_x \approx U_x(z)$). In the absence of transverse circulation, the distributions of temperature are determined by the balance between the heat conduction and the heat convection by $U_x$. Examples of this regime are shown in figure \ref{fig2}$a$ and figure \ref{fig3}. 

The 3D regime observed at $Gr > Gr^{\ast}$($Ha$), when the strength of the magnetic field is insufficient to suppress convection circulation, is characterized by significant transverse flow and fully 2D variations of temperature. (see figure \ref{fig2}$b-e$).

To avoid confusion, it is pertinent to repeat the terminology here. The terms 3D and Q2D are used in this paper to describe the general flow transformation caused by the magnetic field, i.e. suppression of velocity and temperature gradients along the magnetic field lines in the core of the duct and formation of thin Hartmann boundary layers. The base flow, in which streamwise uniformity is also imposed, becomes, respectively, 2D and Q1D. 

We find the Q2D regime in the larger part of the explored range of $Ha$ and $Gr$ including the most interesting cases of large $Ha$ (see the rightmost column in table \ref{tablebase1}). The total friction force increases at stronger magnetic fields. The visible effect of convection is the asymmetry of the velocity profile, with $U_x$ larger in the bottom than in the top half (see, e.g., figure \ref{fig3}$b,d$). The cause of the asymmetry has been explained in section \ref{physmod1}.
 

\begin{table}
  \begin{center}
\def~{\hphantom{0}}
  \begin{tabular}{clcccccccc}
      $Ha$ & $Gr$ & $Gr/{\it Ha^2}$ & $-d\hat{p}/dx$ & $E_x$ & $E_t$ & $E_{\theta}$ & $u_{x,min}$ & $u_{x,max}$ & $\it {Regime}$ \\[3pt]
        \hline \\ [3pt]
     
    $1000$ & $10^8$ & $1.00 \times 10^2$ & $0.222$ & $1.050$ & $1.62 \times 10^{-8}$ & $3.59 \times 10^{-3}$ & $0.0$ & $1.139$ & $Q2D$ \\
    $1000$ & $10^9$ & $1.00 \times 10^3$ & $0.224$ & $1.056$ & $6.32 \times 10^{-3}$ & $9.31 \times 10^{-3}$ & $0.0$ & $1.249$ & $3D$ \\
    $1000$ & $10^{10}$ & $1.00 \times 10^4$ & $0.254$ & $1.892$ & $3.11 \times 10^{-2}$ & $7.11 \times 10^{-4}$ & $-0.402$ & $2.601$ & $3D$ \\[3pt]
    
    $2000$ & $10^8$ & $2.50 \times 10^1$ & $0.430$ & $1.035$ & $5.42 \times 10^{-9}$ & $3.61 \times 10^{-3}$ & $0.0$ & $1.079$ & $Q2D$  \\
    $2000$ & $10^9$ & $2.50 \times 10^2$ & $0.431$ & $1.037$ & $4.96 \times 10^{-7}$ & $3.34 \times 10^{-3}$ & $0.0$ & $1.153$ & $Q2D$  \\
    $2000$ & $10^{10}$ & $2.50 \times 10^3$ & $0.447$ & $1.313$ & $8.30 \times 10^{-3}$ & $2.19 \times 10^{-3}$ & $0.0$ & $1.968$ & $3D$  \\[3pt]
    
    $3000$ & $10^8$ & $1.11 \times 10^1$ & $0.636$ & $1.029$ & $1.76 \times 10^{-9}$ & $3.62 \times 10^{-3}$ & $0.0$ & $1.063$ & $Q2D$   \\
    $3000$ & $10^9$ & $1.11 \times 10^2$ & $0.637$ & $1.030$ & $1.66 \times 10^{-7}$ & $3.45 \times 10^{-3}$ & $0.0$ & $1.114$ & $Q2D$   \\
    $3000$ & $10^{10}$ & $1.11 \times 10^3$ & $0.643$ & $1.172$ & $1.58 \times 10^{-2}$ & $4.48 \times 10^{-3}$ & $0.0$ & $1.683$ & $3D$  \\[3pt]
    
    $4000$ & $10^8$ & $6.25 \times 10^0$ & $0.842$ & $1.025$ & $3.85 \times 10^{-10}$ & $3.63 \times 10^{-3}$ & $0.0$ & $1.054$ & $Q2D$   \\
    $4000$ & $10^9$ & $6.25 \times 10^1$ & $0.842$ & $1.026$ & $3.70 \times 10^{-8}$ & $3.53 \times 10^{-3}$ & $0.0$ & $1.093$ & $Q2D$   \\
    $4000$ & $10^{10}$ & $6.25 \times 10^2$ & $0.847$ & $1.109$ & $2.75 \times 10^{-6}$ & $2.96 \times 10^{-3}$ & $0.0$ & $1.522$ & $Q2D$ \\[3pt]
    
    $5000$ & $10^8$ & $4.00 \times 10^0$ & $1.046$ & $1.022$ & $1.13 \times 10^{-10}$ & $3.63 \times 10^{-3}$ & $0.0$ & $1.048$ & $Q2D$   \\
    $5000$ & $10^9$ & $4.00 \times 10^1$ & $1.046$ & $1.023$ & $1.10 \times 10^{-8}$ & $3.57 \times 10^{-3}$ & $0.0$ & $1.080$ & $Q2D$   \\
    $5000$ & $10^{10}$ & $4.00 \times 10^2$ & $1.049$ & $1.078$ & $8.55 \times 10^{-7}$ & $3.16 \times 10^{-3}$ & $0.0$ & $1.425$ & $Q2D$ \\[3pt]
    
    $6000$ & $10^8$ & $2.78 \times 10^0$ & $1.251$ & $1.020$ & $1.00 \times 10^{-10}$ & $3.63 \times 10^{-3}$ & $0.0$ & $1.043$ & $Q2D$   \\
    $6000$ & $10^9$ & $2.78 \times 10^1$ & $1.251$ & $1.021$ & $9.79 \times 10^{-9}$ & $3.57 \times 10^{-3}$ & $0.0$ & $1.071$ & $Q2D$   \\
    $6000$ & $10^{10}$ & $2.78 \times 10^2$ & $1.254$ & $1.060$ & $7.75 \times 10^{-7}$ & $3.18 \times 10^{-3}$ & $0.0$ & $1.362$ & $Q2D$ \\[3pt]
    
    $7000$ & $10^8$ & $2.04 \times 10^0$ & $1.454$ & $1.019$ & $9.00 \times 10^{-11}$ & $3.63 \times 10^{-3}$ & $0.0$ & $1.040$ & $Q2D$   \\
    $7000$ & $10^9$ & $2.04 \times 10^1$ & $1.454$ & $1.019$ & $8.80 \times 10^{-9}$ & $3.58 \times 10^{-3}$ & $0.0$ & $1.064$ & $Q2D$   \\
    $7000$ & $10^{10}$ & $2.04 \times 10^2$ & $1.457$ & $1.048$ & $7.12 \times 10^{-7}$ & $3.20 \times 10^{-3}$ & $0.0$ & $1.315$ & $Q2D$ \\[3pt]
    
    $8000$ & $10^8$ & $1.56 \times 10^0$ & $1.658$ & $1.018$ & $8.12 \times 10^{-11}$ & $3.63 \times 10^{-3}$ & $0.0$ & $1.037$ & $Q2D$   \\
    $8000$ & $10^9$ & $1.56 \times 10^1$ & $1.658$ & $1.018$ & $7.96 \times 10^{-9}$ & $3.58 \times 10^{-3}$ & $0.0$ & $1.058$ & $Q2D$   \\
    $8000$ & $10^{10}$ & $1.56 \times 10^2$ & $1.661$ & $1.040$ & $6.56 \times 10^{-7}$ & $3.22 \times 10^{-3}$ & $0.0$ & $1.280$ & $Q2D$ \\[3pt]
    
    $9000$ & $10^8$ & $1.23 \times 10^0$ & $1.862$ & $1.017$ & $2.50 \times 10^{-11}$ & $3.63 \times 10^{-3}$ & $0.0$ & $1.035$ & $Q2D$   \\
    $9000$ & $10^9$ & $1.23 \times 10^1$ & $1.862$ & $1.017$ & $2.46 \times 10^{-9}$ & $3.60 \times 10^{-3}$ & $0.0$ & $1.054$ & $Q2D$   \\
    $9000$ & $10^{10}$ & $1.23 \times 10^2$ & $1.864$ & $1.035$ & $2.11 \times 10^{-7}$ & $3.34 \times 10^{-3}$ & $0.0$ & $1.252$ & $Q2D$ \\[3pt]
    
    $10000$ & $10^8$ & $1.00 \times 10^0$ & $2.067$ & $1.016$ & $2.31 \times 10^{-11}$ & $3.63 \times 10^{-3}$ & $0.0$ & $1.033$ & $Q2D$   \\
    $10000$ & $10^9$ & $1.00 \times 10^1$ & $2.067$ & $1.016$ & $2.28 \times 10^{-9}$ & $3.60 \times 10^{-3}$ & $0.0$ & $1.050$ & $Q2D$   \\
    $10000$ & $10^{10}$ & $1.00 \times 10^2$ & $2.068$ & $1.031$ & $1.97 \times 10^{-7}$ & $3.36 \times 10^{-3}$ & $0.0$ & $1.229$ & $Q2D$ \\[3pt]

  \end{tabular}
  \caption{Integral characteristics and type of the computed base flow states.}
 \label{tablebase1}
  \end{center}
\end{table}

The 3D regimes are only found in a limited range of moderate values of $Ha$ at $Gr = 10^9$ and $10^{10}$ (see table \ref{tablebase1}). The transverse circulation consist of a single roll (see figure \ref{fig2}$b,e$) or two symmetric rolls (see figure \ref{fig2}$c,d$). The single roll has no preferred circulation direction and may appear in the solution either as shown in the figure \ref{fig2}$b$ or as a symmetric reflection with respect to the vertical midplane (see figure \ref{fig2}$e$ for an example). The circulation causes a visibly 2D distributions of $\Theta$ and $U_x$, and, at the same values of $Gr$, a decrease of $E_\theta$ as a result of mixing (see table \ref{tablebase1}). We observe stronger top–bottom asymmetries or even formation of reverse flow in the top portion of the duct as the strength of convection increases at fixed $Ha$ (see figure \ref{fig2}$a-c$ for an example). As an illustration of the asymmetry the minimum and maximum values of $u_x$ are shown in table \ref{tablebase1}.

\begin{figure}
	\centering 
	
\begin{tikzpicture}

\node (img4) {\includegraphics[scale=0.375]{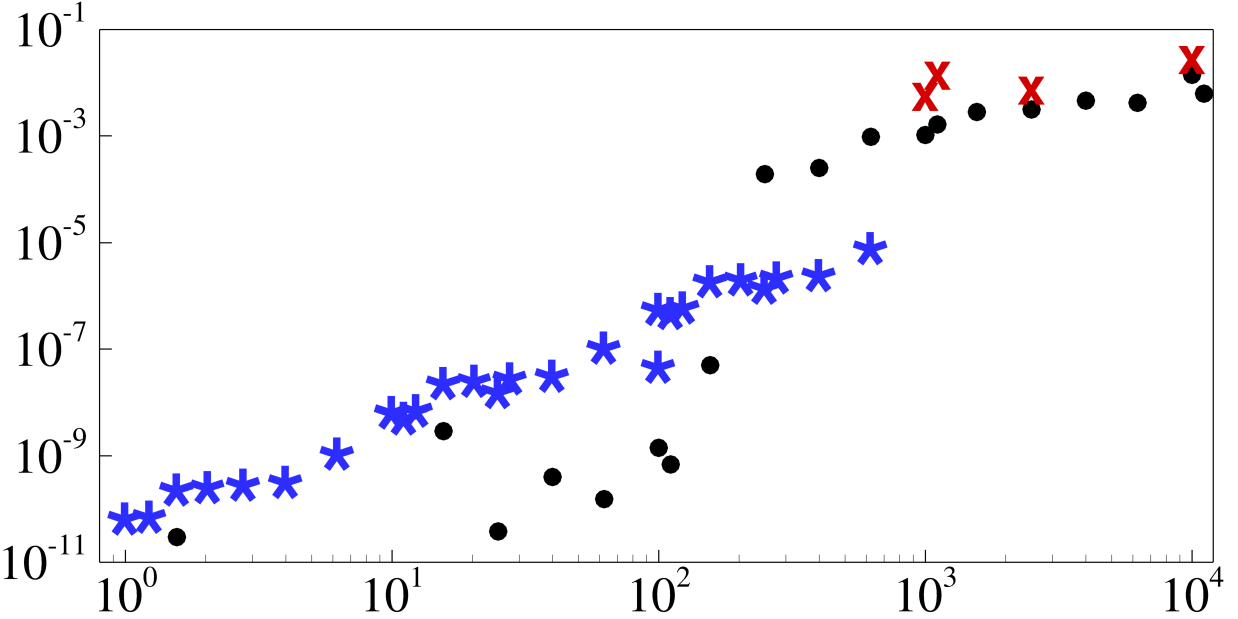}}; 
  	\node[left=of img4, xshift=1.20cm ,yshift=0.15cm,rotate=0,font=\color{black}] {$E_t$};
	\node[left=of img4, xshift=5.75cm ,yshift=-2.25cm,rotate=0,font=\color{black}] {$Gr/Ha^2$};
	
\node [left=of img4, xshift=2.75cm, yshift=0.75cm]  (img4_l)  {\includegraphics[scale=0.225]{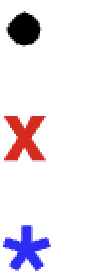}}; 
	\node[right=of img4_l, xshift=-1.25cm ,yshift=0.45cm,rotate=0,font=\color{black}] {\citet{Zhang14}};
	\node[right=of img4_l, xshift=-1.25cm ,yshift=0.05cm,rotate=0,font=\color{black}] {3D};
	\node[right=of img4_l, xshift=-1.25cm ,yshift=-0.40cm,rotate=0,font=\color{black}] {Q2D};

\end{tikzpicture}

  \caption{The average kinetic energy of transverse circulation in the base flow $E_t$ as a function of $Gr/Ha^2$. Circles indicate the numerical results of \citet{Zhang14} at $\Gamma = 1.0$. Stars and crosses indicate, respectively, Q2D and 3D regimes found in this work for the flow at $\Gamma = 3.5$. Values of $E_t$ and $Gr/Ha^2$ for each computed flow can be found in table \ref{tablebase1}.}

\label{fig4}
\end{figure}
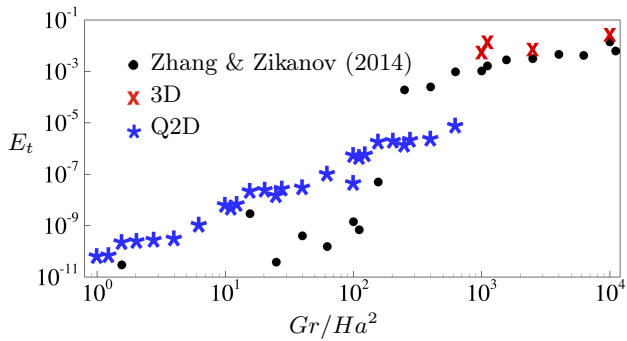

The classification of the flow regimes into Q2D and 3D can also be described in terms of the values of the average kinetic energy of transverse circulation $E_t$. Our data shown in table \ref{tablebase1} and figure \ref{fig4} are in a good qualitative agreement with the results of \citet{Zhang14}. The observed differences can be attributed to the substantially different studied ranges of $Ha$ and $Gr$ and different aspect ratio. 

We see in figure \ref{fig4} that, at a fixed Reynolds number considered in this work, the intensity of the transverse circulation is well approximated by a function of the single control parameter - the combination $Gr/Ha^2$. Analyzing the flow structures at various values of $E_t$ we find a clear demarcation between 3D and Q2D regimes. $E_t$ is greater than, approximately, $10^{-4}$ in 3D regimes. Q2D flows all have values of $E_t$ less than $10^{-6}$. 

Our interest in this study is primarily in the Q2D regimes. The 3D regimes are not considered in the rest of the paper.


 \subsection{Applicability of the SM82 model} \label{results2}
 
In this section, we investigate the applicability of the SM82 model to analysis of magnetoconvection instability.

\subsubsection{Base flow} 

For a streamwise-uniform, unidirectional, steady-state base flow, the SM82 model equations (\ref{eqsm1})$-$(\ref{eqsm3}) are reduced to a system of linear ordinary differential equations. The solution satisfying the boundary conditions is:
 
\begin{equation} \label{eqsm4}
	 {U_x}(z) = 
	\frac{Re}{Ha} \bigg( Cc_1 + \frac{A}{\Gamma}s_1 - Az - C \bigg),
\end{equation}

\begin{equation} \label{eqsm5}
	 \Theta(z) = 
	\frac{Re}{2Ha^2} \bigg( C\Gamma c_1 + A s_1 \bigg) - \frac{Re}{Ha} \frac{\Gamma}{4} \bigg( \frac{A}{3} z^3 + C z^2 \bigg) + a_1z,
\end{equation} where 

\begin{equation} \label{eqsm6}
	 a_1 = 
	- \frac{Re}{Ha^{3/2}} \frac{\Gamma}{2} \bigg( C t_1 + A t_1^{-1} \bigg) + \frac{Re}{2Ha} \bigg( \frac{A}{2\Gamma} + C \bigg),
\end{equation}

\begin{equation} \label{eqsm7}
	c_1 = \frac{ cosh(\sqrt{Ha}z) } { cosh(\sqrt{Ha}/\Gamma) } , \ \ \  s_1 = \frac{ sinh(\sqrt{Ha}z) } { sinh(\sqrt{Ha}/\Gamma) }, \ \ \ t_1 = tanh( \sqrt{Ha}/\Gamma ),
\end{equation}

\begin{equation} \label{eqsm7}
	A = \frac{\Gamma}{2RePr} \frac{Gr}{Re^2}, \ \ \ C = \frac{Ha}{Re\Gamma} \frac{1}{t_1/\sqrt{Ha} - 1/\Gamma}.
\end{equation}

\begin{figure}
	\centering 

	
\begin{tikzpicture}


\node (img5a1) {\includegraphics[scale=0.2]{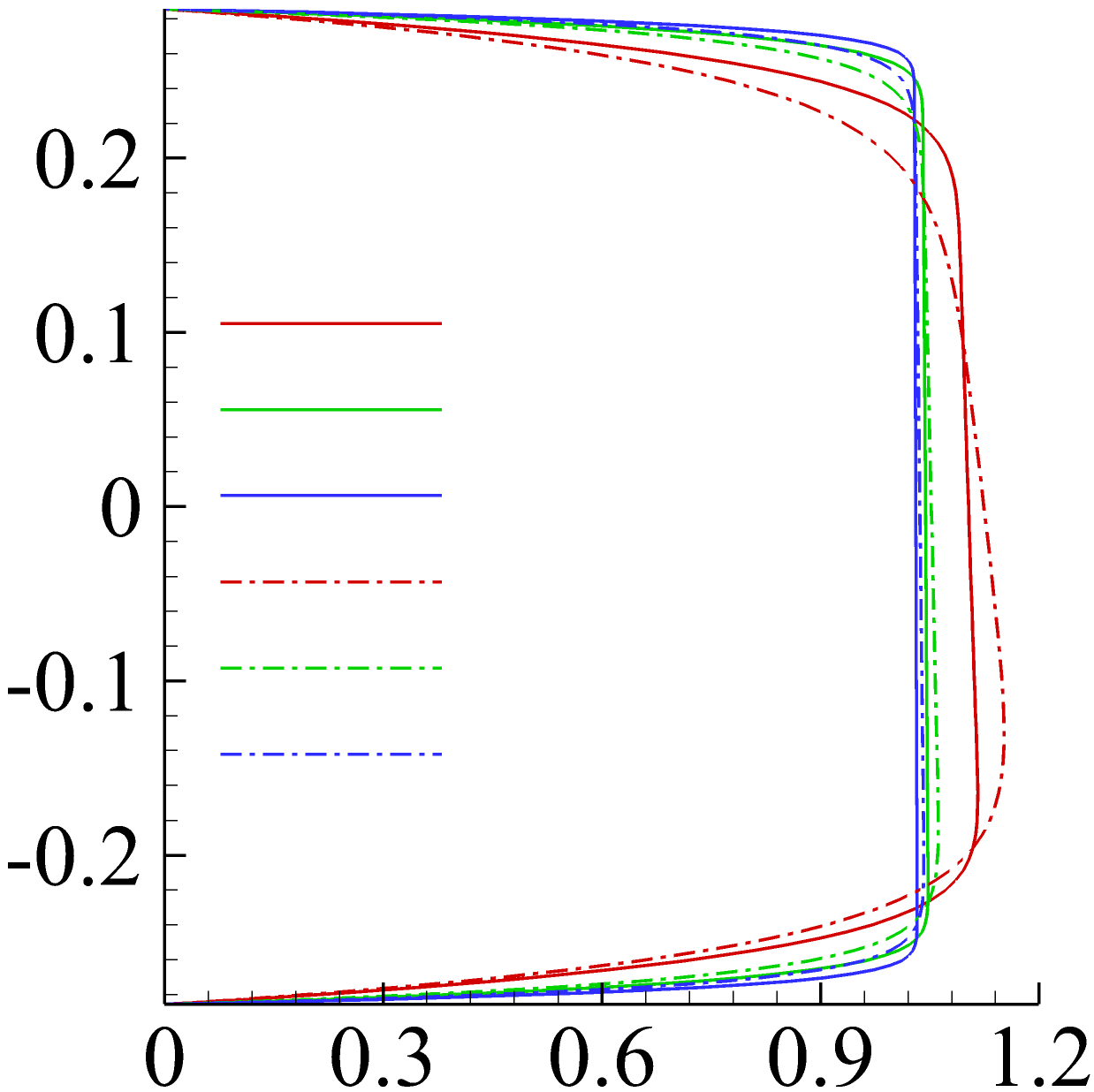}}; 
  	\node[left=of img5a1, xshift=1.25cm ,yshift=2.00cm,rotate=0,font=\color{black}] {({\it a})};
	
	 \node[left=of img5a1, xshift=4.25cm ,yshift=2.50cm,rotate=0,font=\color{black}] {$Gr = 10^{8}$};
	
	\node[left=of img5a1, xshift=1.15cm ,yshift=0.15cm,rotate=0,scale=1.15,font=\color{black}] {{\it z}};
	
	\node[below=of img5a1, xshift=0.25cm ,yshift=1.0cm,rotate=0,scale=1.15,font=\color{black}] {${\it U_x}$};
	
	\node[left=of img5a1, xshift=4.525cm ,yshift=0.855cm,rotate=0,scale=0.75,font=\color{black}] {${\it Ha} = 1000, 3D$};		
	\node[left=of img5a1, xshift=4.525cm ,yshift=0.5cm,rotate=0,scale=0.75,font=\color{black}] {${\it Ha} = 5000, 3D$};
	\node[left=of img5a1, xshift=4.65cm ,yshift=0.175cm,rotate=0,scale=0.75,font=\color{black}] {${\it Ha} = 10000, 3D$};
	
	\node[left=of img5a1, xshift=4.525cm ,yshift=-0.155cm,rotate=0,scale=0.75,font=\color{black}] {${\it Ha} = 1000, 2D$};		
	\node[left=of img5a1, xshift=4.525cm ,yshift=-0.475cm,rotate=0,scale=0.75,font=\color{black}] {${\it Ha} = 5000, 2D$};
	\node[left=of img5a1, xshift=4.650cm ,yshift=-0.8cm,rotate=0,scale=0.75,font=\color{black}] {${\it Ha} = 10000, 2D$};		
	
	
\node [below=of img5a1, xshift=0.0cm, yshift=0.00cm]  (img5a2)  {\includegraphics[scale=0.2]{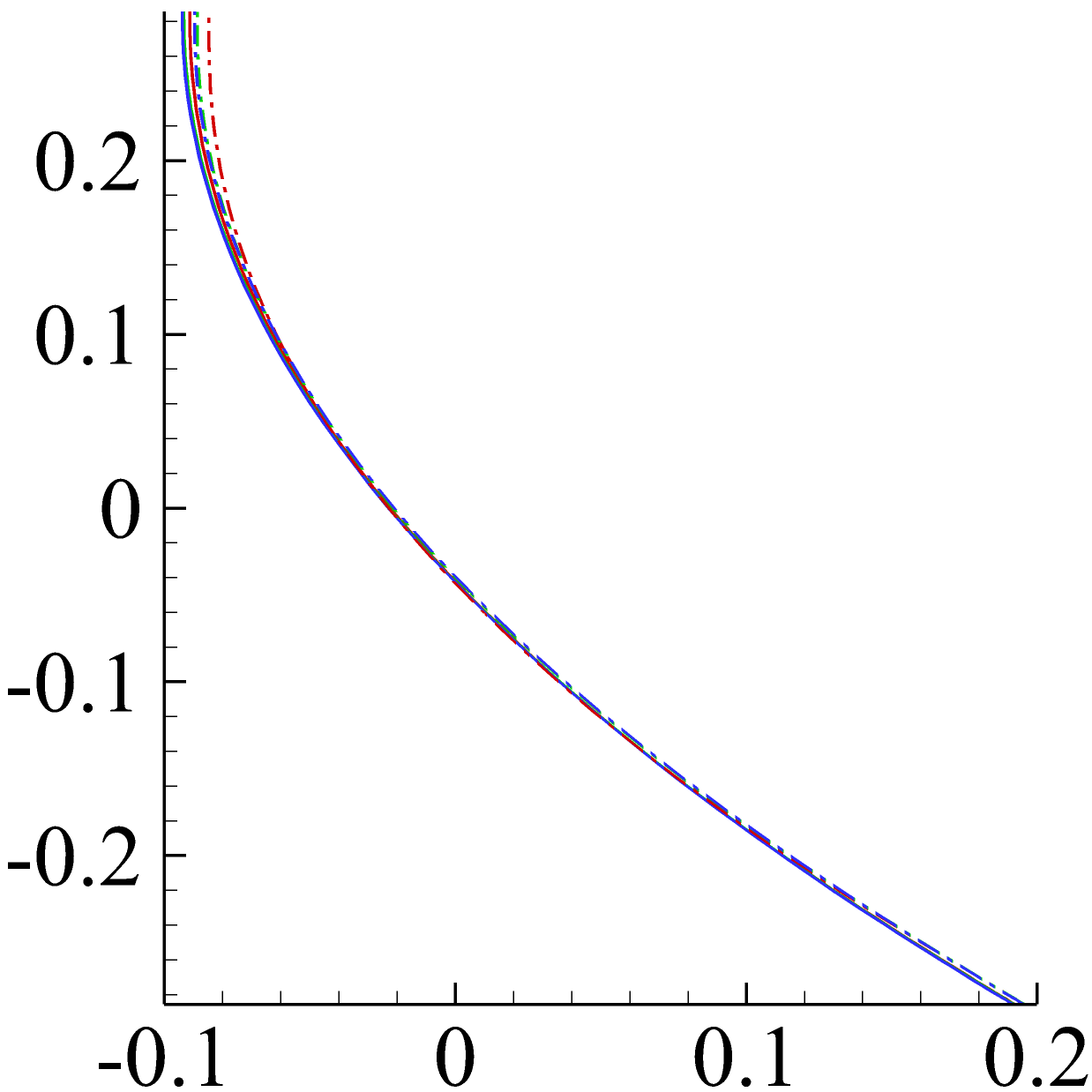}}; 
  	\node[left=of img5a2, xshift=1.25cm ,yshift=2.00cm,rotate=0,font=\color{black}]{({\it d})};
	
	\node[left=of img5a2, xshift=1.15cm ,yshift=0.15cm,rotate=0,scale=1.15,font=\color{black}] {{\it z}};
	
	\node[below=of img5a2, xshift=0.25cm ,yshift=1.0cm,rotate=0,scale=1.15,font=\color{black}] {${\it \Theta}$};
	
	
\node [right=of img5a2, xshift=-3.55cm, yshift=1.00cm] (img5a22)  {\includegraphics[scale=0.1]{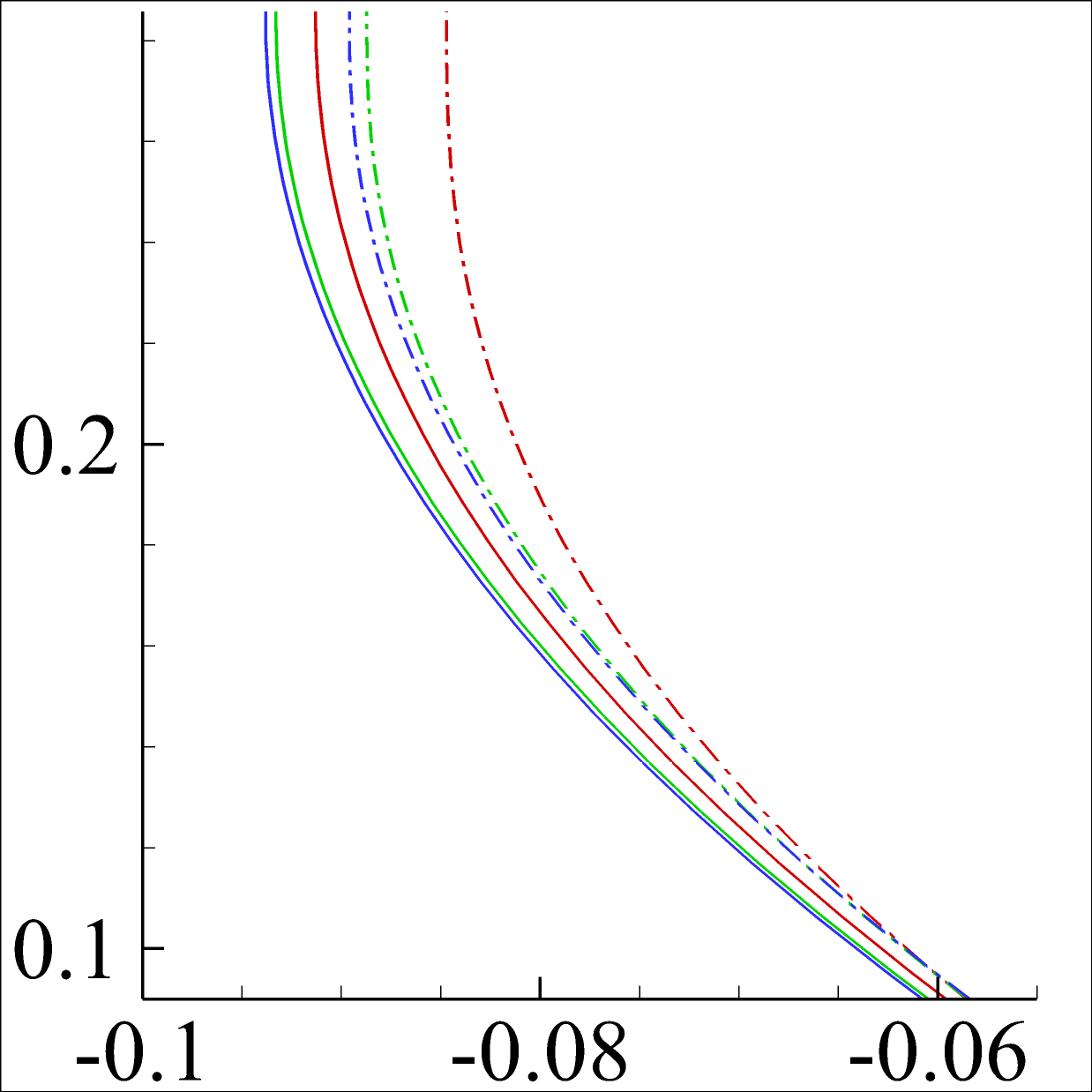}}; 


\node [right=of img5a1, xshift=-1.15cm, yshift=0.00cm]  (img5b1) {\includegraphics[scale=0.2]{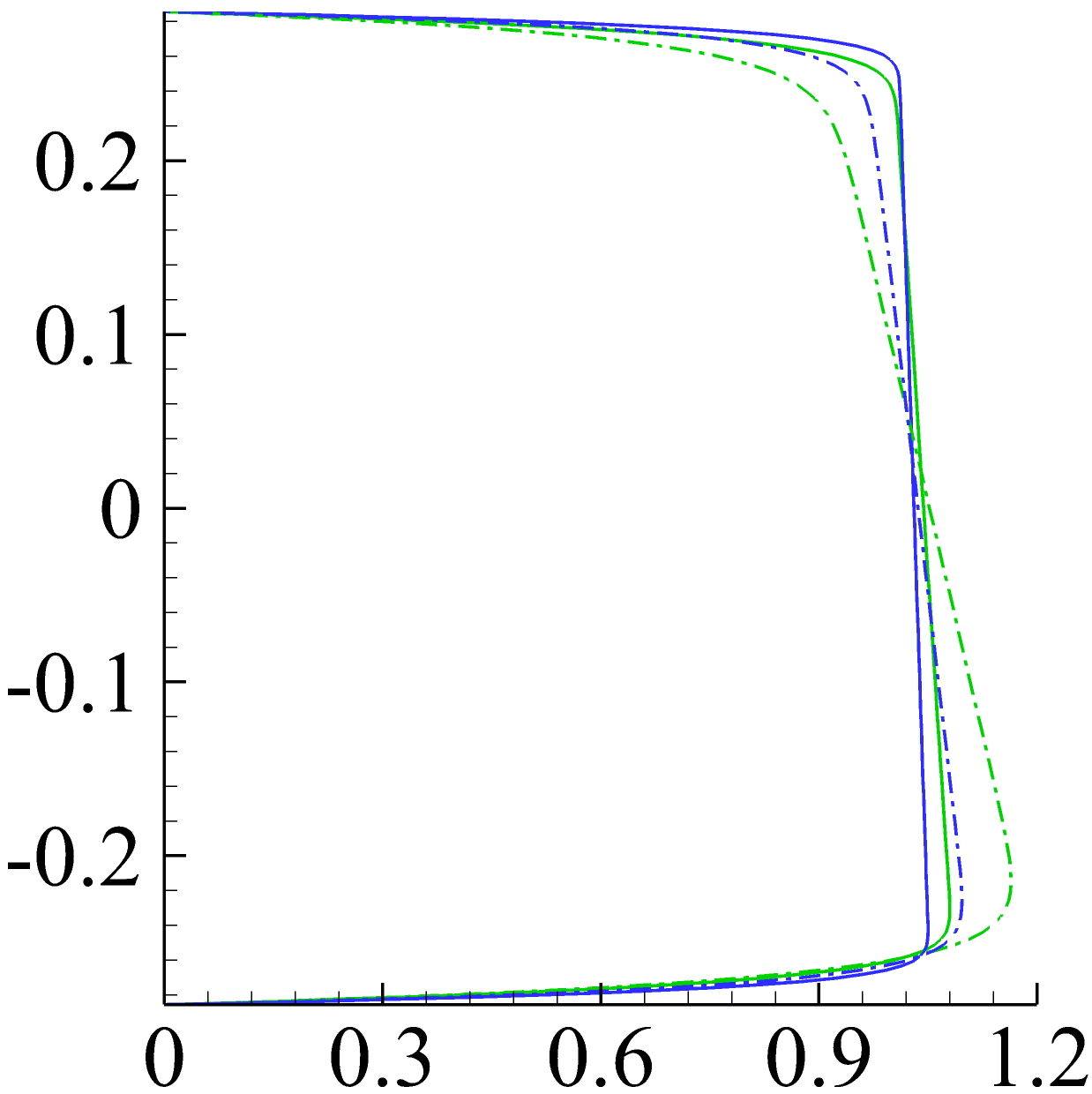}}; 
  	\node[left=of img5b1, xshift=1.40cm ,yshift=2.00cm,rotate=0,font=\color{black}] {({\it b})};
	
	 \node[left=of img5b1, xshift=4.25cm ,yshift=2.50cm,rotate=0,font=\color{black}] {$Gr = 10^{9}$};
	
	\node[below=of img5b1, xshift=0.25cm ,yshift=1.0cm,rotate=0,scale=1.15,font=\color{black}] {${\it U_x}$};
	
	
\node [below=of img5b1, xshift=0.0cm, yshift=0.00cm]  (img5b2)  {\includegraphics[scale=0.2]{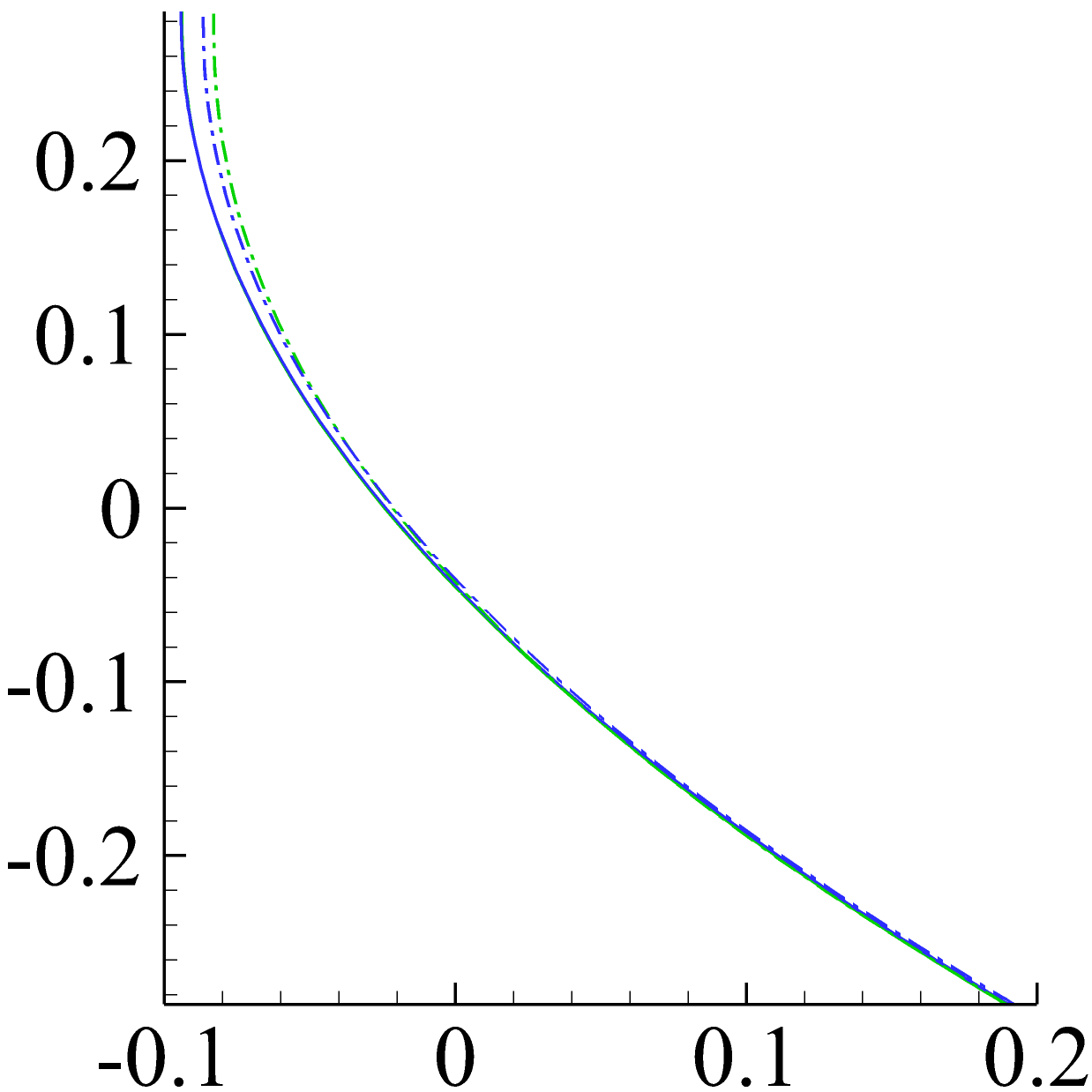}}; 
  	\node[left=of img5b2, xshift=1.40cm ,yshift=2.00cm,rotate=0,font=\color{black}]{({\it e})};
	
	\node[below=of img5b2, xshift=0.25cm ,yshift=1.0cm,rotate=0,scale=1.15,font=\color{black}] {${\it \Theta}$};
	
	
\node [right=of img5b2, xshift=-3.55cm, yshift=1.00cm] (img5b22)  {\includegraphics[scale=0.1]{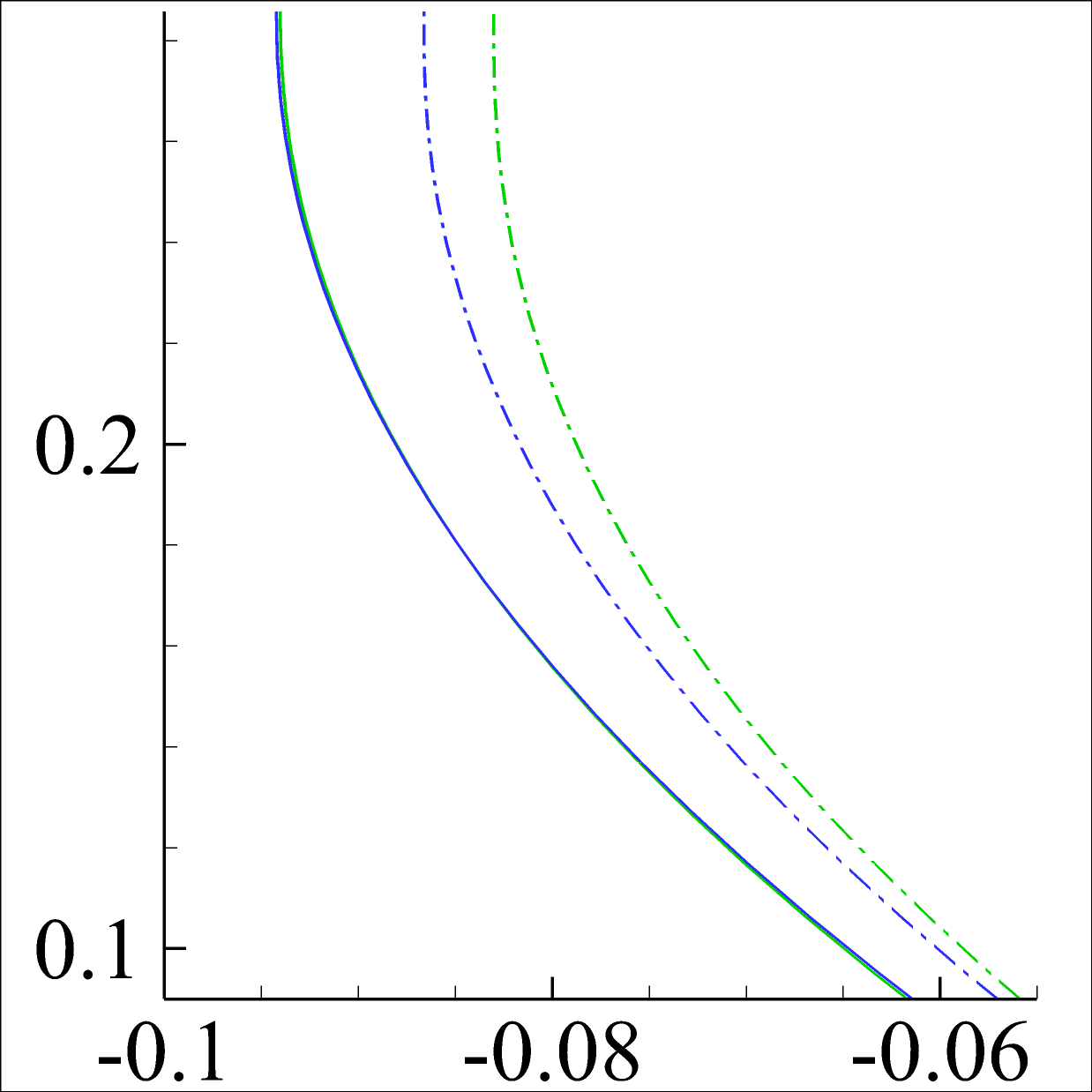}}; 


\node [right=of img5b1, xshift=-1.15cm, yshift=0.00cm]  (img5c1) {\includegraphics[scale=0.2]{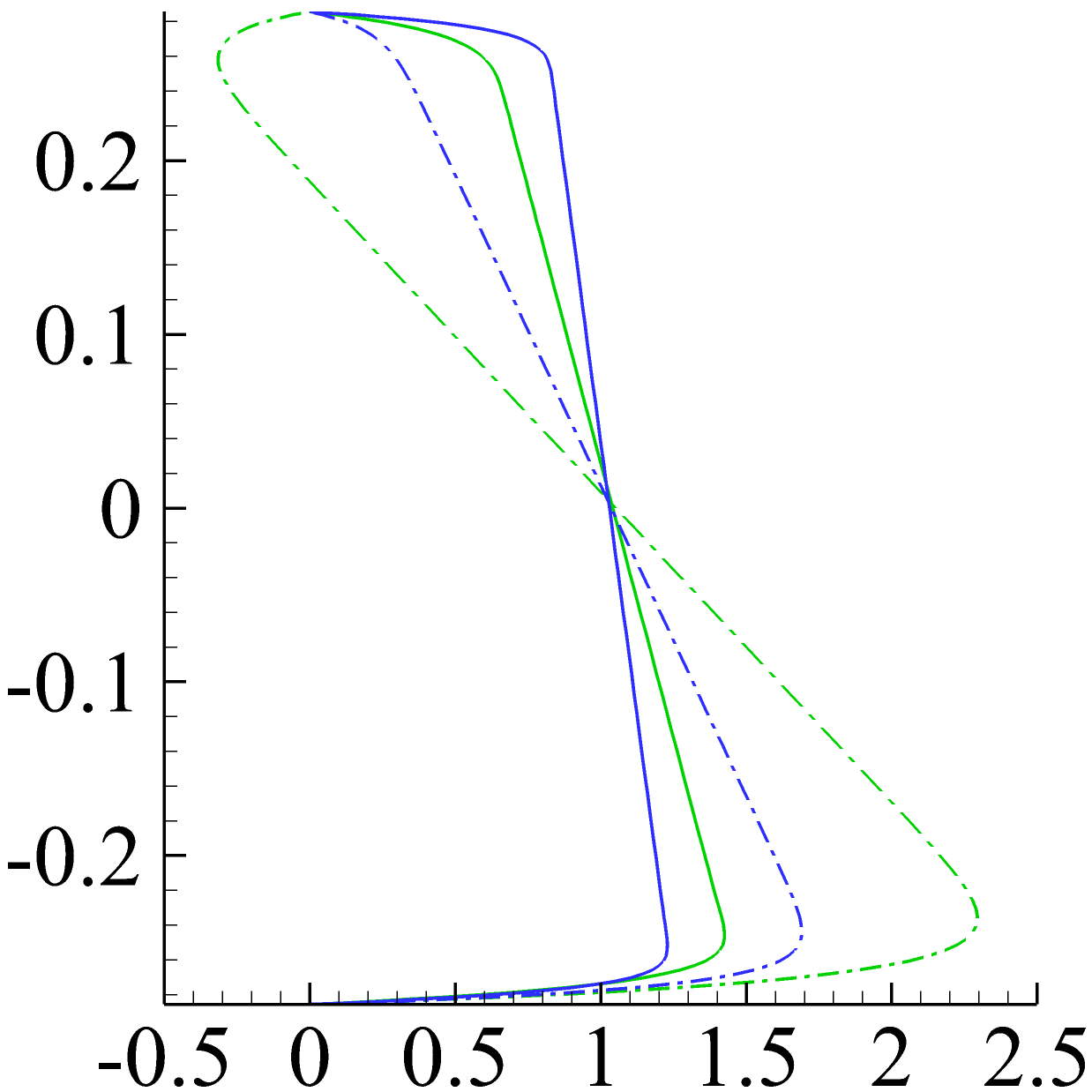}}; 
  	\node[left=of img5c1, xshift=1.40cm ,yshift=2.00cm,rotate=0,font=\color{black}] {({\it c})};
	
	 \node[left=of img5c1, xshift=4.40cm ,yshift=2.50cm,rotate=0,font=\color{black}] {$Gr = 10^{10}$};
	
	\node[below=of img5c1, xshift=0.25cm ,yshift=1.0cm,rotate=0,scale=1.15,font=\color{black}] {${\it U_x}$};
	
	
\node [below=of img5c1, xshift=0.0cm, yshift=0.00cm]  (img5c2)  {\includegraphics[scale=0.2]{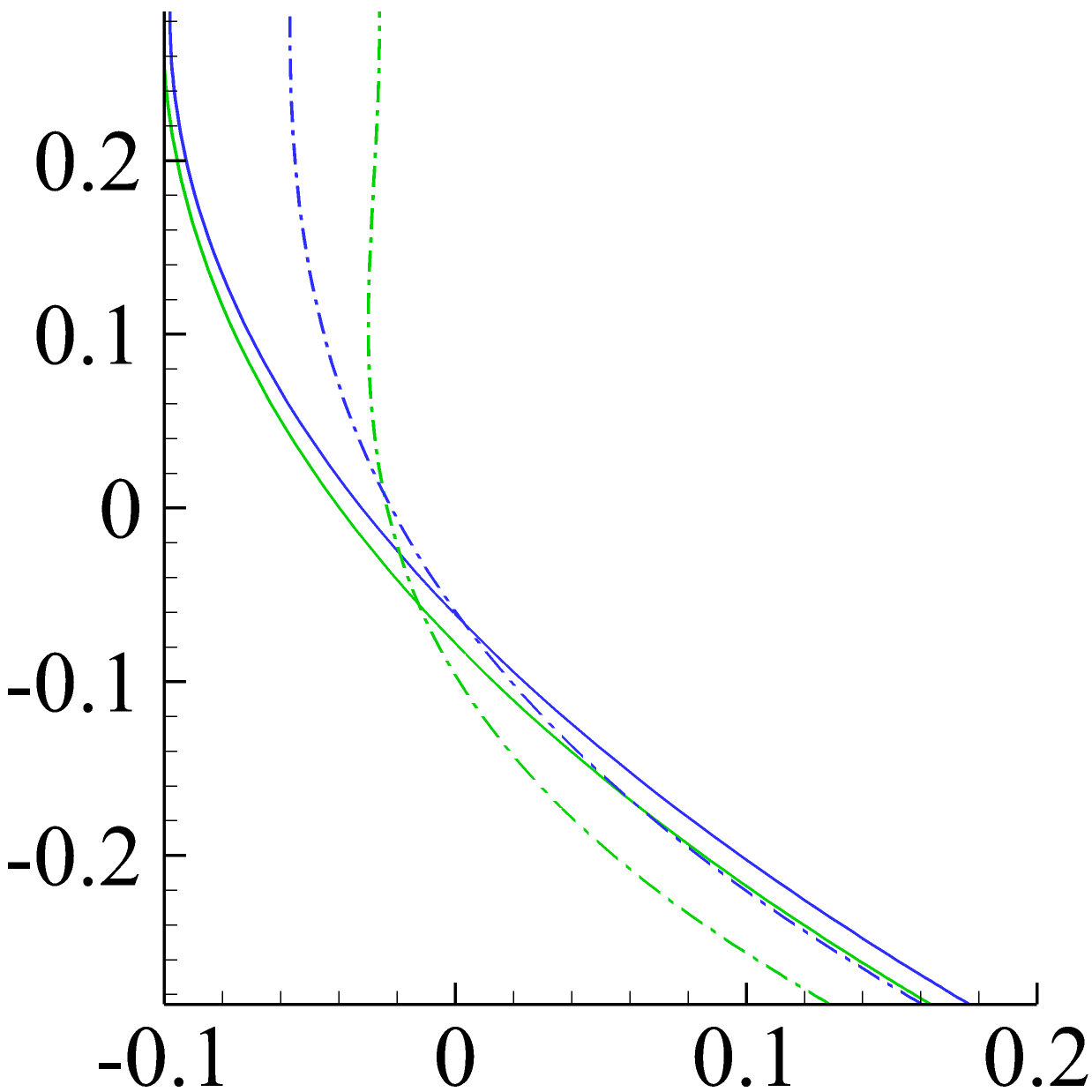}}; 
  	\node[left=of img5c2, xshift=1.40cm ,yshift=2.00cm,rotate=0,font=\color{black}]{({\it f})};
	
	\node[below=of img5c2, xshift=0.25cm ,yshift=1.0cm,rotate=0,scale=1.15,font=\color{black}] {${\it \Theta}$};

\end{tikzpicture}

  \caption{Base flow profiles at $Gr = 10^8$ ($a,d$), $Gr = 10^9$ ($b,e$), and $Gr = 10^{10}$ ($c,f$). The SM82 model solutions (\ref{eqsm4}), (\ref{eqsm5}) and the distributions along the midplane $y=0$ obtained in the full numerical solutions of section \ref{results1} are denoted, respectively, as 2D and 3D. The curves for $Ha = 10^3$ are only shown at $Gr = 10^8$ ($a,d$), since at higher $Gr$ the base flow is not Q2D (see section \ref{results1}). The top and bottom rows show the profiles of streamwise velocity and temperature, respectively. The insets in ($d$) and ($e$) show zoomed-in illustrations of the temperature profiles near the top wall.}
  
\label{fig5}
\end{figure}
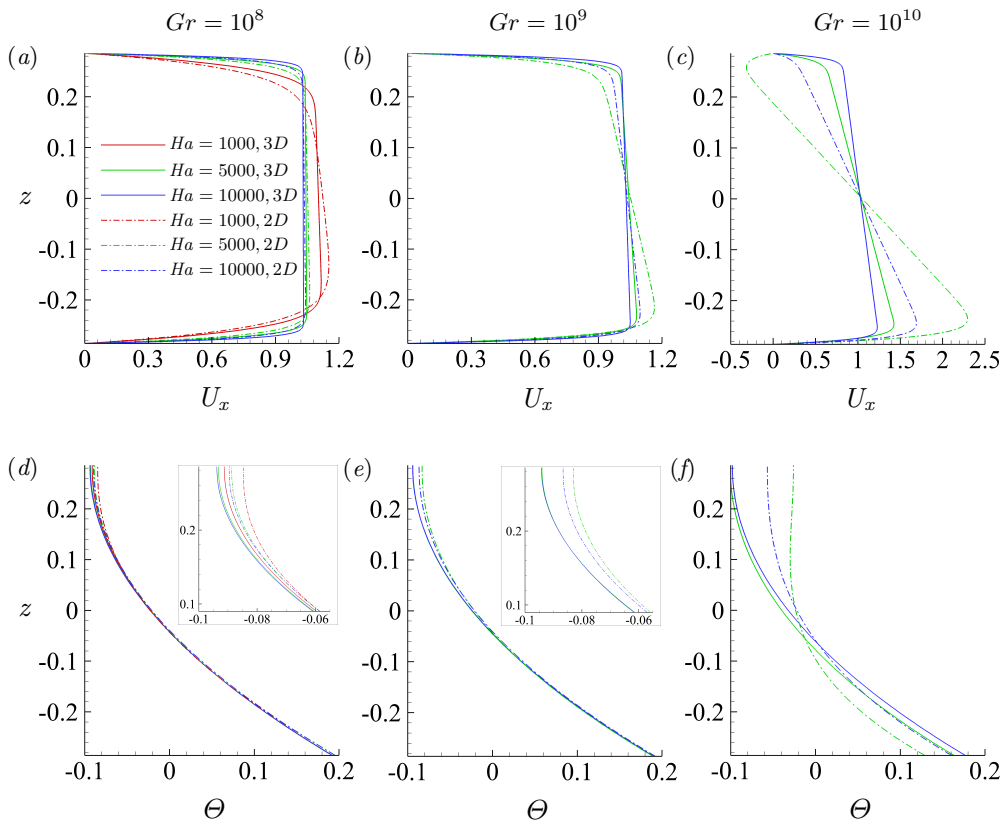

The profiles (\ref{eqsm4}) and (\ref{eqsm5}) are shown for $Gr = 10^8, 10^9, 10^{10}$ and several values of $Ha$ in figure \ref{fig5}. For comparison, solutions of the 3D equations $U_x$ and $\Theta$ obtained for the computed base flow solutions described in section \ref{results1} are shown for the midplane $y=0$. The symbols 2D and 3D correspond to, respectively, approximate SM82 and computed solutions.


Good agreement between the computed solutions and the solutions of the SM82 model is evident at $Gr = 10^8$ and $10^9$. There are some deviations between the computed and model profiles of $U_x$, but they are small and decrease with increasing $Ha$. The agreement is significantly worse in the case of the flows at $Gr = 10^{10}$. Here we observe large deviations between the computed and model curves at $Ha=5000$ and smaller, yet still significant deviations at $Ha=10000$.

The main reason for the discrepancy between the results of 2D model and 3D calculations at such a high Ha is the geometry of the flow. The duct has a large aspect ratio $\Gamma$ and the magnetic field oriented along the long side. Deviations from two-dimensionality become more pronounced in such geometries. To verify this explanation, we performed additional simulations, which revealed that velocity and temperature profiles in 2D and 3D solutions are almost indistinguishable from each other at $\Gamma \le 1$.

\subsubsection{Linear stability analysis}

\begin{figure}
	\centering 

	
\begin{tikzpicture}


\node (img7a1) {\includegraphics[scale=0.35]{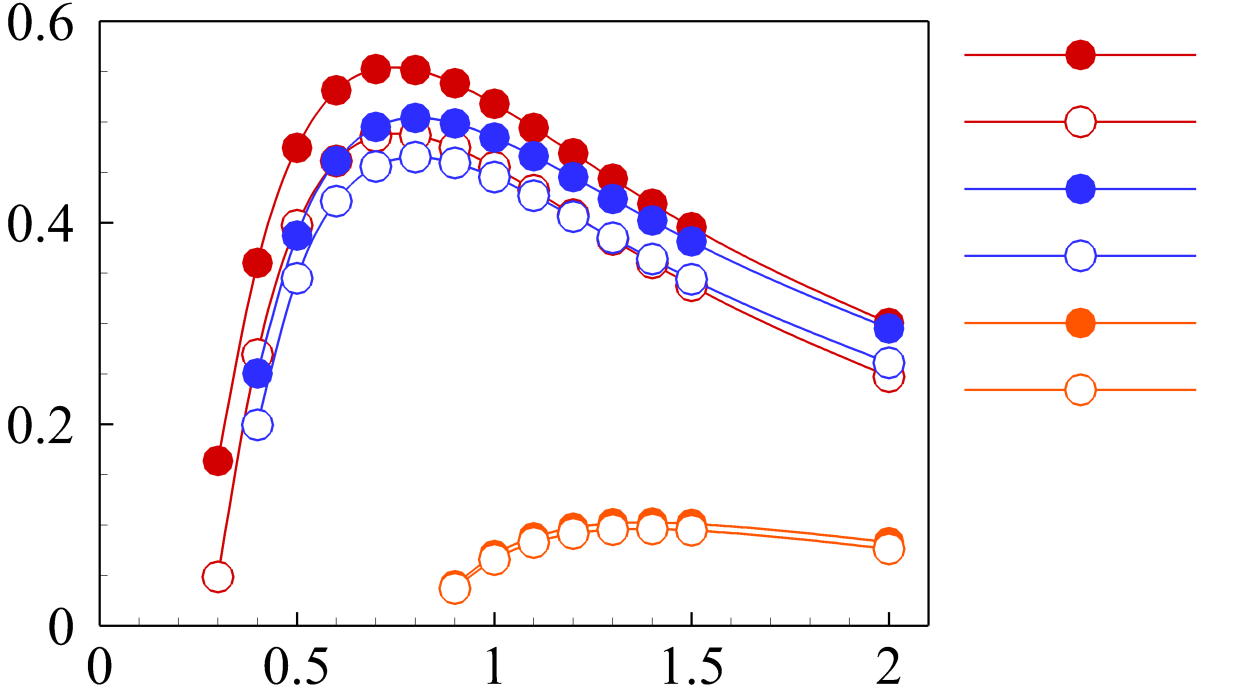}}; 
  	\node[left=of img7a1, xshift=1.00cm ,yshift=2.25cm,rotate=0,font=\color{black}] {$({\it a})$};
	\node[left=of img7a1, xshift=5.00cm ,yshift=2.25cm,rotate=0,font=\color{black}] {$Gr = 10^8$};
	
	\node[left=of img7a1, xshift=1.00cm ,yshift=0.10cm,rotate=0,scale=1.1,font=\color{black}] {${\it \gamma}$};
	\node[below=of img7a1, xshift=-0.45cm ,yshift=1.15cm,rotate=0,scale=1.1,font=\color{black}] {${\it \lambda}$};
	
	\node[right=of img7a1, xshift=-1.3cm, yshift=1.70cm, rotate=0, scale=1.0, font=\color{black}]  {${\it Ha} = 1000, 3D$};
	\node[right=of img7a1, xshift=-1.3cm, yshift=1.30cm,rotate=0,scale=1.0,font=\color{black}]  {${\it Ha} = 1000, 2D$};
	\node[right=of img7a1, xshift=-1.3cm, yshift=0.90cm,rotate=0,scale=1.0,font=\color{black}]  {${\it Ha} = 2000, 3D$};
	\node[right=of img7a1, xshift=-1.3cm, yshift=0.50cm,rotate=0,scale=1.0,font=\color{black}]  {${\it Ha} = 2000, 2D$};
	\node[right=of img7a1, xshift=-1.3cm, yshift=0.10cm,rotate=0,scale=1.0,font=\color{black}]  {${\it Ha} = 10000, 3D$};
	\node[right=of img7a1, xshift=-1.3cm, yshift=-0.30cm,rotate=0,scale=1.0,font=\color{black}] {${\it Ha} = 10000, 2D$};
	
	
\node [below=of img7a1, xshift=0.0cm, yshift=0.35cm]  (img7a2)  {\includegraphics[scale=0.35]{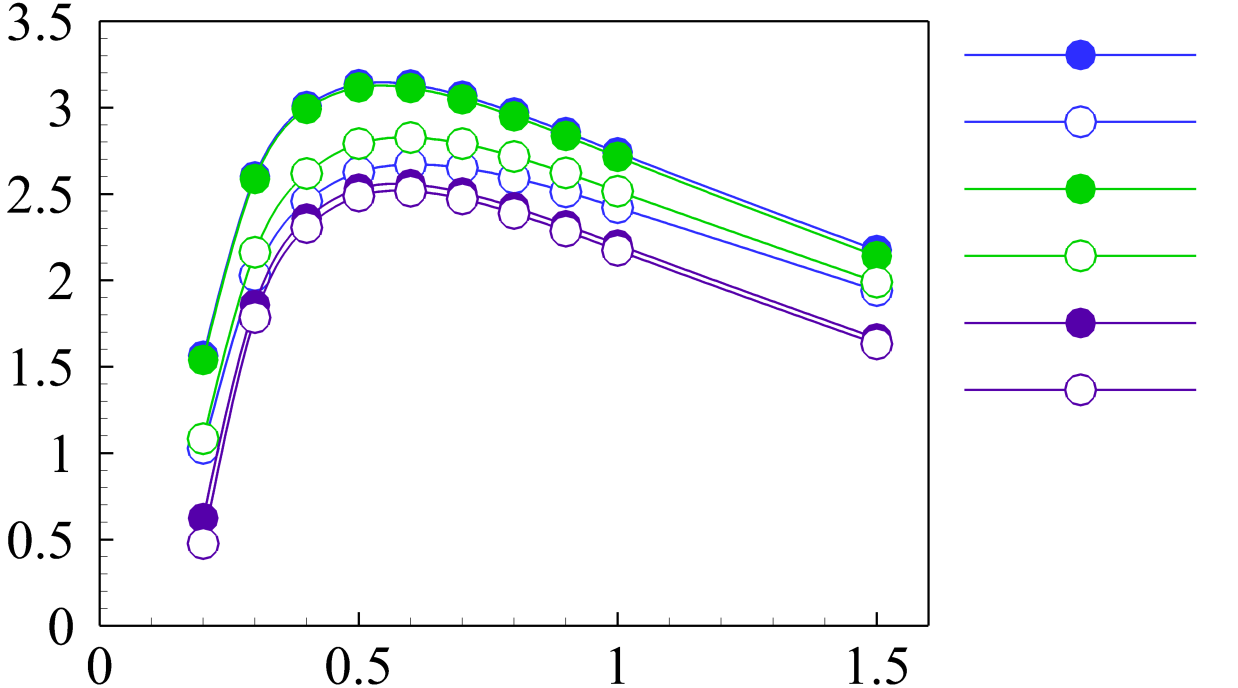}}; 
  	\node[left=of img7a2, xshift=1.00cm ,yshift=2.25cm,rotate=0,font=\color{black}] {$({\it b})$};
	\node[left=of img7a2, xshift=5.00cm ,yshift=2.25cm,rotate=0,font=\color{black}] {$Gr = 10^9$};
	
	\node[left=of img7a2, xshift=1.00cm ,yshift=0.10cm,rotate=0,scale=1.1,font=\color{black}] {${\it \gamma}$};
	\node[below=of img7a2, xshift=-0.45cm ,yshift=1.15cm,rotate=0,scale=1.1,font=\color{black}] {${\it \lambda}$};
	
	\node[right=of img7a2, xshift=-1.3cm, yshift=1.70cm, rotate=0, scale=1.0, font=\color{black}]  {${\it Ha} = 2000, 3D$};
	\node[right=of img7a2, xshift=-1.3cm, yshift=1.30cm,rotate=0,scale=1.0,font=\color{black}]   {${\it Ha} = 2000, 2D$};s
	\node[right=of img7a2, xshift=-1.3cm, yshift=0.90cm,rotate=0,scale=1.0,font=\color{black}]  {${\it Ha} = 3000, 3D$};
	\node[right=of img7a2, xshift=-1.3cm, yshift=0.50cm,rotate=0,scale=1.0,font=\color{black}]   {${\it Ha} = 3000, 2D$};
	\node[right=of img7a2, xshift=-1.3cm, yshift=0.10cm,rotate=0,scale=1.0,font=\color{black}]  {${\it Ha} = 10000, 3D$};
	\node[right=of img7a2, xshift=-1.3cm, yshift=-0.30cm,rotate=0,scale=1.0,font=\color{black}] {${\it Ha} = 10000, 2D$};
	
	
\node [below=of img7a2, xshift=0.0cm, yshift=0.35cm]  (img7a3)  {\includegraphics[scale=0.35]{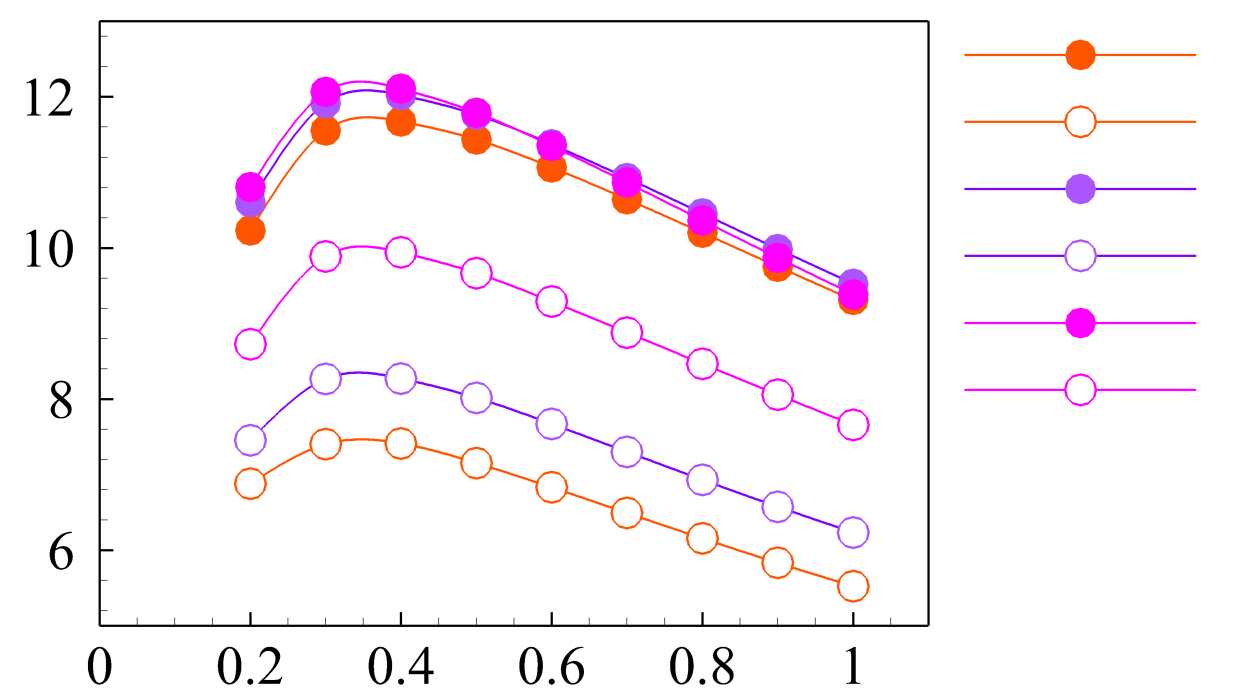}}; 
  	\node[left=of img7a3, xshift=1.00cm ,yshift=2.25cm,rotate=0,font=\color{black}] {$({\it c})$};
	\node[left=of img7a3, xshift=5.00cm ,yshift=2.25cm,rotate=0,font=\color{black}] {$Gr = 10^{10}$};
	
	\node[left=of img7a3, xshift=1.00cm ,yshift=0.10cm,rotate=0,scale=1.1,font=\color{black}] {${\it \gamma}$};
	\node[below=of img7a3, xshift=-0.45cm ,yshift=1.15cm,rotate=0,scale=1.1,font=\color{black}] {${\it \lambda}$};
	
	\node[right=of img7a3, xshift=-1.3cm, yshift=1.70cm, rotate=0, scale=1.0, font=\color{black}]  {${\it Ha} = 4000, 3D$};
	\node[right=of img7a3, xshift=-1.3cm, yshift=1.30cm,rotate=0,scale=1.0,font=\color{black}]  {${\it Ha} = 4000, 2D$};
	\node[right=of img7a3, xshift=-1.3cm, yshift=0.90cm,rotate=0,scale=1.0,font=\color{black}]  {${\it Ha} = 5000, 3D$};
	\node[right=of img7a3, xshift=-1.3cm, yshift=0.50cm,rotate=0,scale=1.0,font=\color{black}] {${\it Ha} = 5000, 2D$};
	\node[right=of img7a3, xshift=-1.3cm, yshift=0.10cm,rotate=0,scale=1.0,font=\color{black}]  {${\it Ha} = 10000, 3D$};
	\node[right=of img7a3, xshift=-1.3cm, yshift=-0.30cm,rotate=0,scale=1.0,font=\color{black}] {${\it Ha} = 10000, 2D$};

\end{tikzpicture}

\caption{ Rates of exponential growth $\gamma$ shown as functions of the axial wavelength $\lambda$ at $Ha = 1000, 2000, 10000$, $Gr = 10^8$ ($a$), $Ha = 2000, 3000, 10000$, $Gr = 10^9$ ($b$), and $Ha = 4000, 5000, 10000$, $Gr = 10^{10}$ ($c$). The results of 3D and 2D (SM82) models are denoted as filled and empty circles, respectively.}
  
\label{fig6}
\end{figure}
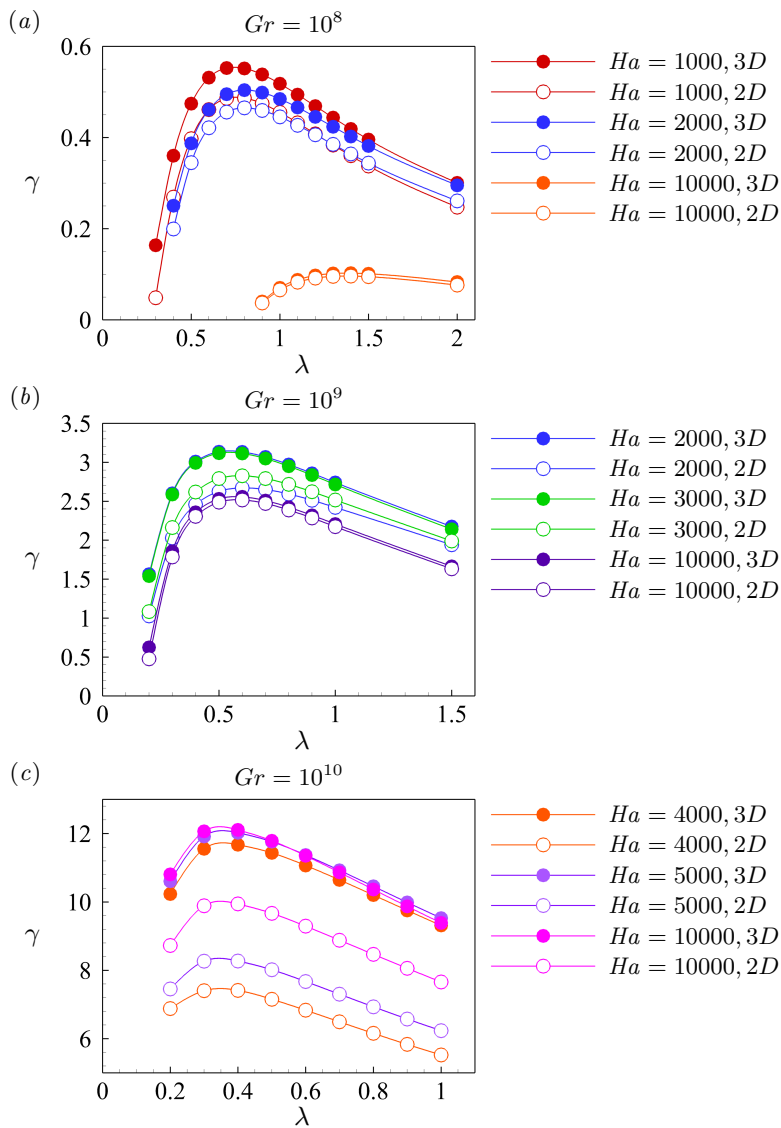

In order to verify applicability of the SM82 model to linear stability analysis of Q2D flows, instability of several high$-Ha$ flows was evaluated twice: once using full 3D model of the base flow and perturbations and once entirely in the framework of the 2D SM82 model. The results are presented in figure \ref{fig6} and table \ref{table4}. We see good agreement between predictions of 2D and 3D models at $Gr=10^8$ and $10^9$. The accuracy improves with growing $Ha$. As an example, the average relative difference between the values of the growth rate $\gamma$ for the two models is $33\%$ at $Ha = 2000$, $Gr = 10^{9}$, $11\%$ at $Ha = 3000$, $Gr = 10^{9}$ and $3\%$ at $Ha = 10000$, $Gr = 10^{9}$ . The situation is less clear for flows at $Gr = 10^{10}$ (see figure \ref{fig6}$c$ and the last six lines of table \ref{table4}). Here we only see a qualitative agreement. The shape of the $\gamma$($\lambda$) curves, the wavelength of the most unstable mode, and the effect of $Ha$ on stability are similar in the 3D and 2D solution. The quantitative agreement is, however, poor, with the difference between the values of $\gamma$ found for the two models being about $50\%$.

The quantitative disagreement between the base flow profiles and, as an evident consequence, stability properties found in the 3D and 2D models is difficult to interpret. The velocity and temperature distributions computed in the framework of the 3D model clearly show that the base flow is Q2D and nearly perfectly unidirectional at $Ha$ higher than approximately $4000$ (see figure \ref{fig3}$b,d$ and values of $E_t$ in table \ref{tablebase1}). As illustrated in section \ref{results3}, fields of growing perturbations also remain Q2D at such high $Ha$. Additional calculations performed with larger grids and longer times of flow evolution did not lead to significant changes. The deviations from quasi-two-dimensionality and inaccuracy of the numerical model are, therefore, excluded as possible reasons.


\begin{table}
  \begin{center}
  \begin{tabular}{lccccccccccccc}
  
    $Ha$ & $Gr$ & $$ & $$ & $$ & $$ & $$ & $\lambda$ & $$ & $$ & $$ & $$ & $$ \\ [3pt]
    $$ & $$ & $0.2$ & $0.3$ & $0.4$ & $0.5$ & $0.6$ & $0.7$ & $0.8$ & $0.9$ & $1.0$ & $1.5$ & $2.0$ \\ [3pt]
      \hline \\ [3pt]

    1000 & $10^8$ & $-$ & 0.164 & 0.360 & 0.474 & 0.532 & 0.553 & 0.552 & 0.538 & 0.518 & 0.396 & 0.301  \\
    $\underline{1000}$ & $\underline{10^8}$ & $-$ & $\underline{0.049}$ & $\underline{0.269}$ & $\underline{0.398}$ & $\underline{0.461}$ & $\underline{0.486}$ & $\underline{0.487}$ & $\underline{0.475}$ & $\underline{0.455}$ & $\underline{0.338}$ & $\underline{0.247}$ \\ [3pt]
    
    2000 & $10^8$ & $-$ & $-$ & 0.251 & 0.387 & 0.461 & 0.495 & 0.504 & 0.499 & 0.485 & 0.381 & 0.295  \\
    $\underline{2000}$ & $\underline{10^8}$ & $-$ & $-$ & $\underline{0.200}$ & $\underline{0.345}$ & $\underline{0.486}$ & $\underline{0.422}$ & $\underline{0.456}$ & $\underline{0.459}$ & $\underline{0.445}$ & $\underline{0.344}$ & $\underline{0.261}$ \\ [3pt]
    
    2000 & $10^9$ & 1.563 & 2.602 & 3.010 & 3.136 & 3.134 & 3.069 & 2.972 & 2.859 & 2.741 & 2.175 & $-$  \\
    $\underline{2000}$ & $\underline{10^9}$ & $\underline{1.028}$ & $\underline{2.029}$ & $\underline{2.459}$ & $\underline{2.626}$ & $\underline{2.671}$ & $\underline{2.651}$ & $\underline{2.593}$ & $\underline{2.513}$ & $\underline{2.421}$ & $\underline{1.941}$ & $-$ \\ [3pt]
    
    3000 & $10^9$ & 1.540 & 2.587 & 2.993 & 3.117 & 3.114 & 3.048 & 2.950 & 2.836 & 2.715 & 2.141 & $-$ \\
    $\underline{3000}$ & $\underline{10^9}$ & $\underline{1.085}$ & $\underline{2.162}$ & $\underline{2.619}$ & $\underline{2.791}$ & $\underline{2.827}$ & $\underline{2.792}$ & $\underline{2.717}$ & $\underline{2.622}$ & $\underline{2.516}$ & $\underline{1.990}$ & $-$ \\ [3pt]
    
    4000 & $10^{10}$ & 10.234 & 11.553 & 11.674 & 11.435 & 11.067 & 10.645 & 10.202 & 9.754 & 9.312 & $-$ & $-$ \\
    $\underline{4000}$ & $\underline{10^{10}}$ & $\underline{6.881}$ & $\underline{7.404}$ & $\underline{7.415}$ & $\underline{7.156}$ & $\underline{6.833}$ & $\underline{6.494}$ & $\underline{6.158}$ & $\underline{5.834}$ & $\underline{5.525}$ & $-$ & $-$ \\ [3pt]
    
    5000 & $10^{10}$ & 10.605 & 11.918 & 12.023 & 11.764 & 11.371 & 10.925 & 10.458 & 9.988 & 9.527 & $-$ & $-$ \\
    $\underline{5000}$ & $\underline{10^{10}}$ & $\underline{7.454}$ & $\underline{8.269}$ & $\underline{8.273}$ & $\underline{8.018}$ & $\underline{7.673}$ & $\underline{7.303}$ & $\underline{6.934}$ & $\underline{6.577}$ & $\underline{6.237}$ & $-$ & $-$ \\ [3pt]
    
    10000 & $10^{10}$ & 10.807 & 12.064 & 12.106 & 11.791 & 11.351 & 10.865 & 10.366 & 9.871 & 9.389 & $-$ & $-$ \\
    $\underline{10000}$ & $\underline{10^{10}}$ & $\underline{8.729}$ & $\underline{9.890}$ & $\underline{9.942}$ & $\underline{9.668}$ & $\underline{9.291}$ & $\underline{8.882}$ & $\underline{8.468}$ & $\underline{8.059}$ & $\underline{7.661}$ & $-$ & $-$ \\ [3pt]
	
  \end{tabular}
  \end{center}
  \caption{Results of the linear stability analysis of 3D and 2D flow solutions. Rates of exponential growth $\gamma$ are shown as functions of the axial wavelength $\lambda$. The results of the SM82 model are underlined. The growth rates are determined as in (\ref{eq_gamma}).}
  \label{table4}
\end{table}

\begin{figure}
	\centering 

	
\begin{tikzpicture}


\node (img5a1) {\includegraphics[scale=0.3]{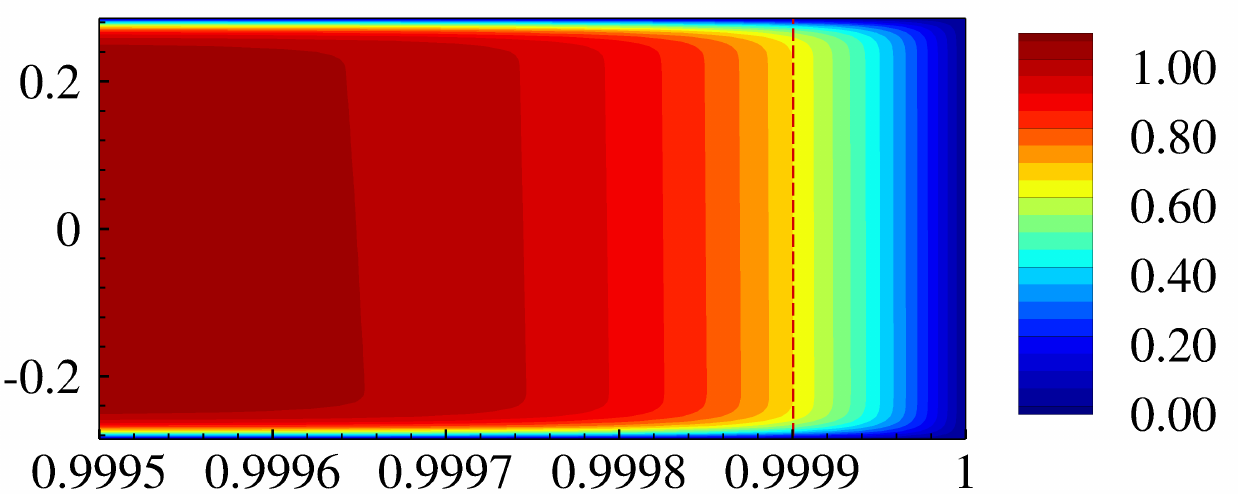}}; 
  	\node[left=of img5a1, xshift=1.15cm ,yshift=1.50cm,rotate=0,font=\color{black}] {({\it a})};
	
	\node[left=of img5a1, xshift=1.15cm ,yshift=0.15cm,rotate=0,scale=1.25,font=\color{black}] {{\it z}};
	
	
\node [right=of img5a1, xshift=-1.00cm, yshift=0.00cm]  (img5a2)  {\includegraphics[scale=0.3]{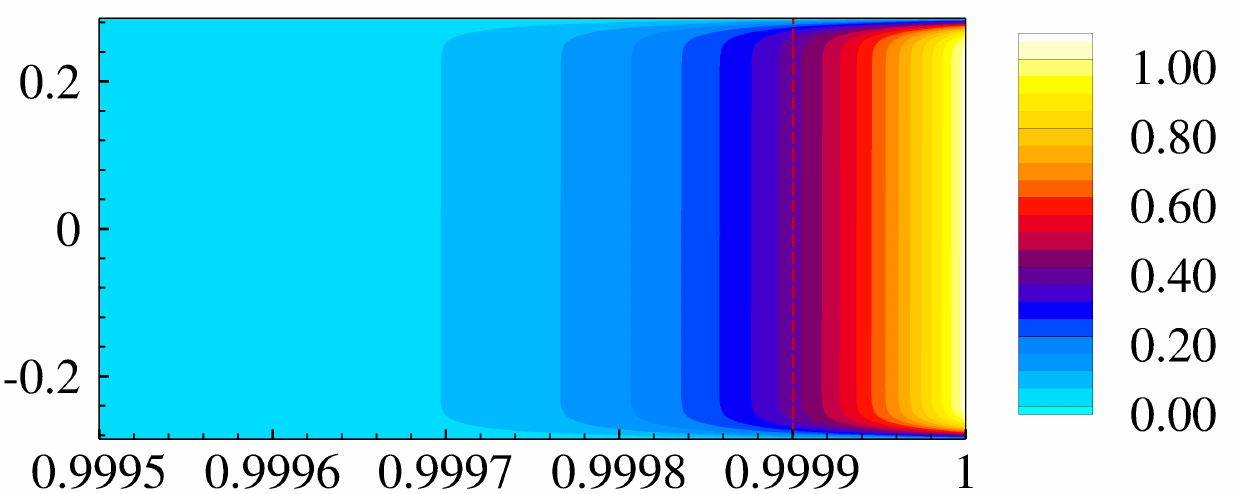}}; 
  	\node[left=of img5a2, xshift=1.60cm ,yshift=1.50cm,rotate=0,font=\color{black}]{({\it b})};


\node [below=of img5a1, xshift=0.0cm, yshift=0.5cm]  (img5b1) {\includegraphics[scale=0.3]{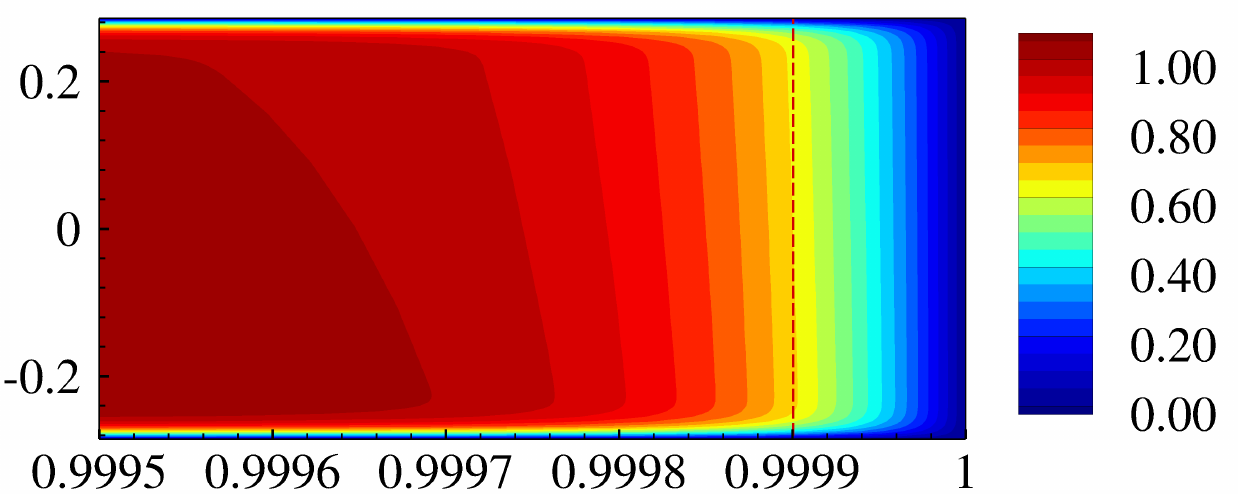}}; 
  	\node[left=of img5b1, xshift=1.15cm ,yshift=1.50cm,rotate=0,font=\color{black}] {({\it c})};
	
	\node[left=of img5b1, xshift=1.15cm ,yshift=0.15cm,rotate=0,scale=1.25,font=\color{black}] {{\it z}};
	
	
\node [right=of img5b1, xshift=-1.00cm, yshift=0.00cm]  (img5b2)  {\includegraphics[scale=0.3]{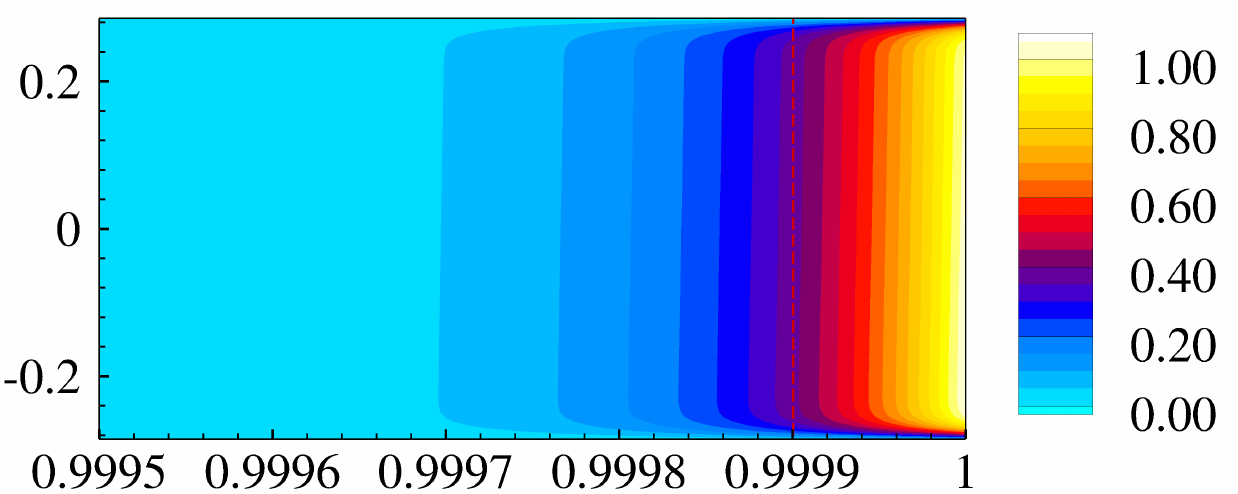}}; 
  	\node[left=of img5b2, xshift=1.60cm ,yshift=1.50cm,rotate=0,font=\color{black}]{({\it d})};
	

\node [below=of img5b1, xshift=0.0cm, yshift=0.50cm]  (img5c1) {\includegraphics[scale=0.3]{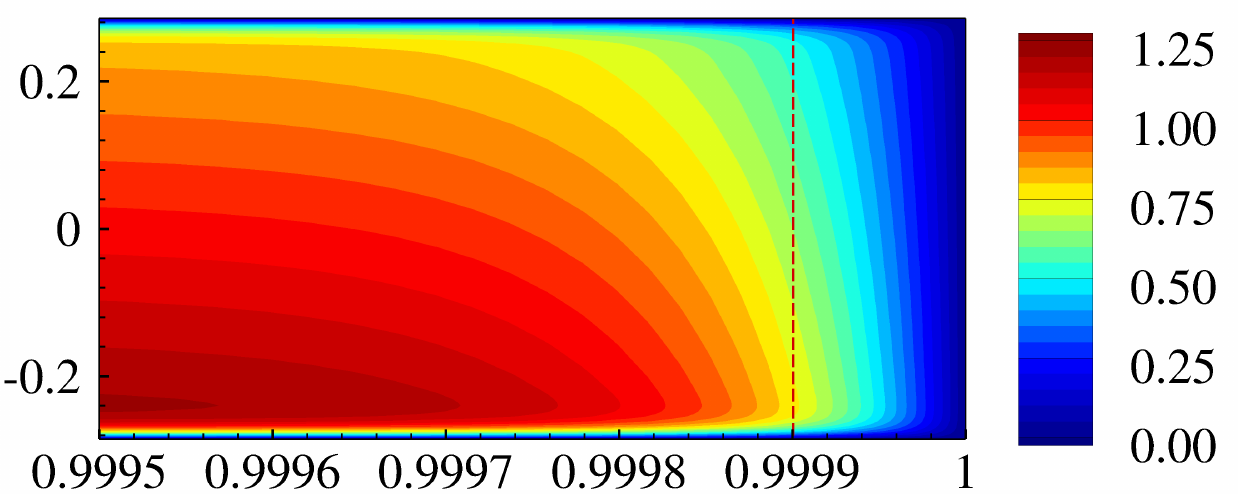}}; 
  	\node[left=of img5c1, xshift=1.15cm ,yshift=1.50cm,rotate=0,font=\color{black}] {({\it e})};
	
	\node[left=of img5c1, xshift=1.15cm ,yshift=0.15cm,rotate=0,scale=1.25,font=\color{black}] {{\it z}};
	
	\node[below=of img5c1, xshift=-0.40cm ,yshift=1.1cm,rotate=0,scale=1.25,font=\color{black}] {{\it y}};
	
	
\node [right=of img5c1, xshift=-1.00cm, yshift=0.00cm]  (img5c2)  {\includegraphics[scale=0.3]{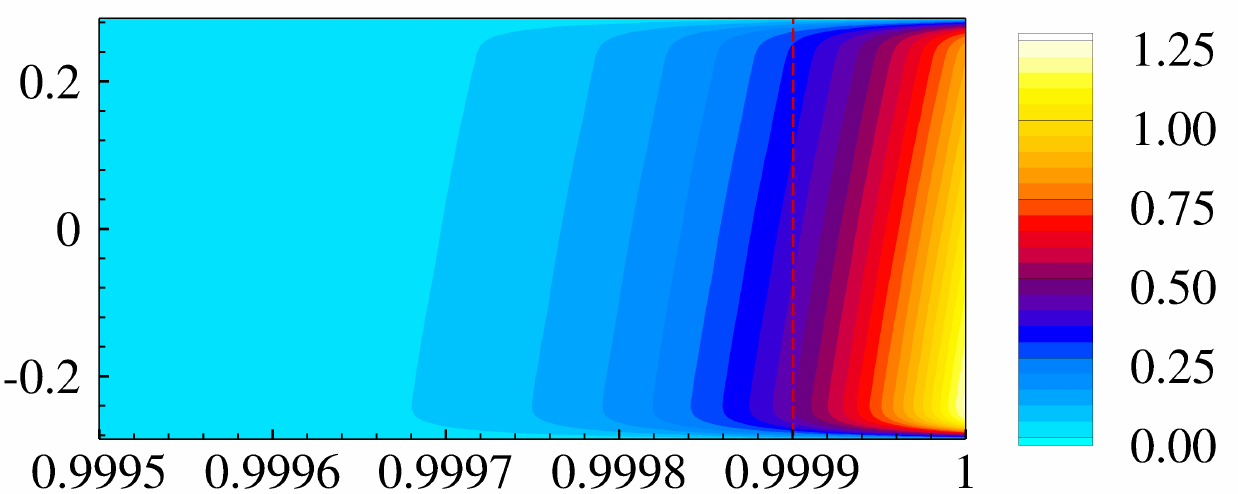}}; 
  	\node[left=of img5c2, xshift=1.60cm ,yshift=1.50cm,rotate=0,font=\color{black}]{({\it f})};
	
	\node[below=of img5c2, xshift=-0.40cm ,yshift=1.1cm,rotate=0,scale=1.25,font=\color{black}] {{\it y}};

\end{tikzpicture}

  \caption{Base flow at $Gr = 10^8$ ($a,b$), $Gr = 10^9$ ($c,d$), and $Gr = 10^{10}$ ($e,f$) for $Ha = 10000$. The left and right columns show, respectively, distributions of streamwise velocity $U_x$ and the streamwise component of the Lorentz force $F_{Lx}$ near the Hartmann wall. The red dashed line shows the boundary of the Hartmann layer of thickness $\delta_{Ha} = {\it Ha}^{-1}$.}
  
\label{fig7}
\end{figure}

A further useful, albeit not fully explaining illustration is provided in figure \ref{fig7}. Computed base flow distributions of $U_x$ and the streamwise component of the Lorentz force $F_{Lx}$ are shown for $Ha = 10000$ and $Gr = 10^8, 10^9$ and $10^{10}$ within and near the Hartmann boundary layer. We see that the strong vertical variation of $U_x$ existing at $Gr = 10^{10}$ extends toward the Hartmann wall and causes a respective variation of the Lorentz force. It must be noted that this picture does not contradict to the identification of the flow as Q2D. The profiles $U_x(y, z = const)$, if taken outside the sidewall layers at the horizontal walls and scaled by the respective maximum values of $U_x$, collapse into one curve with a flat core and Hartmann boundary layers.


\subsubsection{Nonlinear flows} \label{2d3dnonlinear}

\begin{figure}
	\centering 

	
\begin{tikzpicture}


\node (fig8_1) {\includegraphics[scale=0.525]{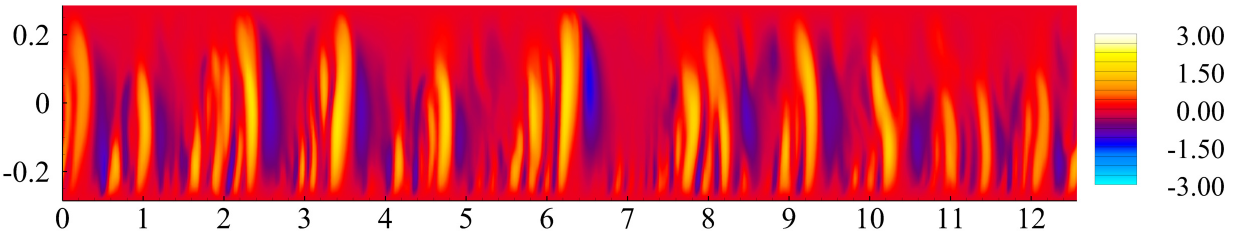}}; 
  	\node[left=of fig8_1, xshift=1.0cm ,yshift=0.90cm,rotate=0,scale=1.0,font=\color{black}] {({\it a})};

	\node[left=of fig8_1, xshift=1.00cm ,yshift=0.1cm,rotate=0,scale=1.0,font=\color{black}] {${\it z}$};
	
	\node[left=of fig8_1, xshift=12.2cm ,yshift=1.00cm,rotate=0,scale=1.0,font=\color{black}] {${\it u_z}$};
	
\node [below=of fig8_1, xshift=0.0cm, yshift=0.85cm]  (fig8_2)  {\includegraphics[scale=0.525]{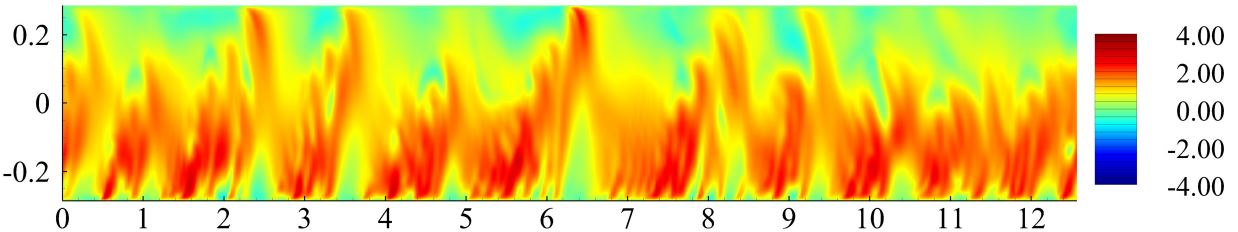}}; 
  	\node[left=of fig8_2, xshift=1.0cm ,yshift=0.90cm,rotate=0,scale=1.0,font=\color{black}] {({\it b})};

	\node[left=of fig8_2, xshift=1.00cm ,yshift=0.1cm,rotate=0,scale=1.0,font=\color{black}] {${\it z}$};
	
	\node[left=of fig8_2, xshift=12.2cm ,yshift=1.00cm,rotate=0,scale=1.0,font=\color{black}] {${\it u_x}$};
	
\node [below=of fig8_1, xshift=0.0cm, yshift=-1.45cm]  (fig8_3)  {\includegraphics[scale=0.525]{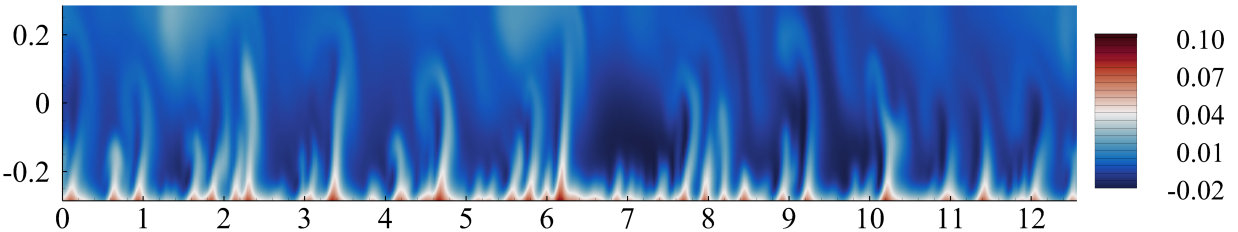}};
  	\node[left=of fig8_3, xshift=1.0cm ,yshift=0.90cm,rotate=0,scale=1.0,font=\color{black}] {({\it c})};

	\node[left=of fig8_3, xshift=1.00cm ,yshift=0.1cm,rotate=0,scale=1.0,font=\color{black}] {${\it z}$};
	\node[below=of fig8_3, xshift=-0.25cm ,yshift=1.0cm,rotate=0,font=\color{black}] {{\it x}};
	
	\node[left=of fig8_3, xshift=12.1cm ,yshift=1.00cm,rotate=0,scale=1.0,font=\color{black}] {${\theta}$};
	

\node [below=of fig8_3, xshift=-2.25cm, yshift=0.50cm]  (fig8_4)  {\includegraphics[scale=0.31]{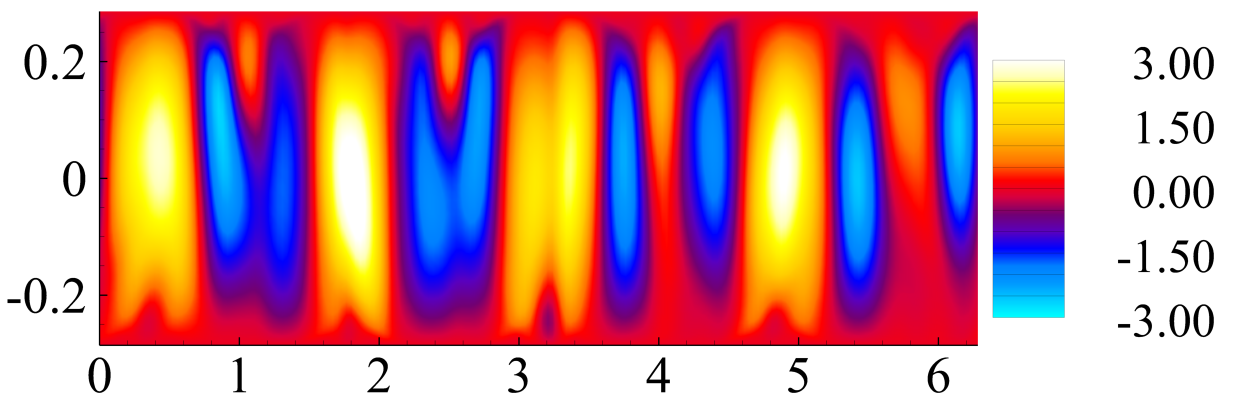}};
  	\node[left=of fig8_4, xshift=1.0cm ,yshift=0.90cm,rotate=0,scale=1.0,font=\color{black}] {({\it d})};

	\node[left=of fig8_4, xshift=1.00cm ,yshift=0.1cm,rotate=0,scale=1.0,font=\color{black}] {${\it z}$};
	
	\node[left=of fig8_4, xshift=7.6cm ,yshift=1.10cm,rotate=0,scale=1.0,font=\color{black}] {${\it u_z}$};

\node [below=of fig8_3, xshift=-2.25cm, yshift=-2.00cm]  (fig8_5)  {\includegraphics[scale=0.31]{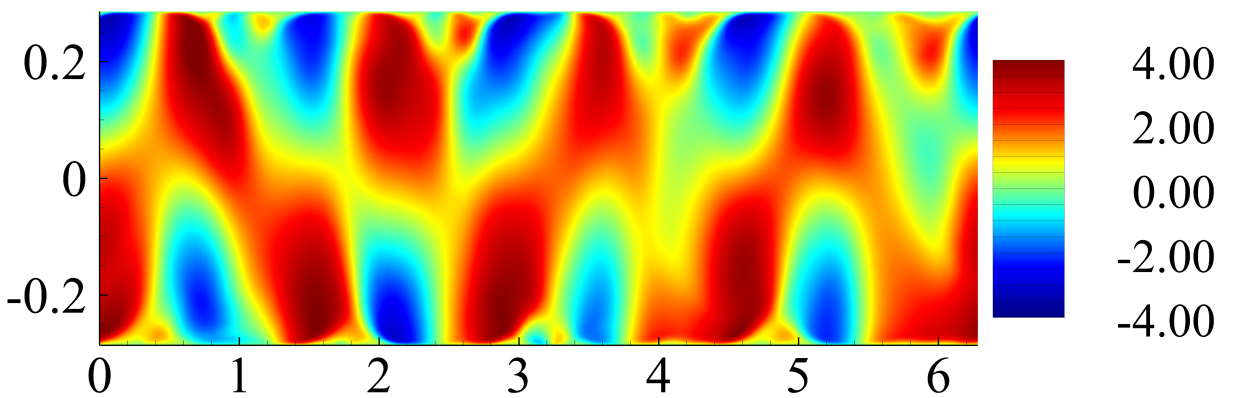}};
  	\node[left=of fig8_5, xshift=1.0cm ,yshift=0.90cm,rotate=0,scale=1.0,font=\color{black}] {({\it e})};

	\node[left=of fig8_5, xshift=1.00cm ,yshift=0.1cm,rotate=0,scale=1.0,font=\color{black}] {${\it z}$};
	
	\node[left=of fig8_5, xshift=7.6cm ,yshift=1.10cm,rotate=0,scale=1.0,font=\color{black}] {${\it u_x}$};
	
\node [below=of fig8_3, xshift=-2.25cm, yshift=-4.50cm]  (fig8_6)  {\includegraphics[scale=0.31]{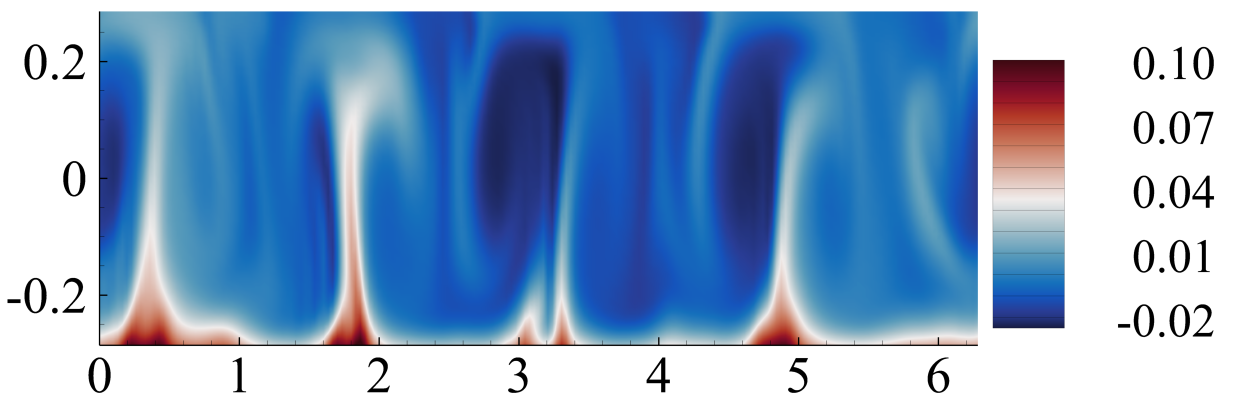}};
  	\node[left=of fig8_6, xshift=1.0cm ,yshift=0.90cm,rotate=0,scale=1.0,font=\color{black}] {({\it f})};

	\node[left=of fig8_6, xshift=1.00cm ,yshift=0.1cm,rotate=0,scale=1.0,font=\color{black}] {${\it z}$};
	\node[below=of fig8_6, xshift=-0.25cm ,yshift=1.0cm,rotate=0,font=\color{black}] {{\it x}};	
	
	\node[left=of fig8_6, xshift=7.6cm ,yshift=1.10cm,rotate=0,scale=1.0,font=\color{black}] {${\theta}$};
	
	
\node [below=of fig8_3, xshift=3.85cm, yshift=0.50cm]  (fig8_7)  {\includegraphics[scale=0.20]{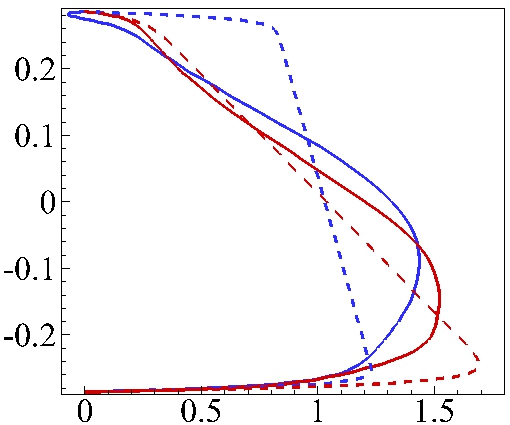}}; 
  	\node[left=of fig8_7, xshift=1.0cm ,yshift=1.40cm,rotate=0,font=\color{black}]{({\it g})};
	
	\node[left=of fig8_7, xshift=1.00cm ,yshift=0.15cm,rotate=0,scale=1.10,font=\color{black}] {$z$};
	\node[below=of fig8_7, xshift=0.25cm ,yshift=1.00cm,rotate=0,scale=1.10,font=\color{black}] {$u_x$};	
	
	
\node [below=of fig8_7, xshift=-0.05cm, yshift=0.35cm]  (fig8_8)  {\includegraphics[scale=0.20]{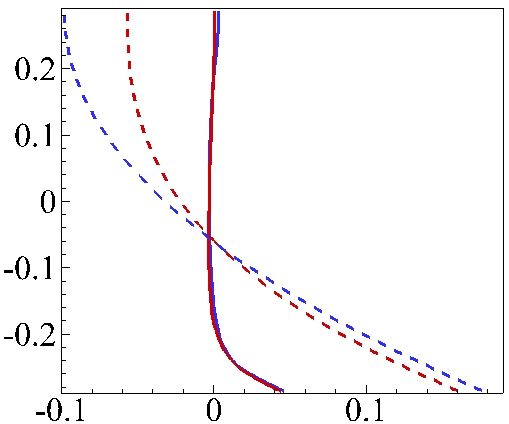}}; 
  	\node[left=of fig8_8, xshift=1.0cm ,yshift=1.40cm,rotate=0,font=\color{black}]{({\it h})};
	
	\node[left=of fig8_8, xshift=1.00cm ,yshift=0.15cm,rotate=0,scale=1.10,font=\color{black}] {$z$};
	\node[below=of fig8_8, xshift=0.25cm ,yshift=1.00cm,rotate=0,scale=1.10,font=\color{black}] {$\theta$};	
	
	
\node [right=of fig8_8, xshift=-3.22cm, yshift=0.58cm]  (fig8_l)  {\includegraphics[scale=0.19]{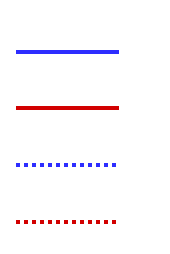}}; 

	\node[right=of fig8_l, xshift=-1.50cm ,yshift=0.55cm,rotate=0,scale=0.8,font=\color{black}]  {$3D$};
	\node[right=of fig8_l, xshift=-1.50cm ,yshift=0.15cm,rotate=0,scale=0.8,font=\color{black}]  {$2D$};	
	\node[right=of fig8_l, xshift=-1.50cm ,yshift=-0.27cm,rotate=0,scale=0.8,font=\color{black}]  {$3D, base$};
	\node[right=of fig8_l, xshift=-1.50cm ,yshift=-0.65cm,rotate=0,scale=0.8,font=\color{black}]  {$2D, base$};

	\end{tikzpicture}

\caption{Flow structure in nonlinear regime at $Gr = 10^{10}$, $Ha = 10000$. The instantaneous distributions of $u_z$ ($a$, $d$), $u_x$ ($b$, $e$), and $\theta$ ($c$, $f$) obtained in 2D and 3D (plotted in the midplane $y = 0$) models are shown in ($a$)-($c$) and ($d$)-($f$), respectively. The profiles of $u_x$ and $\theta$ obtained by averaging over $x$ and time (with the base flow profiles from figure \ref{fig5}$c,f$) are shown, respectively, in ($g$) and ($h$).}
  
\label{fig8}
\end{figure}

\begin{figure}
	\centering 

	
\begin{tikzpicture}


\node (fig9_1) {\includegraphics[scale=0.25]{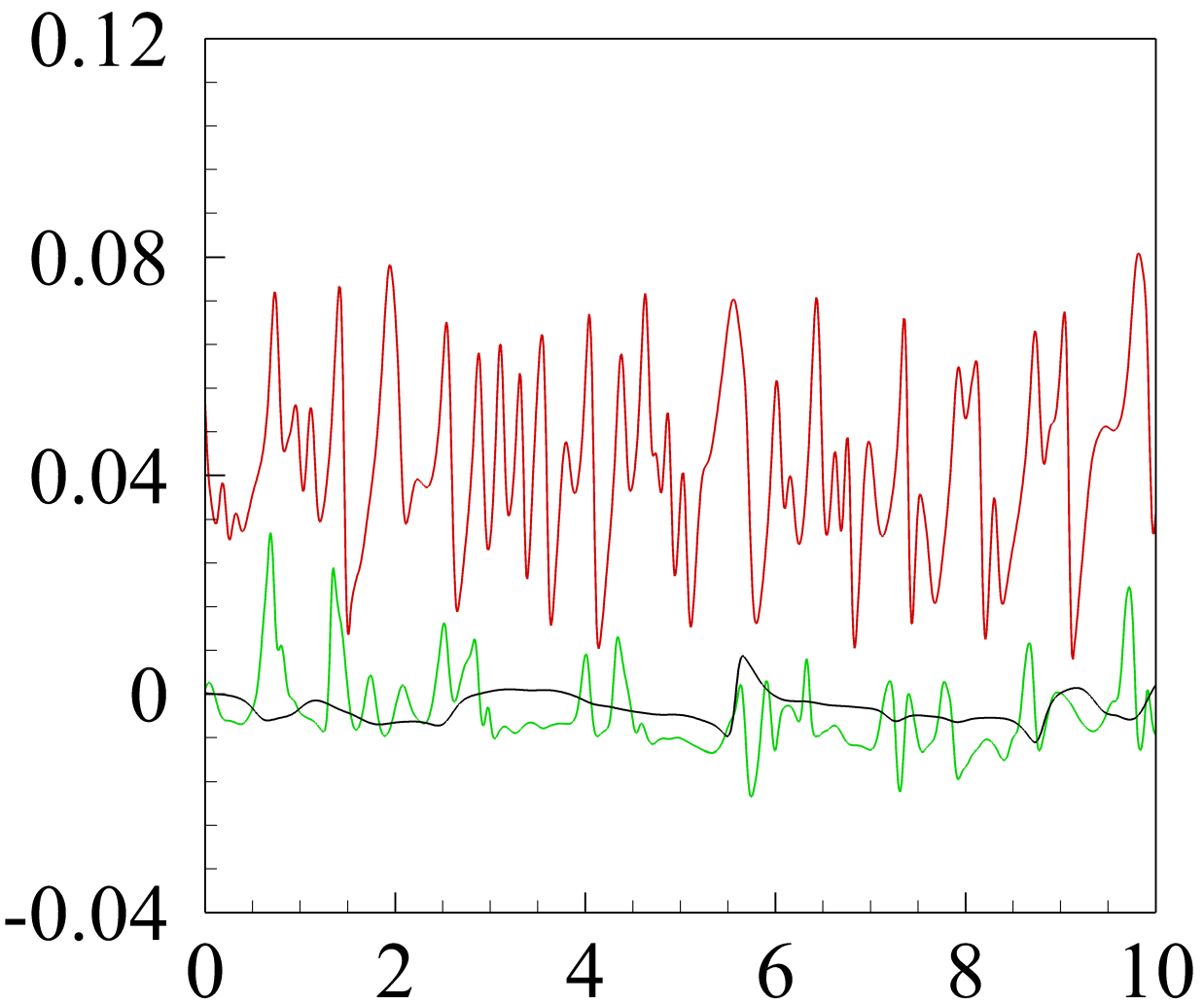}}; 
  	\node[left=of fig9_1, xshift=1.00cm ,yshift=2.00cm,rotate=0,font=\color{black}] {({\it a})};
	
	\node[left=of fig9_1, xshift=1.15cm ,yshift=0.15cm,rotate=0,scale=1.25,font=\color{black}] {$\theta$};
	\node[below=of fig9_1, xshift=0.25cm ,yshift=1.00cm,rotate=0,scale=1.10,font=\color{black}] {$t$};	
	
	
\node [below=of fig9_1, xshift=0.0cm, yshift=0.60cm]  (fig9_2)  {\includegraphics[scale=0.25]{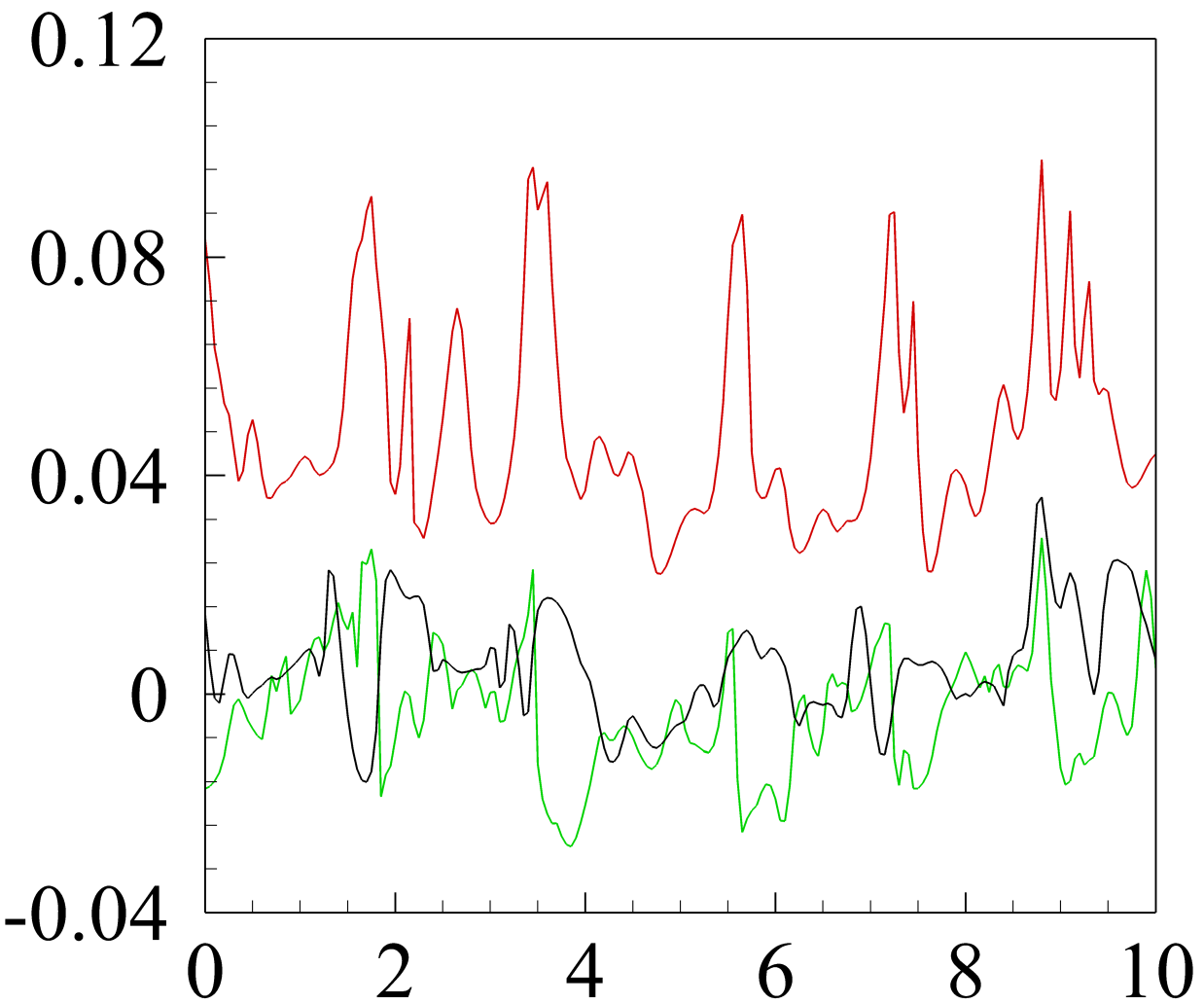}}; 
  	\node[left=of fig9_2, xshift=1.00cm ,yshift=2.00cm,rotate=0,font=\color{black}]{({\it c})};
	
	\node[left=of fig9_2, xshift=1.15cm ,yshift=0.15cm,rotate=0,scale=1.25,font=\color{black}] {$\theta$};
	\node[below=of fig9_2, xshift=0.25cm ,yshift=1.00cm,rotate=0,scale=1.10,font=\color{black}] {$t$};	
	
		
\node [below=of fig9_2, xshift=-1.00cm, yshift=0.65cm]  (fig9_l1)  {\includegraphics[scale=0.25]{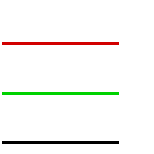}}; 

	\node[right=of fig9_l1, xshift=-1.25cm ,yshift=0.30cm,rotate=0,scale=1.0,font=\color{black}]  {${\it z} = -0.2857$};
	\node[right=of fig9_l1, xshift=-1.25cm ,yshift=-0.10cm,rotate=0,scale=1.0,font=\color{black}]  {${\it z} = 0$};
	\node[right=of fig9_l1, xshift=-1.25cm ,yshift=-0.50cm,rotate=0,scale=1.0,font=\color{black}]  {${\it z} = 0.2857$};


\node [right=of fig9_1, xshift=-0.7cm, yshift=0.0cm]  (fig9_3) {\includegraphics[scale=0.25]{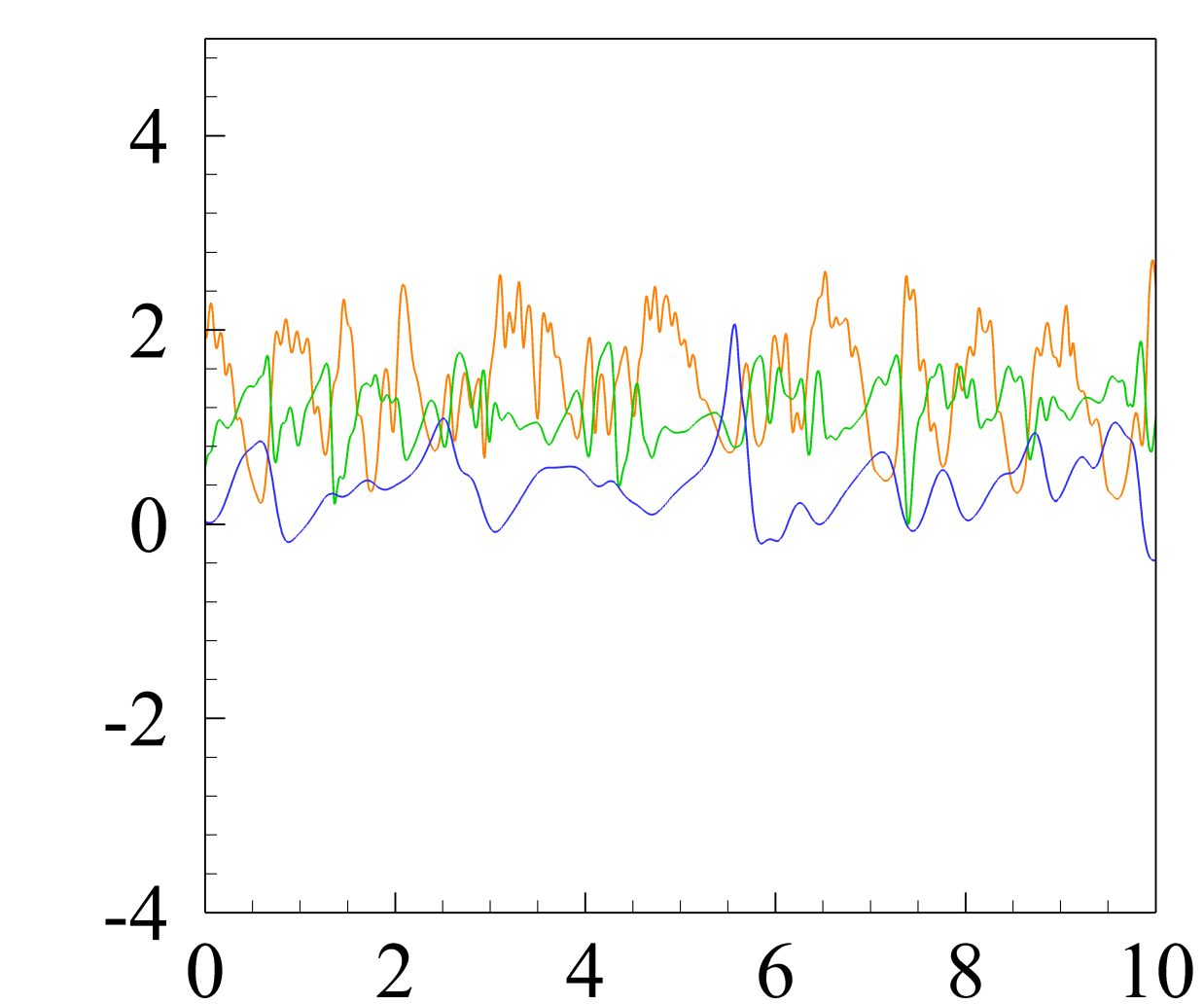}}; 
  	\node[left=of fig9_3, xshift=1.40cm ,yshift=2.00cm,rotate=0,font=\color{black}] {({\it b})};
	
	\node[left=of fig9_3, xshift=1.55cm ,yshift=0.15cm,rotate=0,scale=1.10,font=\color{black}] {$u_x$};
	\node[below=of fig9_3, xshift=0.25cm ,yshift=1.00cm,rotate=0,scale=1.10,font=\color{black}] {$t$};	
	
	
\node [below=of fig9_3, xshift=0.0cm, yshift=0.60cm]  (fig9_4)  {\includegraphics[scale=0.25]{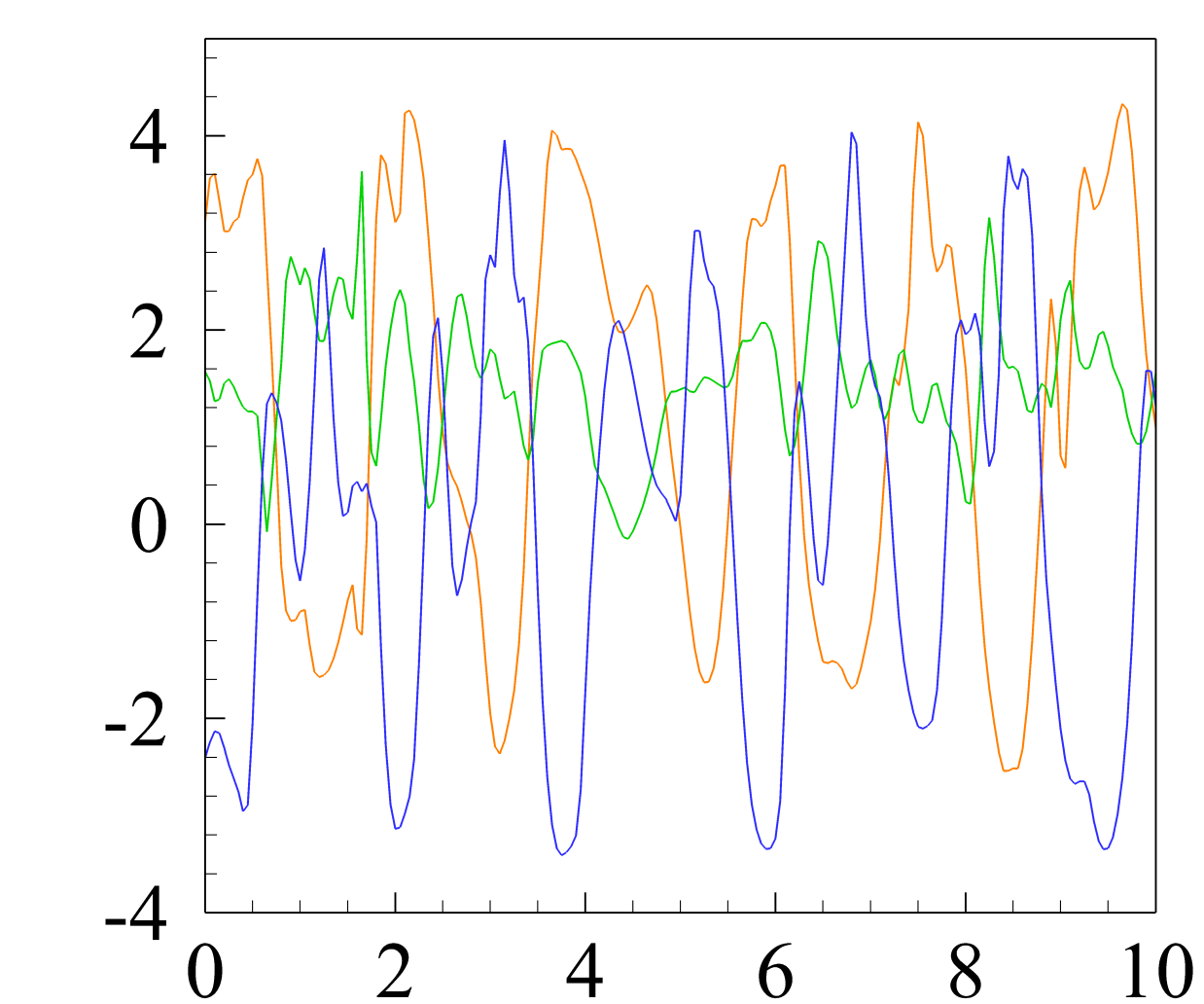}}; 
  	\node[left=of fig9_4, xshift=1.40cm ,yshift=2.00cm,rotate=0,font=\color{black}]{({\it d})};
	
	\node[left=of fig9_4, xshift=1.55cm ,yshift=0.15cm,rotate=0,scale=1.10,font=\color{black}] {$u_x$};
	\node[below=of fig9_4, xshift=0.25cm ,yshift=1.00cm,rotate=0,scale=1.10,font=\color{black}] {$t$};	

		
\node [below=of fig9_4, xshift=-1.00cm, yshift=0.65cm]  (fig9_l2)  {\includegraphics[scale=0.25]{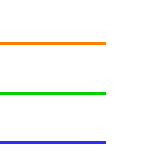}}; 

	\node[right=of fig9_l2, xshift=-1.25cm ,yshift=0.30cm,rotate=0,scale=1.0,font=\color{black}]  {${\it z} = -0.2789$};
	\node[right=of fig9_l2, xshift=-1.25cm ,yshift=-0.10cm,rotate=0,scale=1.0,font=\color{black}]  {${\it z} = 0$};
	\node[right=of fig9_l2, xshift=-1.25cm ,yshift=-0.50cm,rotate=0,scale=1.0,font=\color{black}]  {${\it z} = 0.2789$};

\end{tikzpicture}

  \caption{Time signals of temperature ($a, c$) and streamwise velocity ($b, d$) at $y=0$ (in the 3D flow) in the nonlinear regime at $Ha = 10^4$, $Gr = 10^{10}$ are shown for 2D ($a, b$) and 3D ($c, d$) models.}
  
\label{fig9}
\end{figure}

The results presented so far in this section indicate that the SM82 2D model may also inaccurately describe the nonlinear flow regimes developing as a result of the instability at $Gr = 10^{10}$. As a test of this possibility, comparison between the results of 2D and 3D models at $Gr = 10^{10}$, $Ha = 10000$ is illustrated in figures \ref{fig8} and \ref{fig9} and discussed below. The procedure of computing nonlinear flows is described in section \ref{results4}. Here we only mention that the same numerical resolution is used in 2D and 3D models. The shorter wavelength domain length $L_x = 2\pi$ is used in the 3D model. This rather small length has no significant effect on the flow evolution as it has been confirmed in the additional 2D simulations conducted with $L_x = 4\pi$ (shown in figure \ref{fig8}) and $L_x = 2\pi$.


The simulations show that the 3D flow remains Q2D at these values of $Ha$ and $Gr$. At the same time, its reproduction by the 2D SM82 model is inaccurate in some aspects. The instantaneous distributions of velocity components and temperature shown in figure \ref{fig8} clearly illustrate the difference. 2D approximations (see figure \ref{fig8}$a$-$c$) are similar to 3D flows (see figure \ref{fig8}$d$-$f$) in terms of the largest typical streamwise wavelength (about $1.5$) but demonstrate noticeably less regular pattern and higher energy in shorter wavelengths. The difference is reflected by the point signals of velocity and temperature shown in figure \ref{fig9}. It is also observed in the power spectrum density graphs of velocity and temperature (not shown). Comparing figures \ref{fig9}$b$ and \ref{fig9}$d$, we also see that the 2D approximation substantially underestimates the typical amplitude of velocity fluctuations. As an example, the standard deviation for the signal of $u_x$ in the middle of the duct ($z = 0$) is $0.33$ for the 2D model and $0.71$ for the 3D model. Computed values of volume-average kinetic energy of the fluctuations (not shown) confirm this conclusion. The values of energy found in the 2D approximation are about an order of magnitude lower than in the actual flow computed in the framework of the 3D model. Interestingly, the misrepresentation of the structure of velocity and temperature fluctuations by the 2D model does not lead to a similarly strong inaccuracy in the prediction of the effect of mixing by fluctuations. Profiles of average streamwise velocity and temperature in figures \ref{fig8}($g, h$) show strong change in comparison with the base flow, but only moderate differences between the 2D and 3D results. 

\subsubsection{Applicability of the SM82 model: summary} \label{2d3dsummary}

We conclude that the SM82 approximation accurately represents Q2D flows at moderately large $Gr$ ($10^8$ and $10^9$ in our system). The accuracy deteriorates at higher $Gr$ even though the flow remains Q2D. An example of this is observed at $Ha \gtrsim 4000$ and $Gr = 10^{10}$. The base flow profiles are clearly different between the 2D and 3D models (see figures \ref{fig5}$c,f$). The 2D linear stability analysis is qualitatively correct in the sense that it correctly predicts the principal type of the unstable perturbations and the streamwise wavelength of the most unstable mode (see figure \ref{fig6}$c$). The values of the growth rate $\gamma$ are, however, substantially underestimated by the 2D model (see figure \ref{fig6}$c$ and table \ref{table4}). The nonlinear flow states resulting from the instability are predicted incorrectly by the 2D model, which adds artificial irregularity and short-wave fluctuations and underestimates the amplitude of velocity fluctuations (see figures \ref{fig8} and \ref{fig9}). It should be noted that the discrepancy is not due to irregularities of the model or our computational procedure. Calculations carried out at lower Gr and high Ha reveal a perfectly good agreement in the bulk region between the 2D and 3D models. 

We do not have a satisfactory explanation to this effect and leave its further exploration for future studies. It should be mentioned that the quality of the 2D approximation improves with increasing $Ha$. As an example, the values of the linear instability growth rate $\gamma$ shown in table \ref{table4} are underpredicted by the 2D model by about $35 \%$ at $Ha = 4000$ and by about only $18 \%$ at $Ha = 10000$.

It must be mentioned that this is not the only example of the model's breakdown. The model is known to break down when any 3D structures are present in the flow for which the diffusion length is shorter than the size of the domain \citep{Potherat14, Potherat17}.

In the remaining part of this paper, the discussion of the instability and nonlinear states is based on the SM82 2D approximation for flows at $Gr = 10^8$ and $10^9$ and on the full 3D solutions for flows at $10^{10}$.

The numerical solution for the 2D approximation is obtained using a modified version of the code which solves (\ref{eqsm1})$-$(\ref{eqsm3}). The code has been verified through comparison of its results with the analytical solution of  (\ref{eqsm4})$-$(\ref{eqsm7}) for the base flow.


 \subsection{Results of linear stability analysis} \label{results3}
 
\begin{figure}
	\centering 
	
\begin{tikzpicture}


\node (fig10a1) {\includegraphics[scale=0.3]{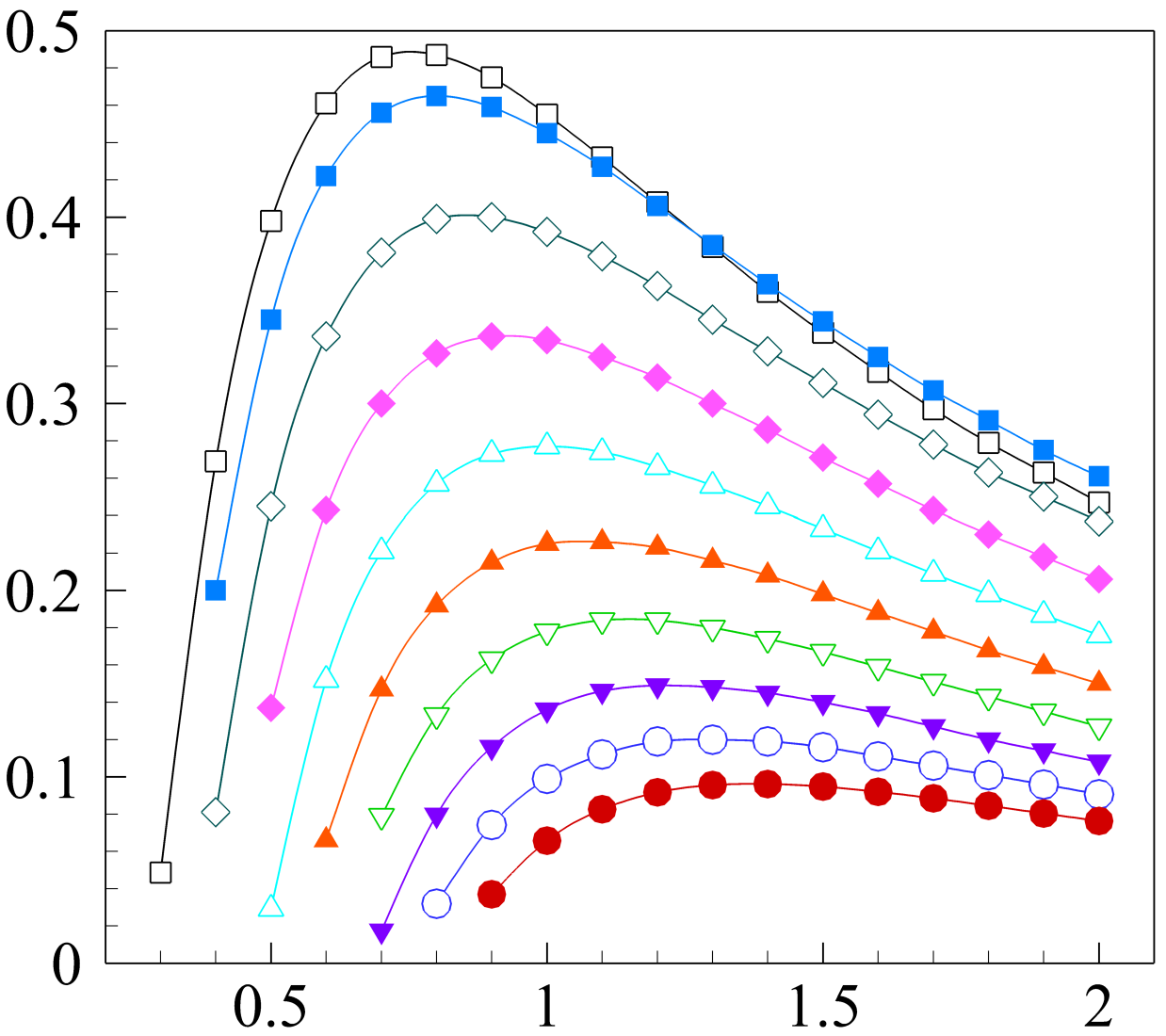}};
  	\node[left=of fig10a1, xshift=1.0cm ,yshift=2.75cm,rotate=0,scale=1.0,font=\color{black}] {({\it a})};

	\node[left=of fig10a1, xshift=1.00cm ,yshift=0.0cm,rotate=0,scale=1.1,font=\color{black}] {${\it \gamma}$};
	\node[below=of fig10a1, xshift=0.15cm ,yshift=1.05cm,rotate=0,scale=1.1,font=\color{black}] {${\it \lambda}$};
	fig11
	\node[right=of fig10a1, xshift=1.4cm ,yshift=1.95cm,rotate=0,scale=1.15,font=\color{black}]  {${\it Ha} = 1000$};
	\node[right=of fig10a1, xshift=1.4cm ,yshift=1.50cm,rotate=0,scale=1.15,font=\color{black}]  {${\it Ha} = 2000$};
	\node[right=of fig10a1, xshift=1.4cm ,yshift=1.05cm,rotate=0,scale=1.15,font=\color{black}]  {${\it Ha} = 3000$};
	\node[right=of fig10a1, xshift=1.4cm ,yshift=0.60cm,rotate=0,scale=1.15,font=\color{black}]  {${\it Ha} = 4000$};
	\node[right=of fig10a1, xshift=1.4cm ,yshift=0.15cm,rotate=0,scale=1.15,font=\color{black}]  {${\it Ha} = 5000$};
	\node[right=of fig10a1, xshift=1.4cm ,yshift=-0.30cm,rotate=0,scale=1.15,font=\color{black}]  {${\it Ha} = 6000$};
	\node[right=of fig10a1, xshift=1.4cm ,yshift=-0.75cm,rotate=0,scale=1.15,font=\color{black}]  {${\it Ha} = 7000$};
	\node[right=of fig10a1, xshift=1.4cm ,yshift=-1.20cm,rotate=0,scale=1.15,font=\color{black}]  {${\it Ha} = 8000$};
	\node[right=of fig10a1, xshift=1.4cm ,yshift=-1.65cm,rotate=0,scale=1.15,font=\color{black}]  {${\it Ha} = 9000$};
	\node[right=of fig10a1, xshift=1.4cm ,yshift=-2.10cm,rotate=0,scale=1.15,font=\color{black}]  {${\it Ha} = 10000$};

\node [right=of fig10a1, xshift=-1.0cm, yshift=0.00cm]  (fig10a2)  {\includegraphics[scale=0.19]{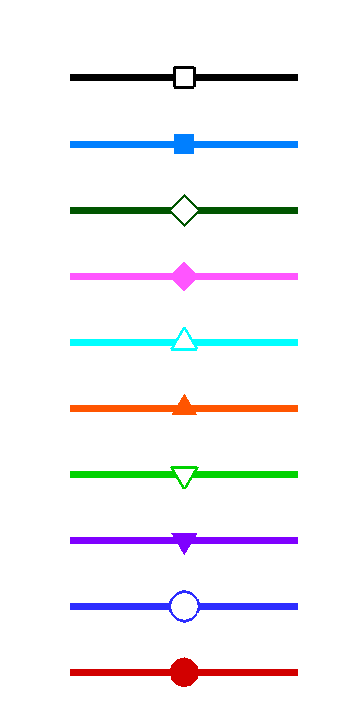}}; 


\node [below=of fig10a1, xshift=0.00cm, yshift=0.5cm]  (fig10b1)  {\includegraphics[scale=0.3]{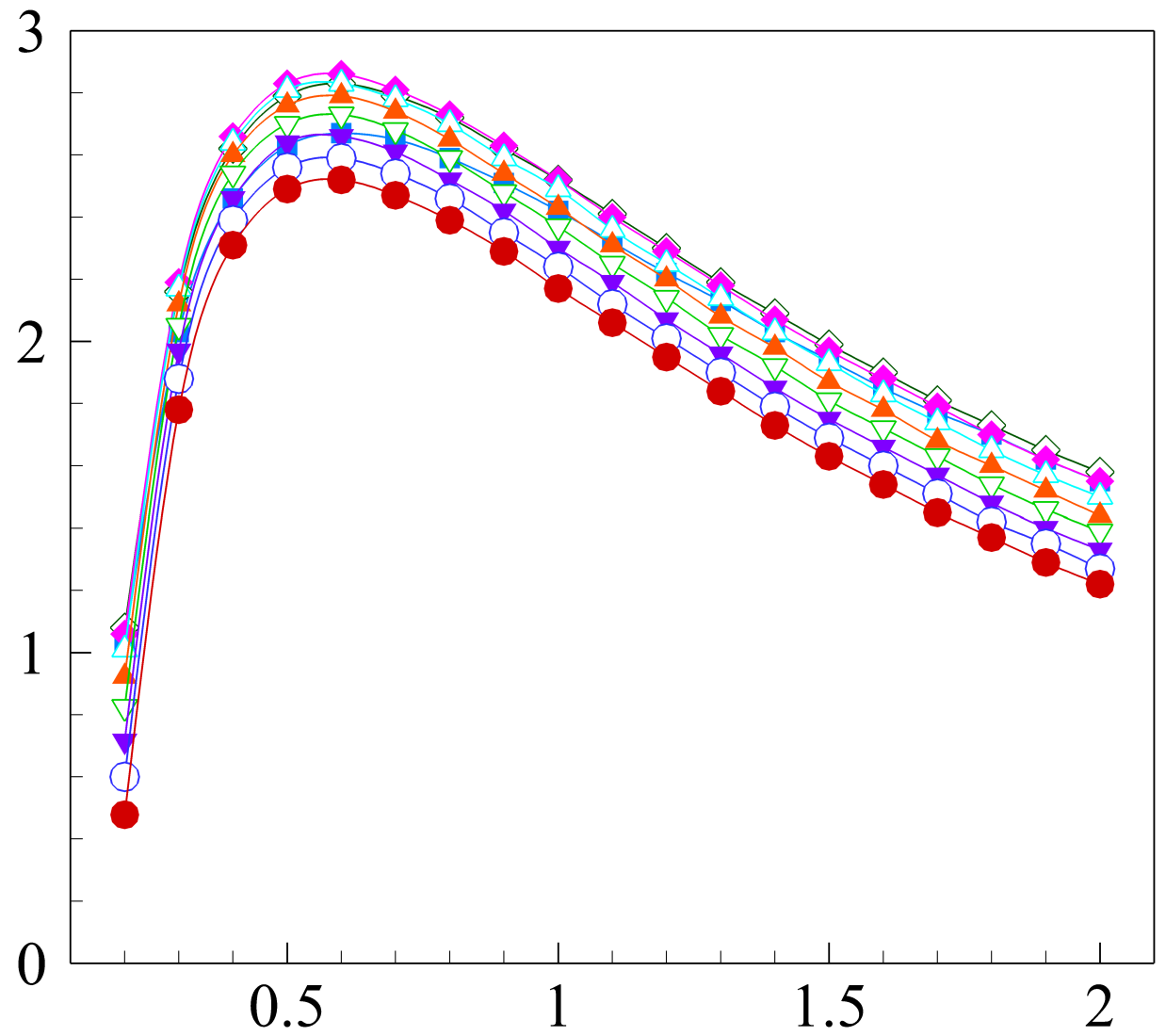}}; 
\node [right=of fig10b1, xshift=-1.00cm, yshift=0.00cm]  (fig10b2)  {\includegraphics[scale=0.14]{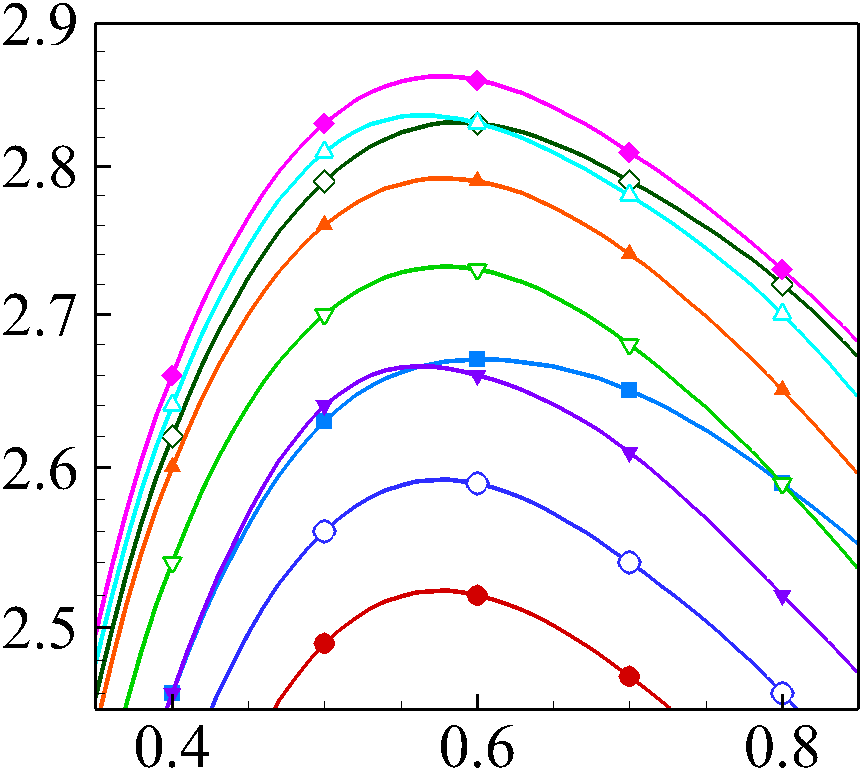}}; 
  	\node[left=of fig10b1, xshift=1.0cm ,yshift=2.75cm,rotate=0,scale=1.0,font=\color{black}] {({\it b})};

	\node[left=of fig10b1, xshift=1.00cm ,yshift=0.0cm,rotate=0,scale=1.1,font=\color{black}] {${\it \gamma}$};
	\node[below=of fig10b1, xshift=0.15cm ,yshift=1.05cm,rotate=0,scale=1.1,font=\color{black}] {${\it \lambda}$};
	
	
\node [below=of fig10b1, xshift=0.00cm, yshift=0.5cm]  (fig10c1)  {\includegraphics[scale=0.3]{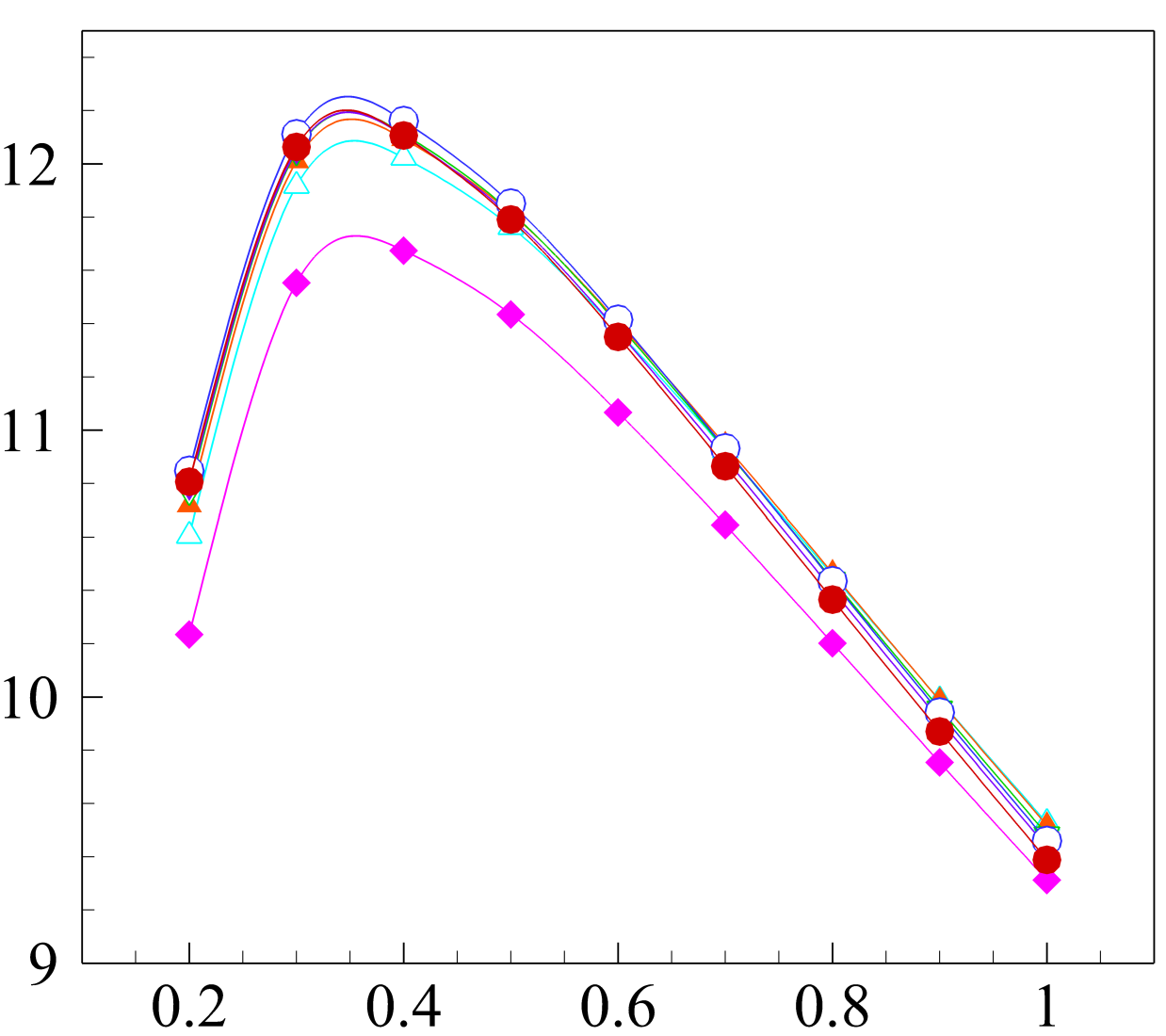}};
\node [right=of fig10c1, xshift=-1.00cm, yshift=0.00cm]  (fig10c2)  {\includegraphics[scale=0.14]{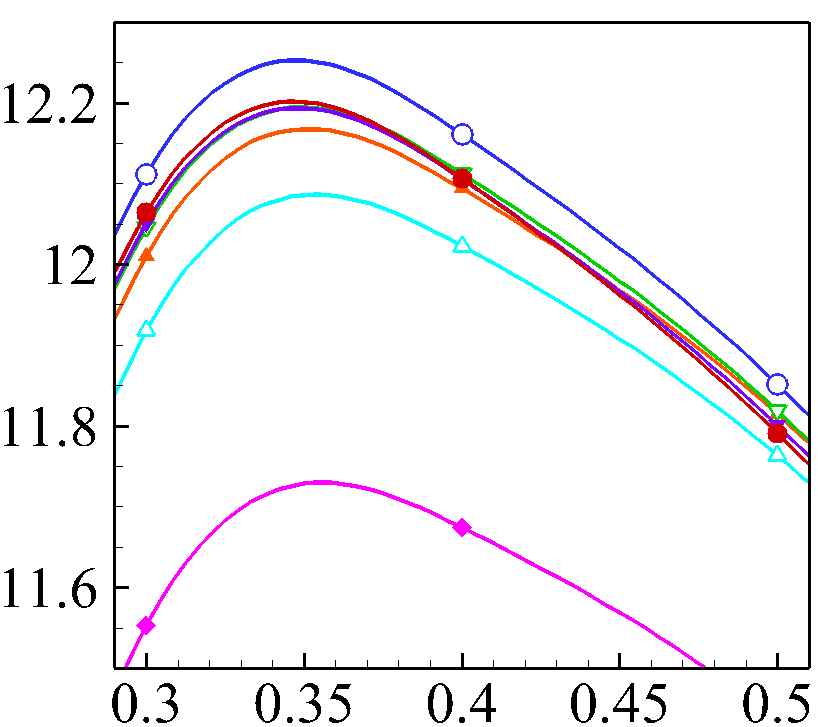}}; 
  	\node[left=of fig10c1, xshift=1.0cm ,yshift=2.75cm,rotate=0,scale=1.0,font=\color{black}] {({\it c})};

	\node[left=of fig10c1, xshift=1.00cm ,yshift=0.0cm,rotate=0,scale=1.1,font=\color{black}] {${\it \gamma}$};
	\node[below=of fig10c1, xshift=0.15cm ,yshift=1.05cm,rotate=0,scale=1.1,font=\color{black}] {${\it \lambda}$};
	
\end{tikzpicture}

\caption{Rates of exponential growth $\gamma$ shown as functions of axial wavelength $\lambda$ at $Gr = 10^8$ ($a$), $Gr = 10^9$ ($b$), and $Gr = 10^{10}$ ($c$) for various values of $Ha$. Results of 2D SM82 approximation are shown for $Gr = 10^8$ and $10^9$. Results of 3D computational analysis are shown for $Gr = 10^{10}$. Additional plots in ($b$) and ($c$) show zoomed-in areas around ($\lambda_{max}, \gamma_{max}$). }
  
\label{fig10}
\end{figure}
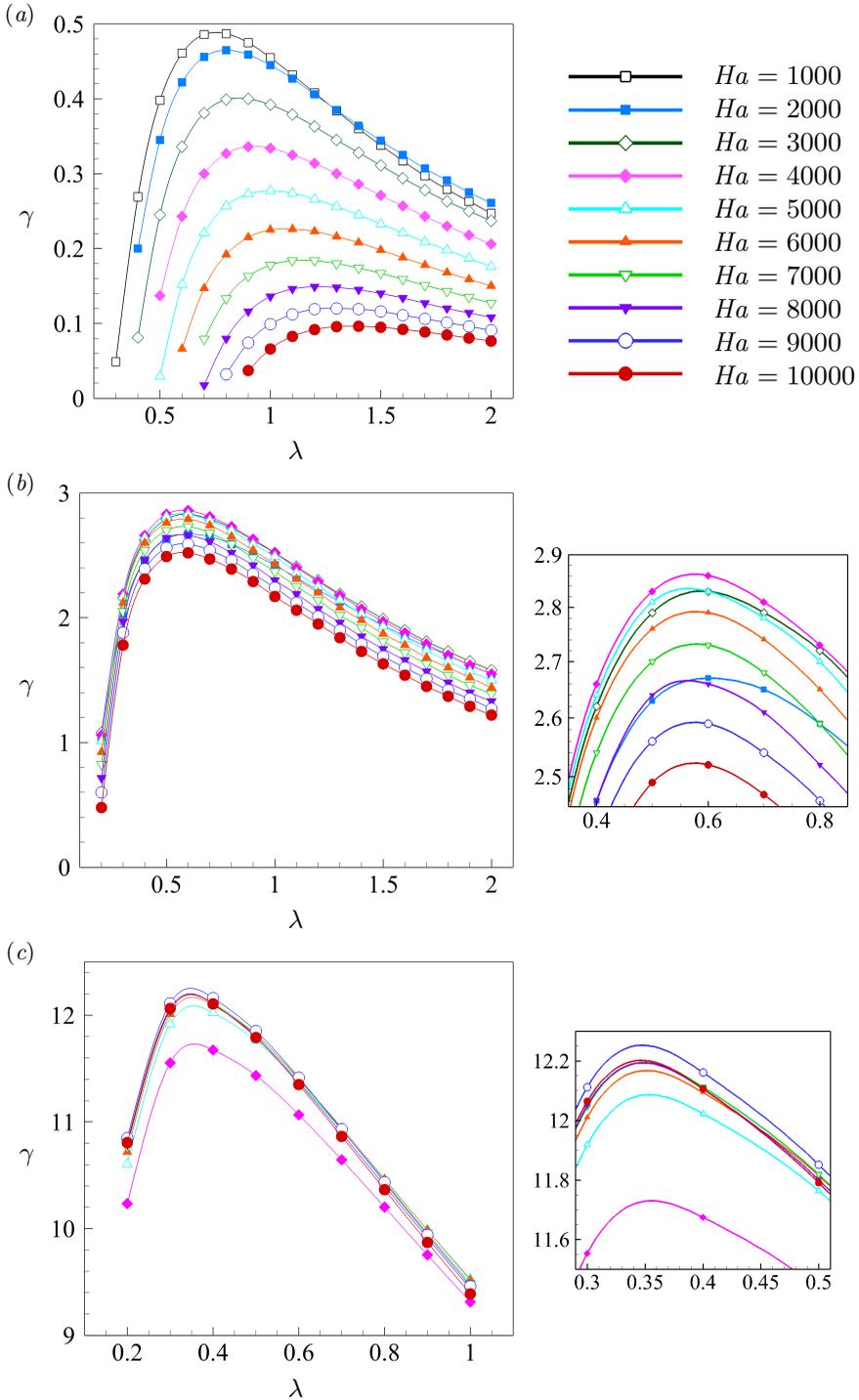

\begin{figure}
	\centering 
	
\begin{tikzpicture}


\node (fig11a) {\includegraphics[scale=0.375]{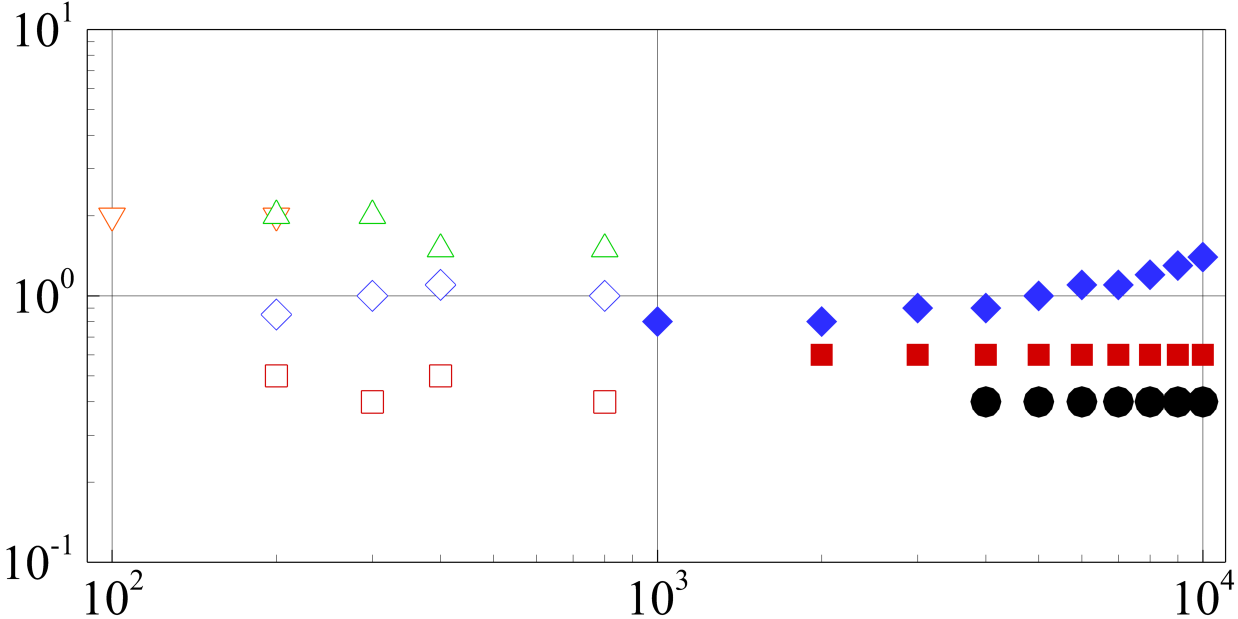}}; 

  	\node[left=of fig11a, xshift=0.90cm ,yshift=1.7cm,font=\color{black}] {({\it a})};

  	\node[left=of fig11a, xshift=0.90cm ,yshift=0.50cm,rotate=90,font=\color{black}] {${\lambda_{max}}$};
	
	\node[left=of fig11a, xshift=5.65cm ,yshift=-2.25cm,rotate=0,font=\color{black}] {$Ha$};
	

\node [below=of fig11a, xshift=0.00cm, yshift=0.75cm] (fig11b)  {\includegraphics[scale=0.375]{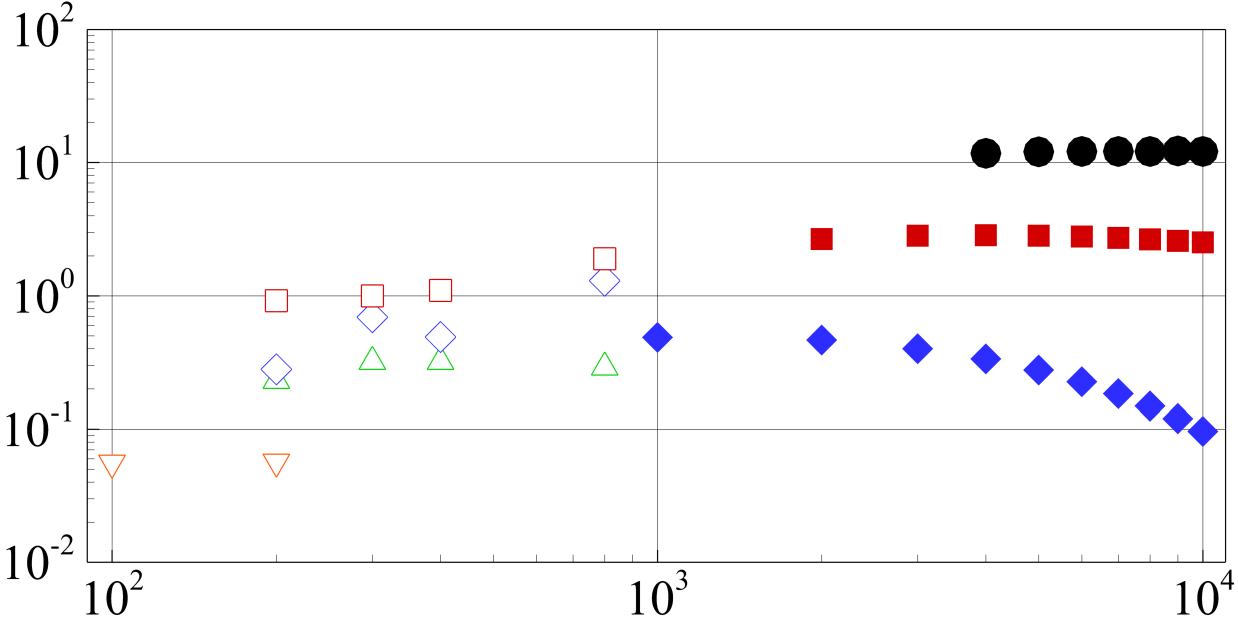}};

  	\node[left=of fig11b, xshift=0.90cm ,yshift=1.7cm,font=\color{black}] {({\it b})};
	
	\node[left=of fig11b, xshift=0.90cm ,yshift=0.50cm,rotate=90,font=\color{black}] {${\gamma_{max}}$};

	\node[left=of fig11b, xshift=5.65cm ,yshift=-2.25cm,rotate=0,font=\color{black}] {$Ha$};

\node [right=of fig11a, xshift=-1.00cm, yshift=0.15cm] (fig11_2)  {\includegraphics[scale=0.175]{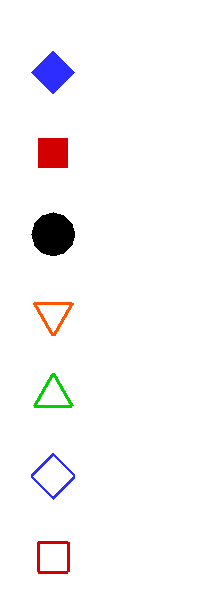}};

  	\node[right=of fig11_2, xshift=-1.75cm ,yshift=1.45cm,rotate=0,scale=1.0,font=\color{black}] {$Gr=10^8$};
  	\node[right=of fig11_2, xshift=-1.75cm ,yshift=0.95cm,rotate=0,scale=1.0,font=\color{black}] {$Gr=10^9$};
  	\node[right=of fig11_2, xshift=-1.75cm ,yshift=0.45cm,rotate=0,scale=1.0,font=\color{black}] {$Gr=10^{10}$};
  	\node[right=of fig11_2, xshift=-1.75cm ,yshift=-0.05cm,rotate=0,scale=1.0,font=\color{black}] {$Gr=10^6$};
  	\node[right=of fig11_2, xshift=-1.75cm ,yshift=-0.55cm,rotate=0,scale=1.0,font=\color{black}] {$Gr=10^7$};
  	\node[right=of fig11_2, xshift=-1.75cm ,yshift=-1.05cm,rotate=0,scale=1.0,font=\color{black}] {$Gr=10^8$};
  	\node[right=of fig11_2, xshift=-1.75cm ,yshift=-1.55cm,rotate=0,scale=1.0,font=\color{black}] {$Gr=10^9$};

\end{tikzpicture}

  \caption{The wavelength $\lambda_{max}$ ($a$) and the exponential growth rate $\gamma_{max}$ ($b$) of the fastest growing perturbations as a function of $Ha$. The data are taken from table \ref{table5}. The results of the current work and that of \citet{Zhang14} are denoted as filled and unfilled elements, respectively. Results of 2D SM82 approximation are shown for $Gr = 10^8$ and $10^9$. Results of 3D computational analysis are shown for $Gr = 10^{10}$.} 

\label{fig11}
\end{figure}

\begin{figure}
	\centering 
	
\begin{tikzpicture}


\node (fig12a) {\includegraphics[scale=0.225]{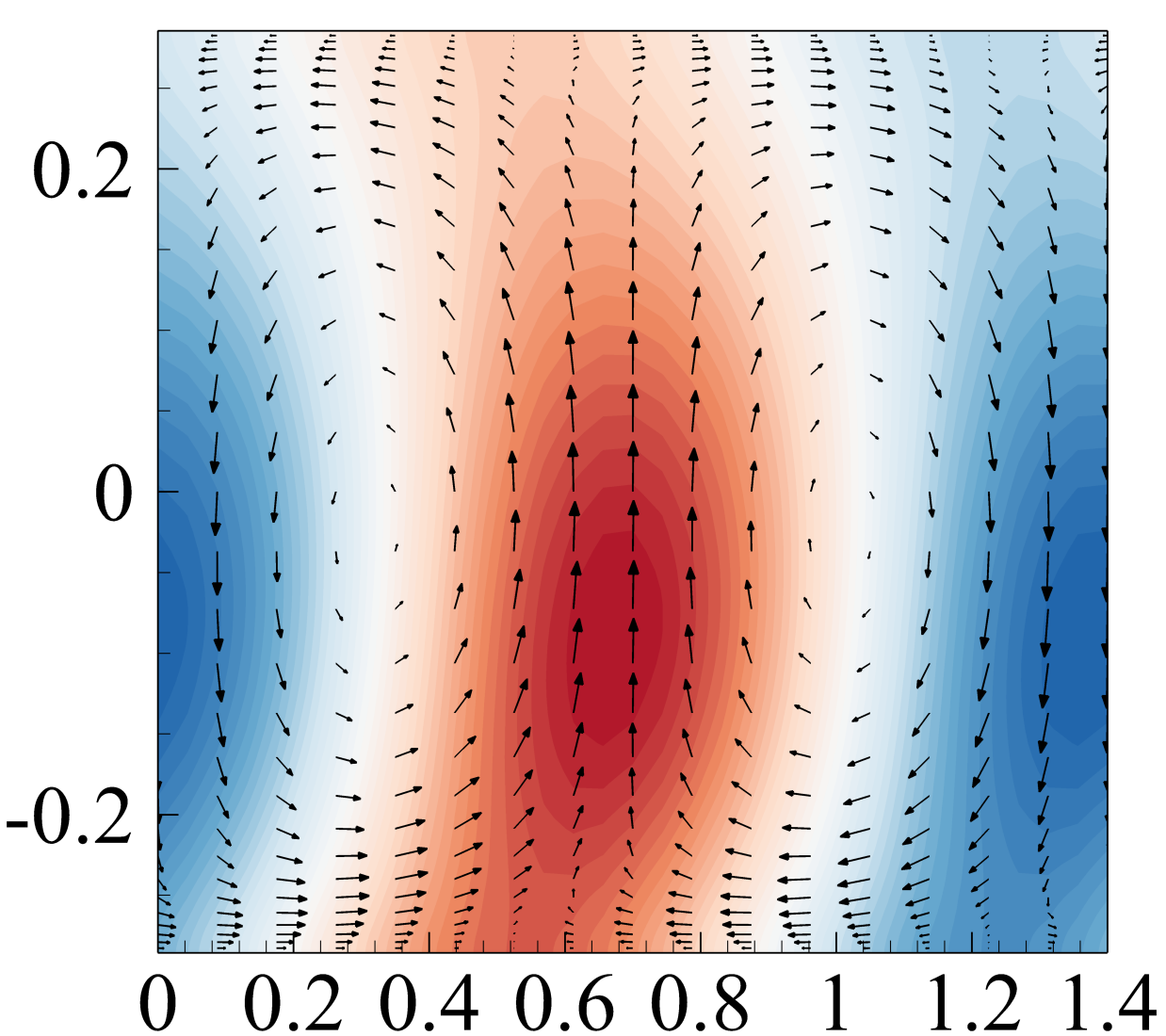}}; 
  	\node[left=of fig12a, xshift=1.25cm ,yshift=2.25cm,rotate=0,scale=1.1,font=\color{black}] {({\it a})};
	
	\node[left=of fig12a, xshift=1.15cm ,yshift=0.10cm,rotate=0,scale=1.1,font=\color{black}] {{\it z}};
	\node[below=of fig12a, xshift=0.25cm ,yshift=1.0cm,rotate=0,scale=1.1,font=\color{black}] {{\it x}};
	
	\node[above=of fig12a, xshift=0.25cm ,yshift=-1.15cm,rotate=0,scale=1.1,font=\color{black}] {${\Theta}$};	
	

\node [below=of fig12a, xshift=0.00cm, yshift=0.00cm]  (fig12b) {\includegraphics[scale=0.225]{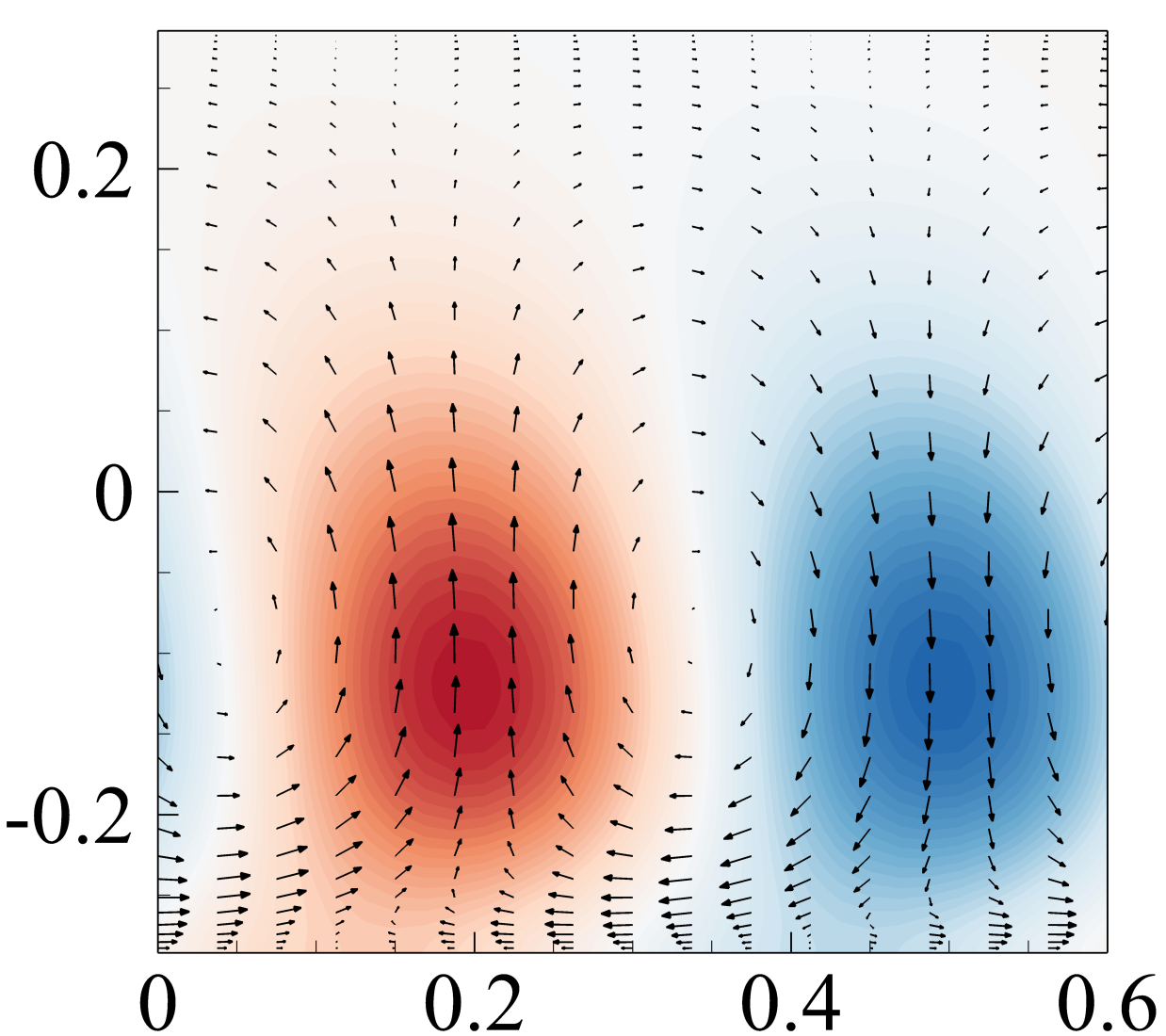}}; 
  	\node[left=of fig12b, xshift=1.25cm ,yshift=2.25cm,rotate=0,scale=1.1,font=\color{black}] {({\it b})};
	
	\node[left=of fig12b, xshift=1.15cm ,yshift=0.10cm,rotate=0,scale=1.1,font=\color{black}] {{\it z}};
	\node[below=of fig12b, xshift=0.25cm ,yshift=1.0cm,rotate=0,scale=1.1,font=\color{black}] {{\it x}};
	
	\node[above=of fig12b, xshift=0.25cm ,yshift=-1.15cm,rotate=0,scale=1.1,font=\color{black}] {${\Theta}$};


\node [right=of fig12a, xshift=-0.50cm, yshift=0.00cm]  (fig12c)  {\includegraphics[scale=0.225]{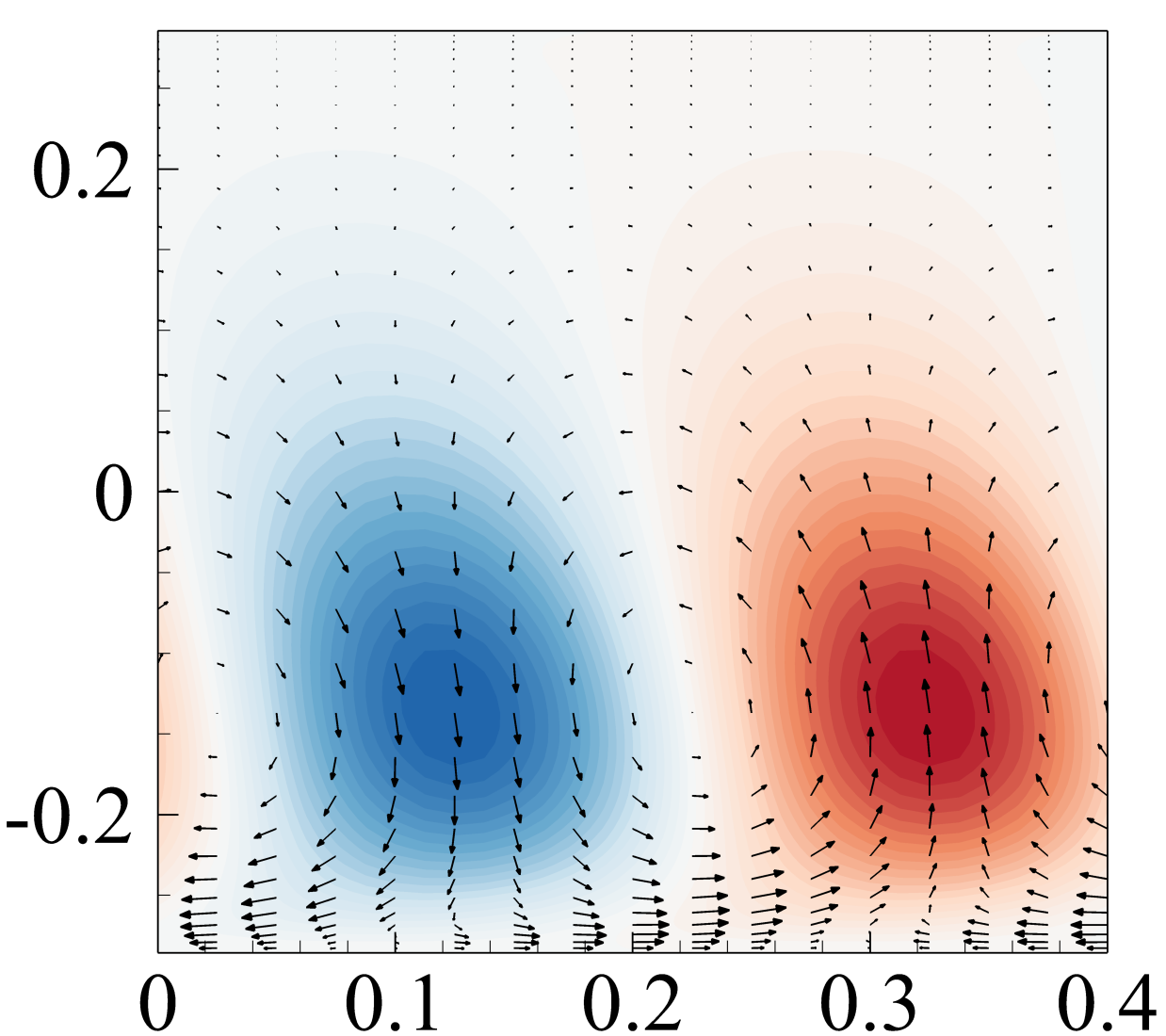}};
  	\node[left=of fig12c, xshift=1.25cm ,yshift=2.25cm,rotate=0,scale=1.1,font=\color{black}] {({\it c})};
	
	\node[left=of fig12c, xshift=1.15cm ,yshift=0.10cm,rotate=0,scale=1.1,font=\color{black}] {{\it z}};
	\node[below=of fig12c, xshift=0.25cm ,yshift=1.0cm,rotate=0,scale=1.1,font=\color{black}] {{\it x}};
	
	\node[above=of fig12c, xshift=0.25cm ,yshift=-1.15cm,rotate=0,scale=1.1,font=\color{black}] {${\Theta}$};

			
\node [below=of fig12c, xshift=0.00cm, yshift=0.00cm]  (fig12d) {\includegraphics[scale=0.225]{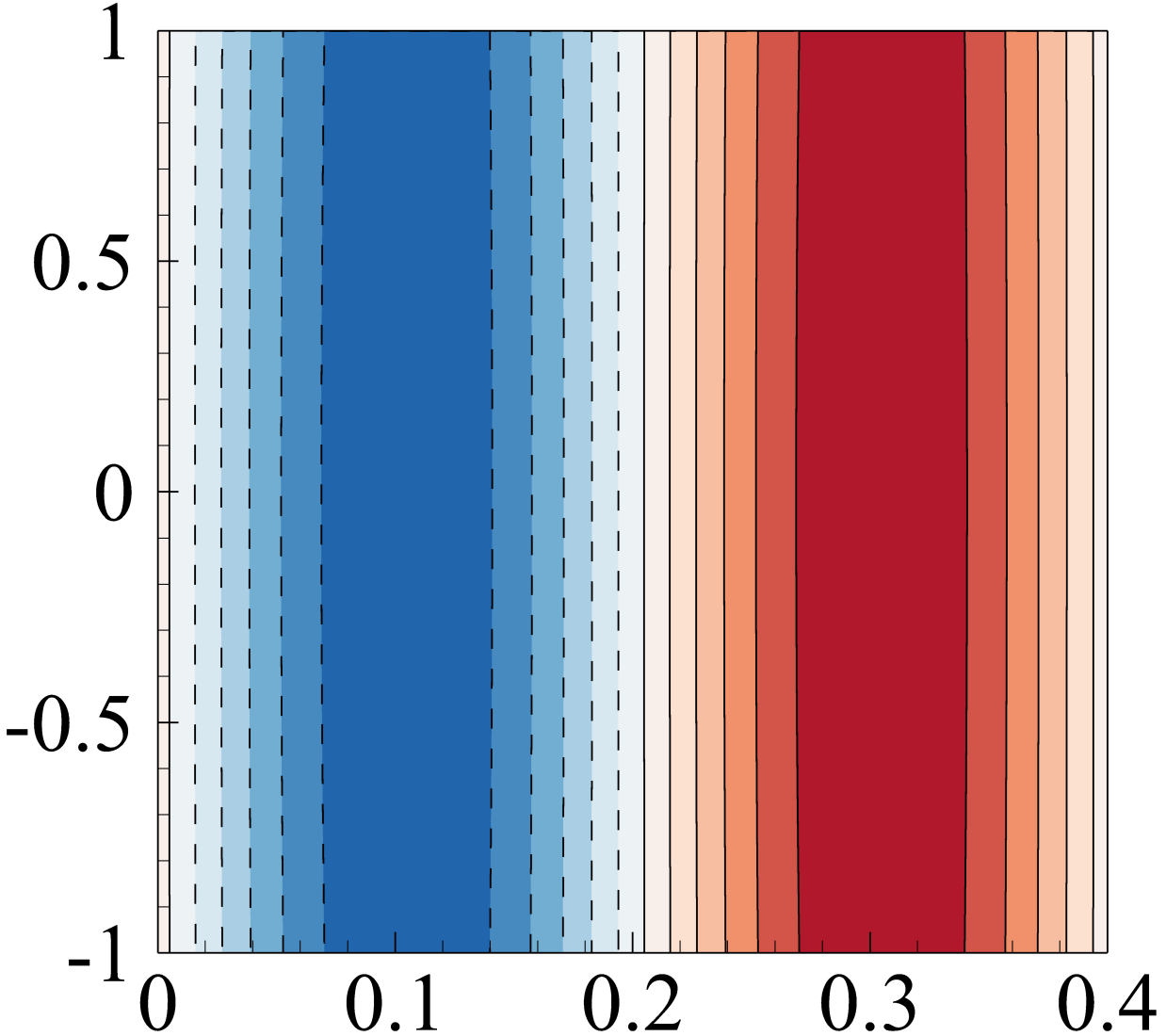}}; 
  	\node[left=of fig12d, xshift=1.25cm ,yshift=2.25cm,rotate=0,scale=1.1,font=\color{black}] {({\it d})};		
	
	\node[above=of fig12d, xshift=0.25cm ,yshift=-1.15cm,rotate=0,scale=1.1,font=\color{black}] {${\it U_z}$};
	
	\node[left=of fig12d, xshift=1.15cm ,yshift=0.10cm,rotate=0,scale=1.1,font=\color{black}] {{\it y}};
	\node[below=of fig12d, xshift=0.25cm ,yshift=1.0cm,rotate=0,scale=1.1,font=\color{black}] {{\it x}};

\end{tikzpicture}

  \caption{Spatial structure of the fastest growing instability modes during the stage of exponential growth at $Gr = 10^8$ ($a$), $Gr = 10^9$ ($b$), and $Gr = 10^{10}$ ($c$),($d$) for $Ha = 10000$. Results of 2D SM82 approximation are shown for $Gr = 10^8$ and $10^9$. Results of 3D computational analysis are shown for $Gr = 10^{10}$. Perturbations of temperature and vector fields of velocity perturbations ($u'$, $w'$) in the vertical midplane ($y=0$) are shown in ($a$)-($c$). Perturbations of vertical velocity in the horizontal midplane section $z=0$ are shown in ($d$).  Solid and dashed isolines indicate positive and negative values, respectively.}

\label{fig12}
\end{figure}

 The results of the linear stability analysis are summarized in figures \ref{fig10}$-$\ref{fig12} and table \ref{table5}.  We need to mention that the wavelength $\lambda$ is varied with step $0.1$ in the simulations. The computed values of the exponential growth rate $\gamma$ as a function of the wavelength $\lambda$ for various combinations of $Ha$ and $Gr$ are shown in figure \ref{fig10}. Two trends of the linear stability behaviour were proposed by \citet{Zhang14}: (1) a higher growth rate and shorter wavelength appear at higher $Gr$; (2) an increase of $Ha$ leads to a higher growth rate. The second, apparently counterintuitive effect was attributed by \citet{Zhang14} to modification of the base flow, namely to suppression of the transverse circulation resulting in weaker mixing and stronger unstable temperature stratification.
 
 In order to investigate these trends for our system and parameter range, we present the exponential growth rate $\gamma_{max}$ of the fastest growing modes and the corresponding wavelengths $\lambda_{max}$ in figure \ref{fig11} and table \ref{table5}. Our results are clearly consistent with the first trend, $\gamma_{max}$ increases and $\lambda_{max}$ decreases with growing $Gr$. However, we find a different behaviour in regard of the second trend. The increase of $Ha$ leads to noticeable or slight decrease of the exponential growth rates at $Gr = 10^8$ or $Gr = 10^9$, respectively (see figure \ref{fig10}$c$). For $Gr = 10^{10}$, $\gamma$ is nearly insensitive to the values of $Ha$.
 
We conclude that the counterintuitive behaviour of stronger instability at higher $Ha$ is not observed in Q2D flows at high $Ha$ considered in our study. It is, nevertheless, noteworthy that $\gamma_{max}$ does not decrease with $Ha$ at $Gr = 10^{10}$. It increases slightly at $Ha \lesssim 6000$ and remains practically constant at higher $Ha$. The enhanced friction in the Hartmann boundary layers is compensated by another effect, the only plausible candidate for which is the strong modification of the streamwise velocity profile visible in figure \ref{fig5}$c$. $U_x$ strongly grows with $Ha$ in the bottom half of the duct, i.e., its part with the strongest unstable temperature gradient. 

The most unstable mode is oscillatory. This was also observed in the earlier works of \citet{Zikanov13, Zhang14}. Point signals of temperature and velocity oscillate in time with constant frequency, This is caused by the transport of the rolls by mean flow. We have computed the phase velocity as the ratio of the axial wavelength to the period of oscillations of a signal at a given point to illustrate this effect. We found that, similarly to findings of \citet{Zikanov13} and \citet{Zhang14}, it varies little with $Ha$ and $Gr$ for the most unstable modes, and has the value close to the mean velocity value 1.

The findings  have critical implications for design and operation of the fusion reactor systems, since they indicate that strength of the convection instability is not diminished by strong magnetic fields at $Gr \geq 10^{10}$ typical for reactor blankets.


\begin{table}
  \begin{center}
  \begin{tabular}{lcccclcccclccccc}
  
    $Gr$ & $Ha$ & $\lambda_{max}$ & $\gamma_{max}$ & $$ & $Gr$ & $Ha$ & $\lambda_{max}$ & $\gamma_{max}$ & $$ & $Gr$ & $Ha$ & $\lambda_{max}$ & $\gamma_{max}$ \\ [3pt]
      \hline \\ [3pt]

    $10^8$ & $1000$   & $0.8$ & 0.487 & $$ & $10^9$ & $1000$ & $-$ & $-$ & $$ &         $10^{10}$ & $1000$ & $-$ & $-$ \\
    $10^8$ & $2000$   & $0.8$ & 0.465 & $$ & $10^9$ & $2000$ & $0.6$ & 2.671 & $$ & $10^{10}$ & $2000$ & $-$ & $-$ \\
    $10^8$ & $3000$   & $0.9$ & 0.400 & $$ & $10^9$ & $3000$ & $0.6$ & 2.827 & $$ & $10^{10}$ & $3000$ & $-$ & $-$ \\
    $10^8$ & $4000$   & $0.9$ & 0.336 & $$ & $10^9$ & $4000$ & $0.6$ & 2.856 & $$ & $10^{10}$ & $4000$ & $0.4$ & 11.674 \\
    $10^8$ & $5000$   & $1.0$ & 0.277 & $$ & $10^9$ & $5000$ & $0.6$ & 2.833 & $$ & $10^{10}$ & $5000$ & $0.4$ & 12.023 \\
    $10^8$ & $6000$   & $1.1$ & 0.226 & $$ & $10^9$ & $6000$ & $0.6$ & 2.786 & $$ & $10^{10}$ & $6000$ & $0.4$ & 12.094 \\
    $10^8$ & $7000$   & $1.1$ & 0.184 & $$ & $10^9$ & $7000$ & $0.6$ & 2.726 & $$ & $10^{10}$ & $7000$ & $0.4$ & 12.112 \\
    $10^8$ & $8000$   & $1.2$ & 0.149 & $$ & $10^9$ & $8000$ & $0.6$ & 2.659 & $$ & $10^{10}$ & $8000$ & $0.4$ & 12.106 \\
    $10^8$ & $9000$   & $1.3$ & 0.120 & $$ & $10^9$ & $9000$ & $0.6$ & 2.589 & $$ & $10^{10}$ & $9000$ & $0.4$ & 12.161 \\
    $10^8$ & $10000$ & $1.4$ & 0.096 & $$ & $10^9$ & $10000$ & $0.6$ & 2.517 & $$ & $10^{10}$ & $10000$ & $0.4$ & 12.106 \\
	
  \end{tabular}
  \end{center}
  \caption{Results of linear stability analysis. Wavelengths $\lambda_{max}$ and exponential growth rates $\gamma_{max}$ of the fastest growing modes are shown as functions of $Ha$ and $Gr$. Only the data for flow regimes identified as Q2D in the analysis of the base flow are shown.}
  \label{table5}
\end{table}
 

 The spatial structure of the unstable modes is illustrated in figure \ref{fig12} for $Gr = 10^8, 10^9, 10^{10}$ and $Ha = 10^4$. The structures are qualitatively similar to those found for Q2D instabilities by \citet{Zhang14}. Consistent with the first trend mentioned above and with the base flow modification illustrated in figure \ref{fig5} is the fact that the energy of growing perturbations becomes contained to the lower part of the duct at higher values of $Gr$.

			
  \subsection{Results of DNS of nonlinear flows} \label{results4}
  
\begin{figure}
	\centering 

	
\begin{tikzpicture}


\node (fig13a1) {\includegraphics[scale=0.2]{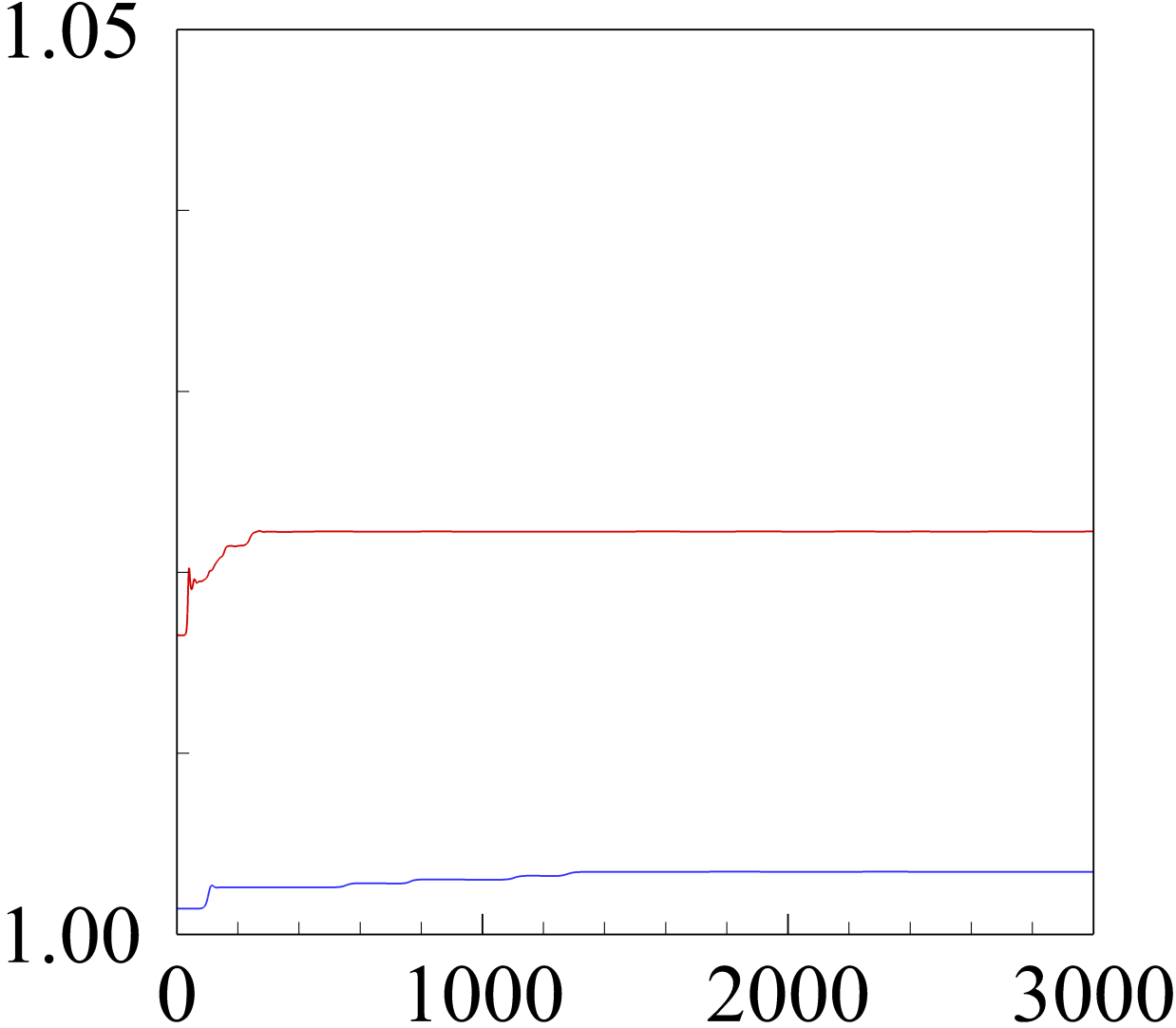}}; 
  	\node[left=of fig13a1, xshift=4.25cm ,yshift=2.25cm,rotate=0,font=\color{black}] {$({\it a})$ $Gr = 10^8$};
	
	\node[left=of fig13a1, xshift=1.3cm ,yshift=0.15cm,rotate=0,scale=1.1,font=\color{black}] {${\it E_x}$};
	\node[below=of fig13a1, xshift=0.15cm ,yshift=1.05cm,rotate=0,scale=1.1,font=\color{black}] {${\it t}$};
	
			
	\node [left=of fig13a1, xshift=3.5cm, yshift=1.25cm]  (fig13_l)  {\includegraphics[scale=0.3]{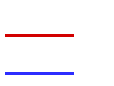}}; 		

	\node[left=of fig13a1, xshift=4.85cm ,yshift=1.35cm,rotate=0,scale=0.9,font=\color{black}]  {${\it Ha} = 5 \times 10^3$};
	\node[left=of fig13a1, xshift=4.35cm ,yshift=0.95cm,rotate=0,scale=0.9,font=\color{black}]  {${\it Ha} = 10^4$};
	
	
\node [right=of fig13a1, xshift=-1.1cm, yshift=0.00cm]  (fig13a2)  {\includegraphics[scale=0.2]{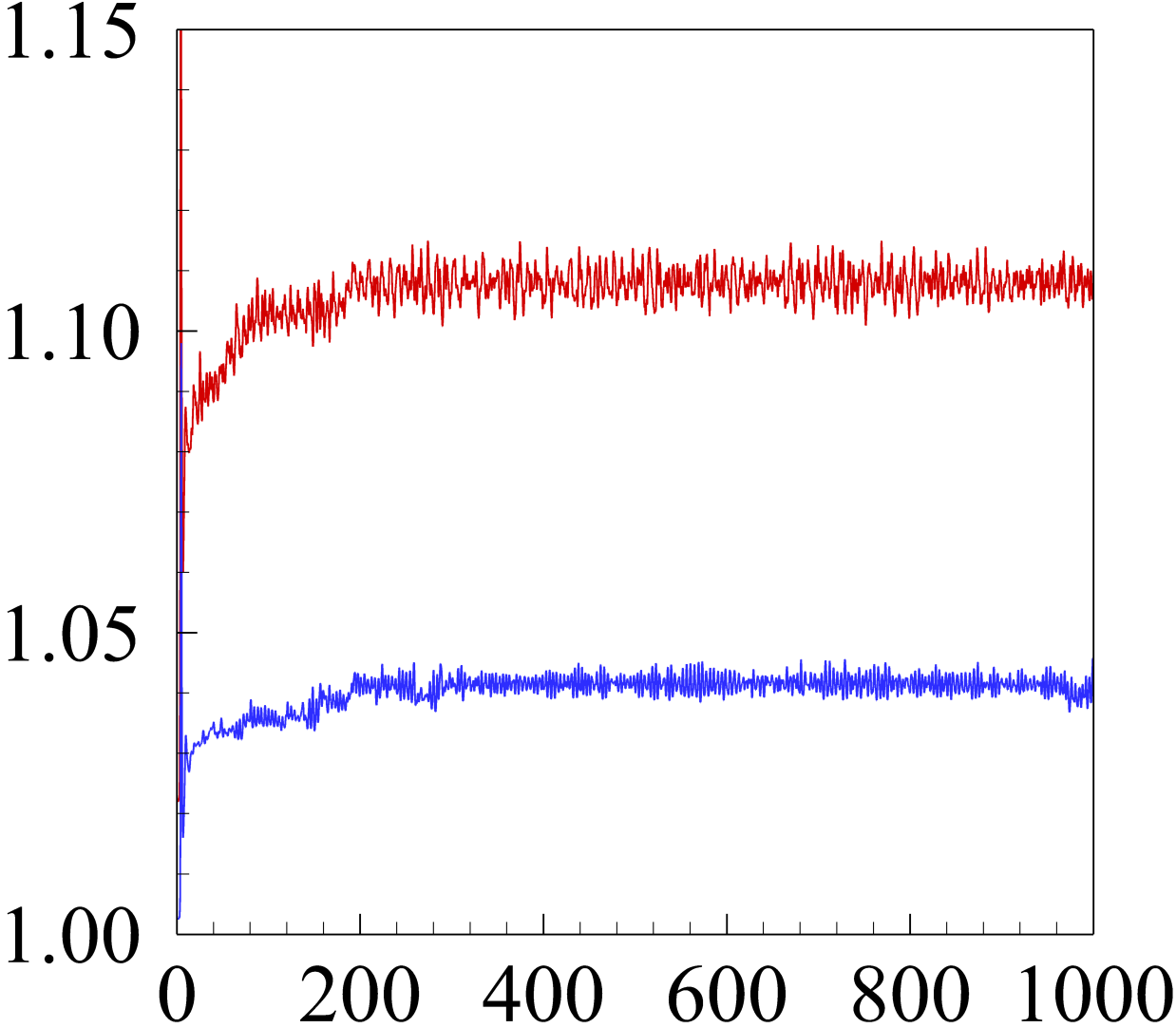}}; 
  	\node[left=of fig13a2, xshift=4.25cm ,yshift=2.25cm,rotate=0,font=\color{black}] {$({\it b})$ $Gr = 10^9$};
	
	\node[below=of fig13a2, xshift=0.15cm ,yshift=1.05cm,rotate=0,scale=1.1,font=\color{black}] {${\it t}$};
	
	
\node [right=of fig13a2, xshift=-1.1cm, yshift=0.00cm]  (fig13a3)  {\includegraphics[scale=0.2]{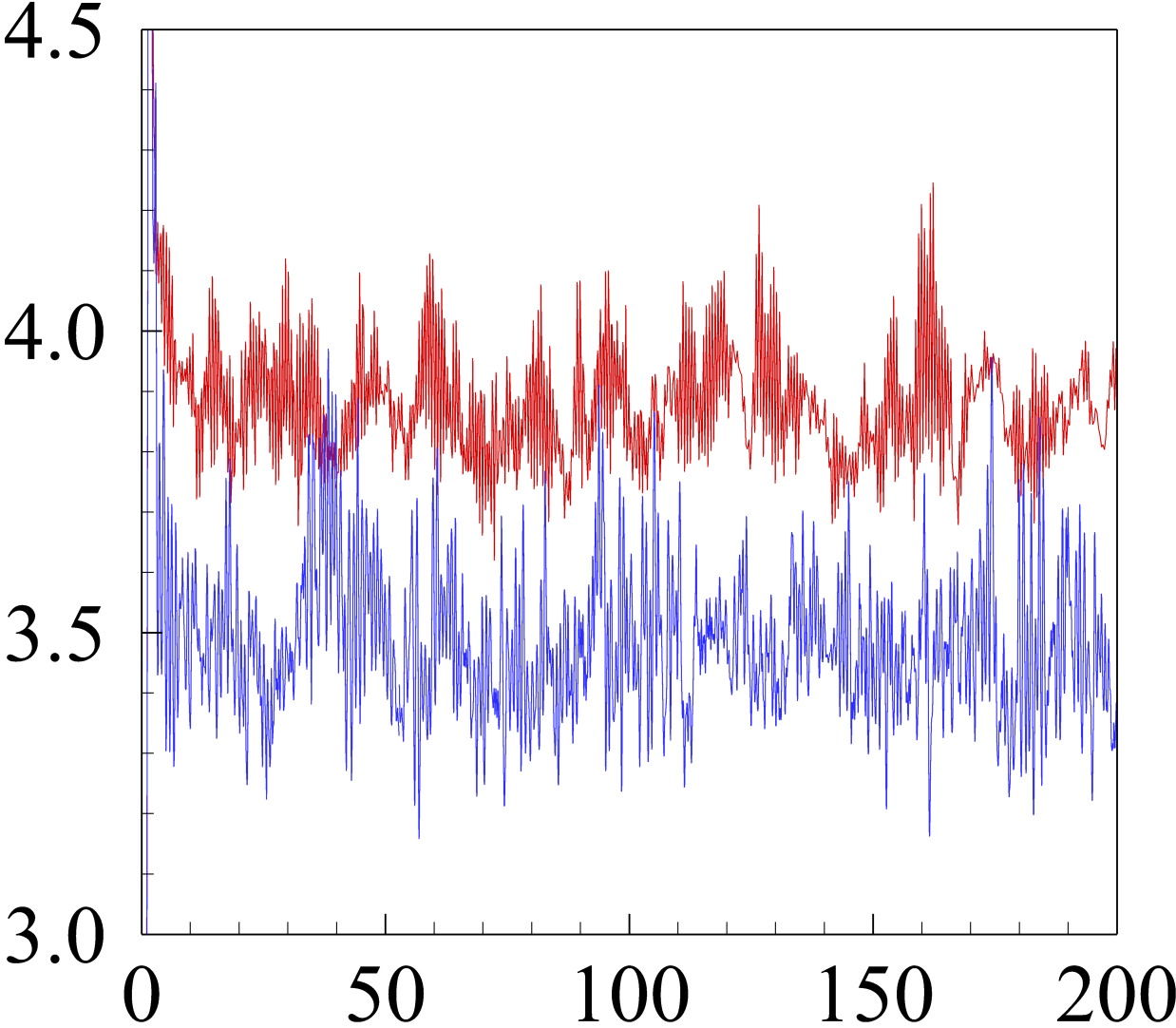}}; 
  	\node[left=of fig13a3, xshift=4.25cm ,yshift=2.25cm,rotate=0,font=\color{black}] {$({\it c})$ $Gr = 10^{10}$};
	
	\node[below=of fig13a3, xshift=0.15cm ,yshift=1.05cm,rotate=0,scale=1.1,font=\color{black}] {${\it t}$};

\end{tikzpicture}

\caption{Time signals of the kinetic energy of the streamwise velocity obtained in the DNS of flows at $Gr = 10^8$ ($a$), $Gr = 10^9$ ($b$), and $Gr = 10^{10}$ ($c$) for $Ha = 5000$ and $Ha = 10000$. Results of 2D SM82 approximation are shown for $Gr = 10^8$ and $10^9$. Results of 3D computational analysis are shown for $Gr = 10^{10}$.} 
  
\label{fig13}
\end{figure}

\begin{figure}
	\centering 

	
\begin{tikzpicture}


\node (fig14_1) {\includegraphics[scale=0.525]{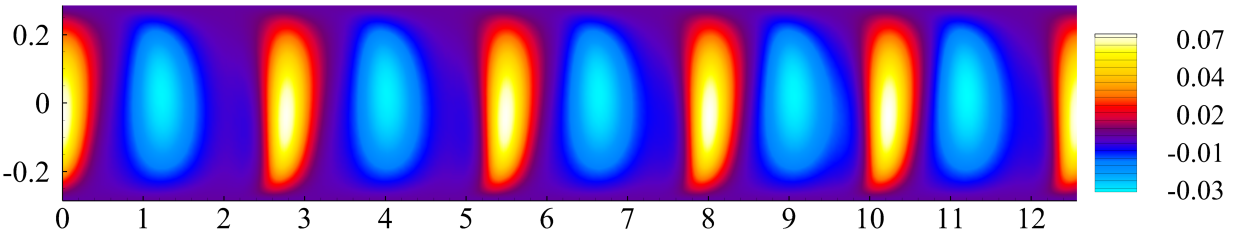}}; 
  	\node[left=of fig14_1, xshift=1.0cm ,yshift=0.90cm,rotate=0,scale=1.0,font=\color{black}] {({\it a})};

	\node[left=of fig14_1, xshift=1.00cm ,yshift=0.1cm,rotate=0,scale=1.0,font=\color{black}] {${\it z}$};
	
	\node[left=of fig14_1, xshift=12.2cm ,yshift=1.00cm,rotate=0,scale=1.0,font=\color{black}] {${\it u_z}$};
	
\node [below=of fig14_1, xshift=0.0cm, yshift=1.00cm]  (fig14_2)  {\includegraphics[scale=0.525]{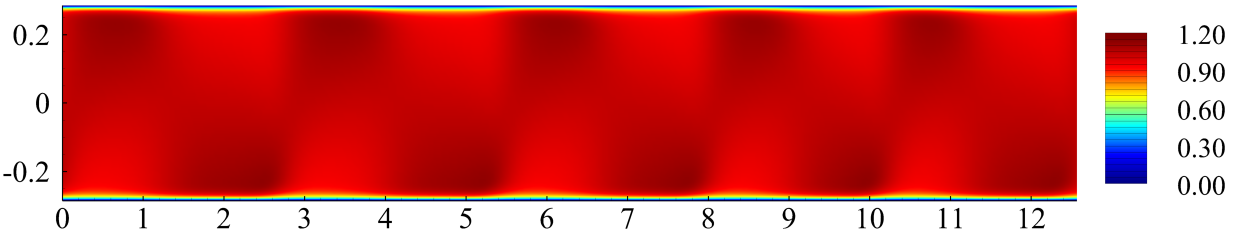}}; 
  	\node[left=of fig14_2, xshift=1.0cm ,yshift=0.90cm,rotate=0,scale=1.0,font=\color{black}] {({\it b})};

	\node[left=of fig14_2, xshift=1.00cm ,yshift=0.1cm,rotate=0,scale=1.0,font=\color{black}] {${\it z}$};
	
	\node[left=of fig14_2, xshift=12.2cm ,yshift=1.00cm,rotate=0,scale=1.0,font=\color{black}] {${\it u_x}$};
	
\node [below=of fig14_2, xshift=0.0cm, yshift=1.00cm]  (fig14_3)  {\includegraphics[scale=0.525]{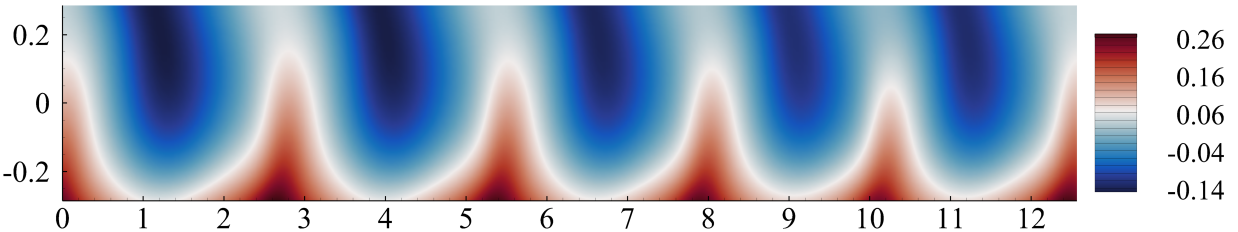}};
  	\node[left=of fig14_3, xshift=1.0cm ,yshift=0.90cm,rotate=0,scale=1.0,font=\color{black}] {({\it c})};

	\node[left=of fig14_3, xshift=1.00cm ,yshift=0.1cm,rotate=0,scale=1.0,font=\color{black}] {${\it z}$};
	
	\node[left=of fig14_3, xshift=12.1cm ,yshift=1.00cm,rotate=0,scale=1.0,font=\color{black}] {${\theta}$};
	

\node [below=of fig14_3, xshift=0.0cm, yshift=1.00cm]  (fig14_4)  {\includegraphics[scale=0.525]{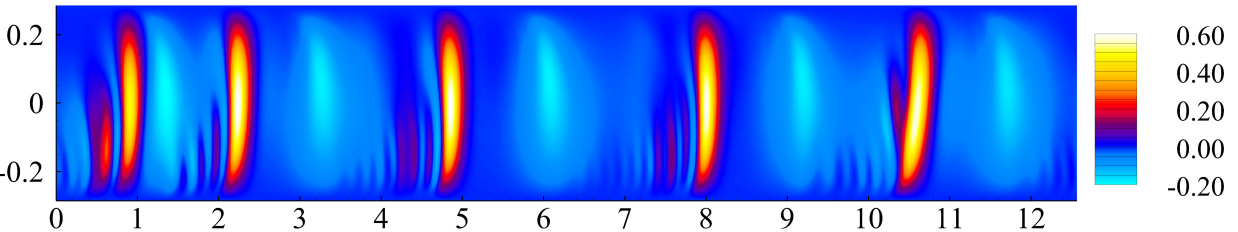}};
  	\node[left=of fig14_4, xshift=1.0cm ,yshift=0.90cm,rotate=0,scale=1.0,font=\color{black}] {({\it d})};

	\node[left=of fig14_4, xshift=1.00cm ,yshift=0.1cm,rotate=0,scale=1.0,font=\color{black}] {${\it z}$};
	
	\node[left=of fig14_4, xshift=12.2cm ,yshift=1.00cm,rotate=0,scale=1.0,font=\color{black}] {${\it u_z}$};

\node [below=of fig14_4, xshift=0.0cm, yshift=1.00cm]  (fig14_5)  {\includegraphics[scale=0.525]{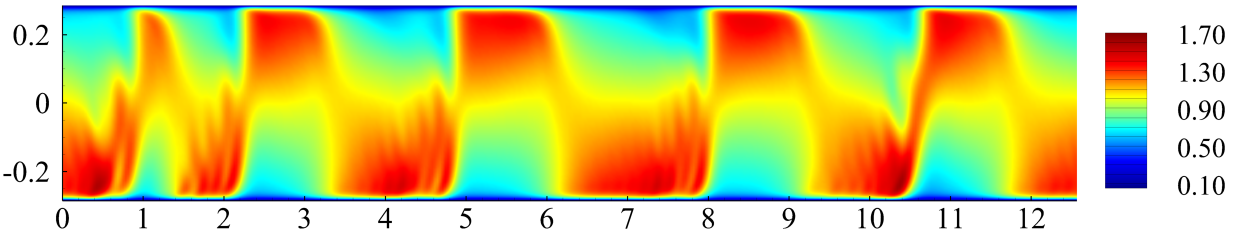}};
  	\node[left=of fig14_5, xshift=1.0cm ,yshift=0.90cm,rotate=0,scale=1.0,font=\color{black}] {({\it e})};

	\node[left=of fig14_5, xshift=1.00cm ,yshift=0.1cm,rotate=0,scale=1.0,font=\color{black}] {${\it z}$};
	
	\node[left=of fig14_5, xshift=12.2cm ,yshift=1.00cm,rotate=0,scale=1.0,font=\color{black}] {${\it u_x}$};
	
\node [below=of fig14_5, xshift=0.0cm, yshift=1.00cm]  (fig14_6)  {\includegraphics[scale=0.525]{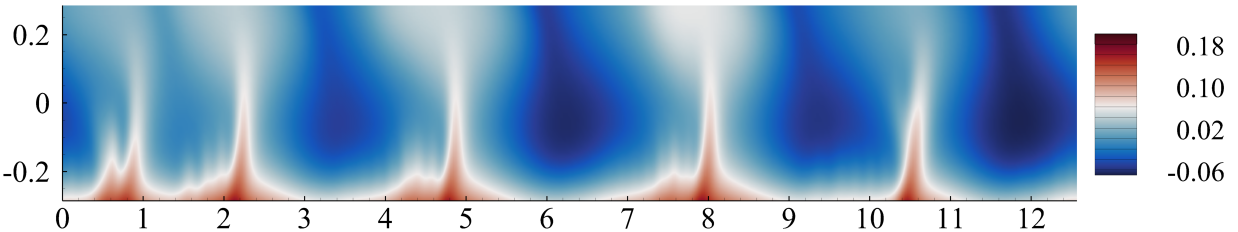}};
  	\node[left=of fig14_6, xshift=1.0cm ,yshift=0.90cm,rotate=0,scale=1.0,font=\color{black}] {({\it f})};

	\node[left=of fig14_6, xshift=1.00cm ,yshift=0.1cm,rotate=0,scale=1.0,font=\color{black}] {${\it z}$};
	\node[below=of fig14_6, xshift=0.25cm ,yshift=1.0cm,rotate=0,font=\color{black}] {{\it x}};	
	
	\node[left=of fig14_6, xshift=12.1cm ,yshift=1.00cm,rotate=0,scale=1.0,font=\color{black}] {${\theta}$};
	
	
\node [below=of fig14_6, xshift=-3.35cm, yshift=0.60cm]  (fig14_7)  {\includegraphics[scale=0.235]{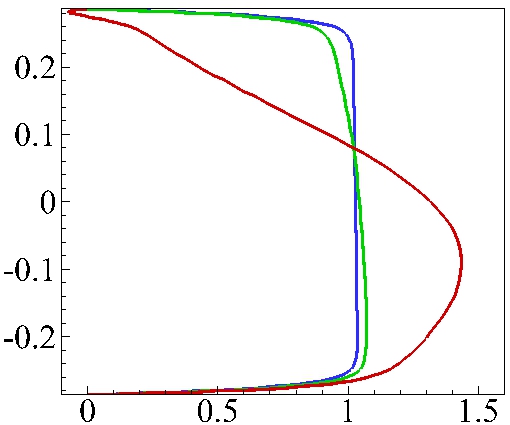}}; 
  	\node[left=of fig14_7, xshift=1.0cm ,yshift=1.50cm,rotate=0,font=\color{black}]{({\it g})};
	
	\node[left=of fig14_7, xshift=1.00cm ,yshift=0.15cm,rotate=0,scale=1.10,font=\color{black}] {$z$};
	\node[below=of fig14_7, xshift=0.25cm ,yshift=1.00cm,rotate=0,scale=1.10,font=\color{black}] {$u_x$};	
	
	
\node [right=of fig14_7, xshift=-0.05cm, yshift=0.00cm]  (fig14_8)  {\includegraphics[scale=0.235]{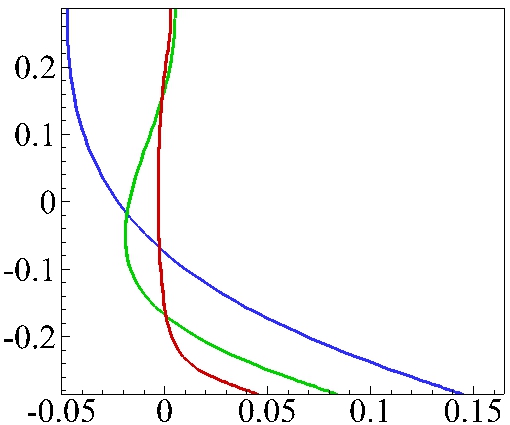}}; 
  	\node[left=of fig14_8, xshift=1.0cm ,yshift=1.50cm,rotate=0,font=\color{black}]{({\it h})};
	
	\node[left=of fig14_8, xshift=1.00cm ,yshift=0.15cm,rotate=0,scale=1.10,font=\color{black}] {$z$};
	\node[below=of fig14_8, xshift=0.25cm ,yshift=1.00cm,rotate=0,scale=1.10,font=\color{black}] {$\theta$};	
	
	
\node [right=of fig14_8, xshift=-3.85cm, yshift=0.75cm]  (fig14_l)  {\includegraphics[scale=0.225]{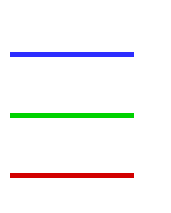}}; 

	\node[right=of fig14_l, xshift=-1.40cm ,yshift=0.45cm,rotate=0,scale=0.9,font=\color{black}]  {$Gr = 10^8$};
	\node[right=of fig14_l, xshift=-1.40cm ,yshift=0.00cm,rotate=0,scale=0.9,font=\color{black}]  {$Gr = 10^9$};
	\node[right=of fig14_l, xshift=-1.40cm ,yshift=-0.45cm,rotate=0,scale=0.9,font=\color{black}]  {$Gr = 10^{10}$};

	\end{tikzpicture}

\caption{Flow structure in nonlinear regime at $Gr = 10^8$ ($a$-$c$) and $Gr = 10^9$ ($d$-$f$) for $Ha = 10000$. The instantaneous distributions of $u_z$ ($a$, $d$), $u_x$ ($b$, $e$), and $\theta$ ($c$, $f$) obtained in the 2D model are shown. The profiles of $u_x$ and $\theta$ obtained by averaging over $x$ and time are shown, respectively, in ($g$) and ($h$)}
  
\label{fig14}
\end{figure}
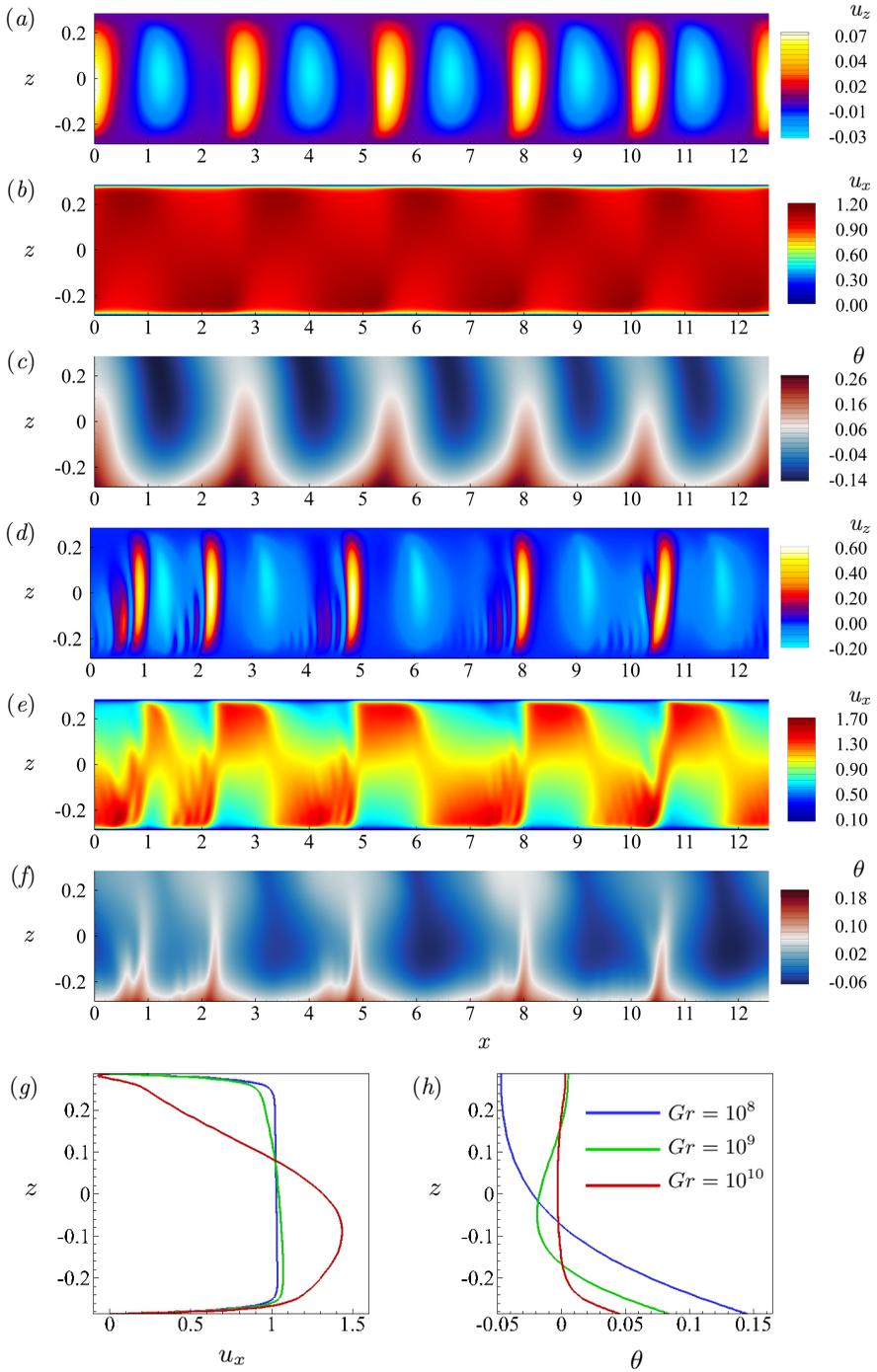

\begin{figure}
	\centering 

\begin{tikzpicture}


\node (fig15_1) {\includegraphics[scale=0.25]{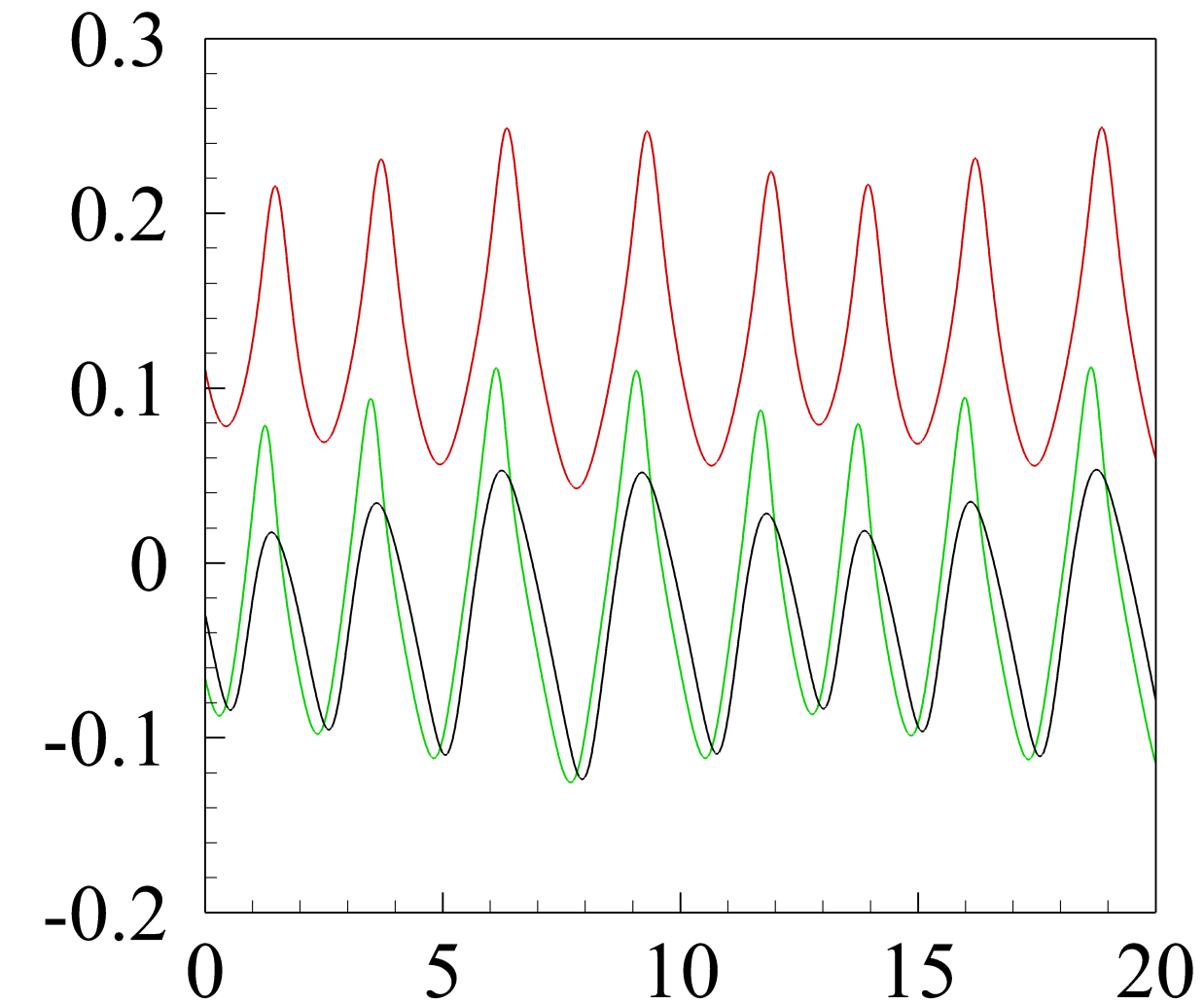}};
 
  	\node[left=of fig15_1, xshift=6.15cm ,yshift=2.5cm,rotate=0,font=\color{black}] {$({\it a})$ $Ha = 5 \times 10^3, Gr = 10^8$};
	
	\node[left=of fig15_1, xshift=1.15cm ,yshift=0.15cm,rotate=0,scale=1.25,font=\color{black}] {$\theta$};
	\node[below=of fig15_1, xshift=0.25cm ,yshift=1.00cm,rotate=0,scale=1.10,font=\color{black}] {$t$};	
	

\node [right=of fig15_1, xshift=-0.4cm, yshift=0.0cm]  (fig15_2) {\includegraphics[scale=0.25]{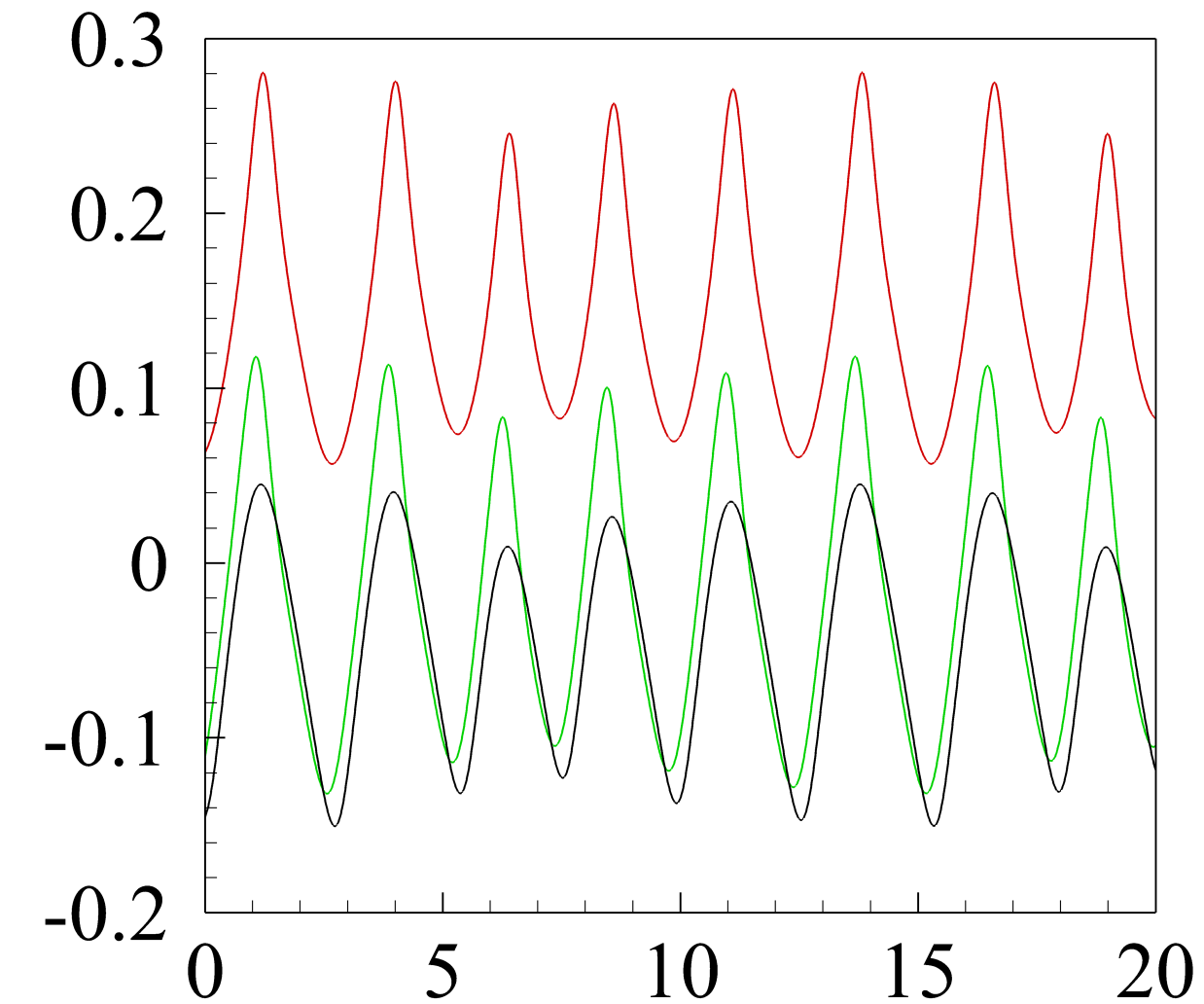}}; 
  	\node[left=of fig15_2, xshift=5.75cm ,yshift=2.5cm,rotate=0,font=\color{black}] {$({\it b})$ $Ha = 10^4, Gr = 10^8$};

	\node[left=of fig15_2, xshift=1.15cm ,yshift=0.15cm,rotate=0,scale=1.25,font=\color{black}] {$\theta$};
	\node[below=of fig15_2, xshift=0.25cm ,yshift=1.00cm,rotate=0,scale=1.10,font=\color{black}] {$t$};	
	

\node [below=of fig15_1, xshift=0.0cm, yshift=0.0cm]  (fig15_3) {\includegraphics[scale=0.25]{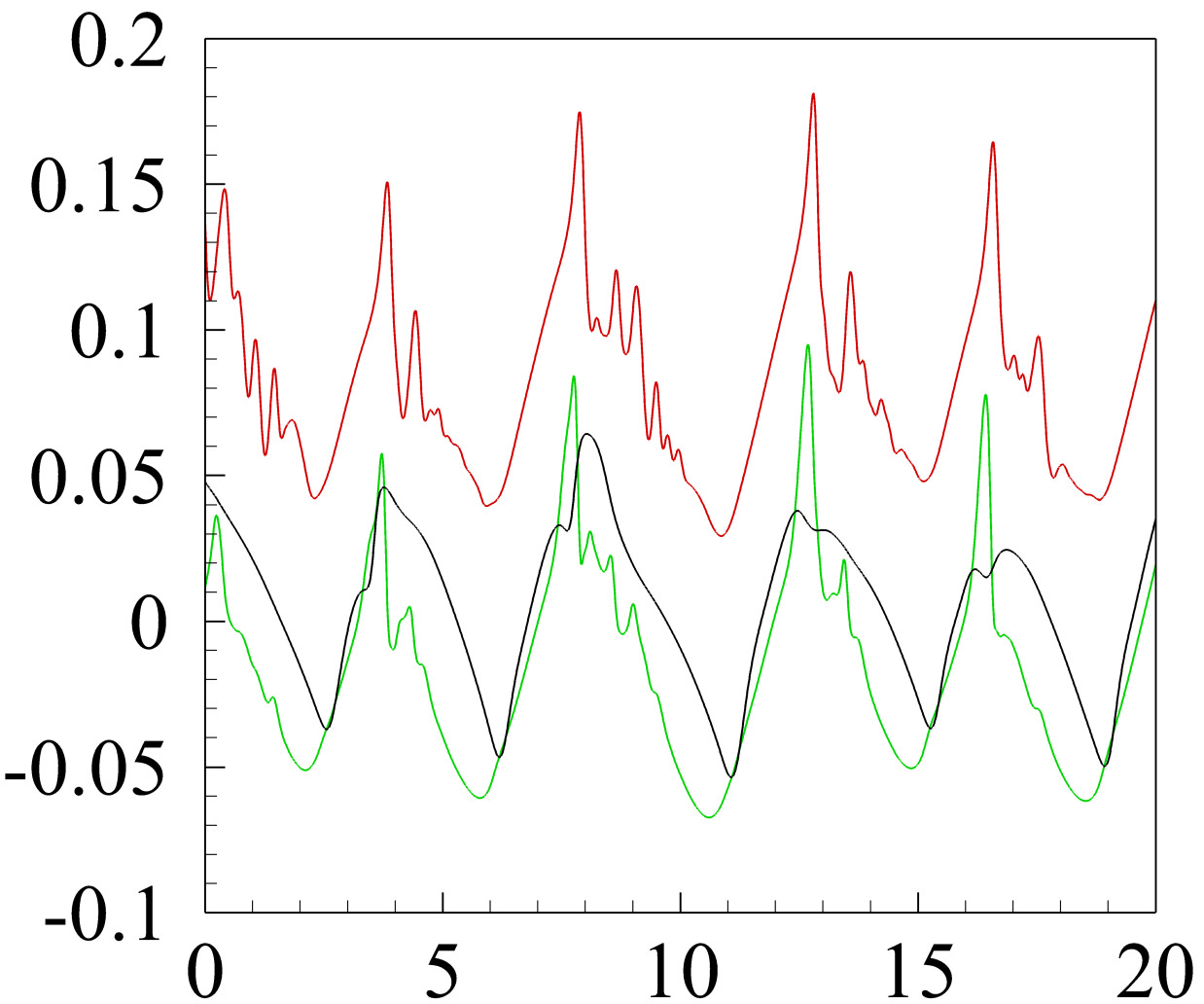}}; 
  	\node[left=of fig15_3, xshift=6.15cm ,yshift=2.5cm,rotate=0,font=\color{black}] {$({\it c})$ $Ha = 5 \times 10^3, Gr = 10^9$};

	\node[left=of fig15_3, xshift=1.15cm ,yshift=0.15cm,rotate=0,scale=1.25,font=\color{black}] {$\theta$};
	\node[below=of fig15_3, xshift=0.25cm ,yshift=1.00cm,rotate=0,scale=1.10,font=\color{black}] {$t$};
	

\node [right=of fig15_3, xshift=-0.4cm, yshift=0.0cm]  (fig15_4) {\includegraphics[scale=0.25]{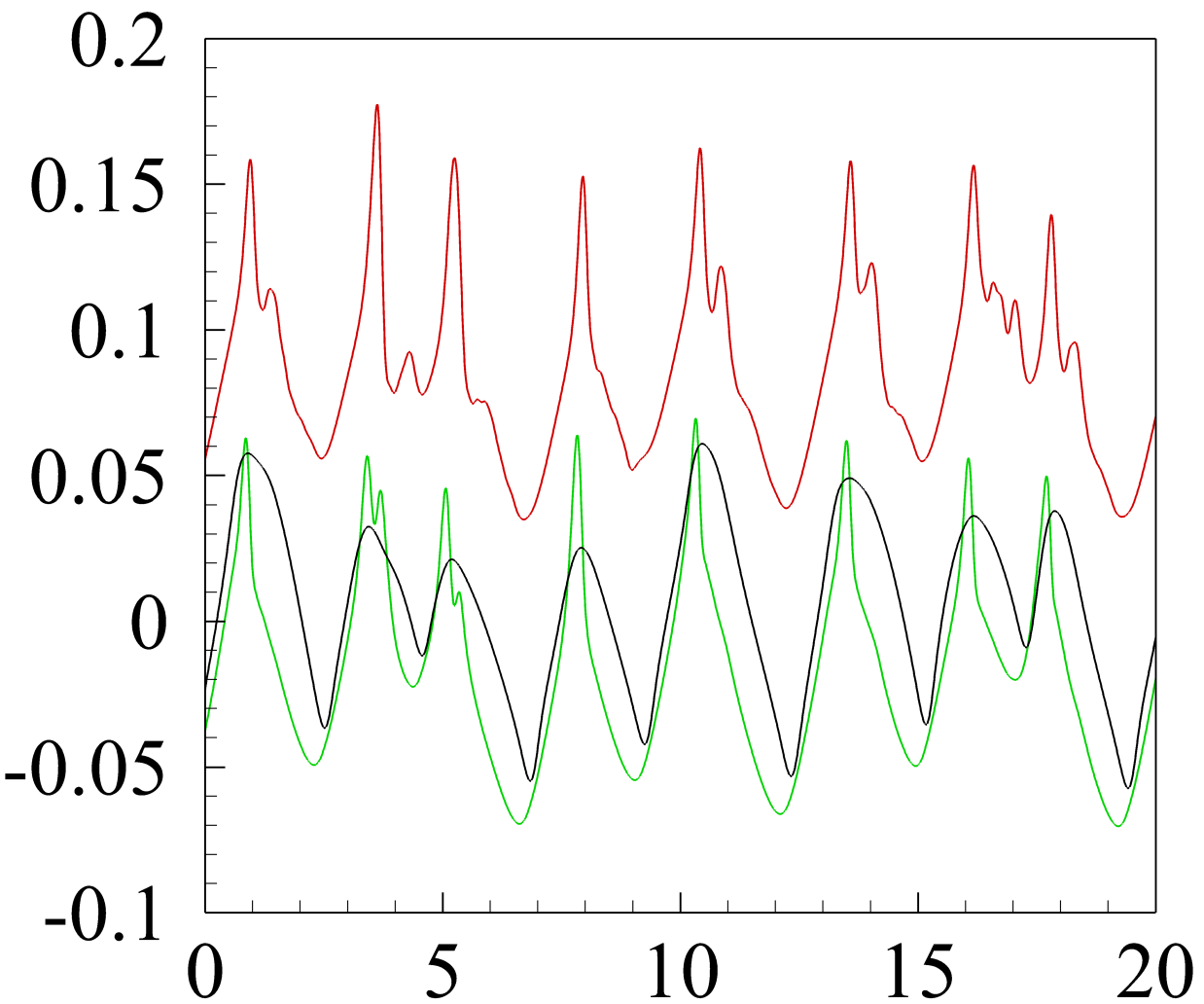}}; 
  	\node[left=of fig15_4, xshift=5.75cm ,yshift=2.5cm,rotate=0,font=\color{black}] {$({\it d})$ $Ha = 10^4, Gr = 10^9$};

	\node[left=of fig15_4, xshift=1.15cm ,yshift=0.15cm,rotate=0,scale=1.25,font=\color{black}] {$\theta$};
	\node[below=of fig15_4, xshift=0.25cm ,yshift=1.00cm,rotate=0,scale=1.10,font=\color{black}] {$t$};		
	

\node [below=of fig15_3, xshift=0.0cm, yshift=0.0cm]  (fig15_5) {\includegraphics[scale=0.25]{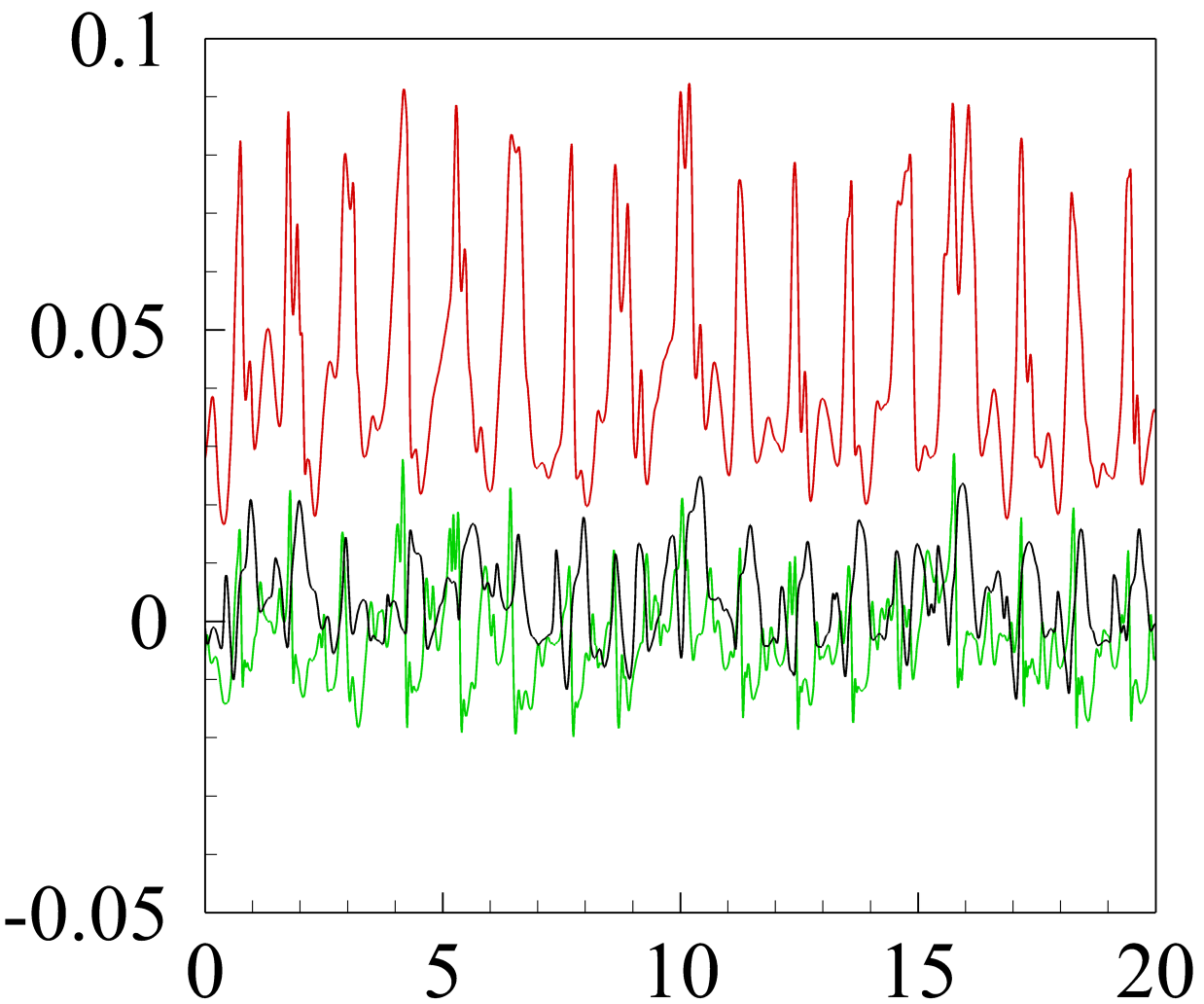}}; 
  	\node[left=of fig15_5, xshift=6.15cm ,yshift=2.5cm,rotate=0,font=\color{black}] {$({\it e})$ $Ha = 5 \times 10^3, Gr = 10^{10}$};

	\node[left=of fig15_5, xshift=1.15cm ,yshift=0.15cm,rotate=0,scale=1.25,font=\color{black}] {$\theta$};
	\node[below=of fig15_5, xshift=0.25cm ,yshift=1.00cm,rotate=0,scale=1.10,font=\color{black}] {$t$};
	

\node [right=of fig15_5, xshift=-0.4cm, yshift=0.0cm]  (fig15_6) {\includegraphics[scale=0.25]{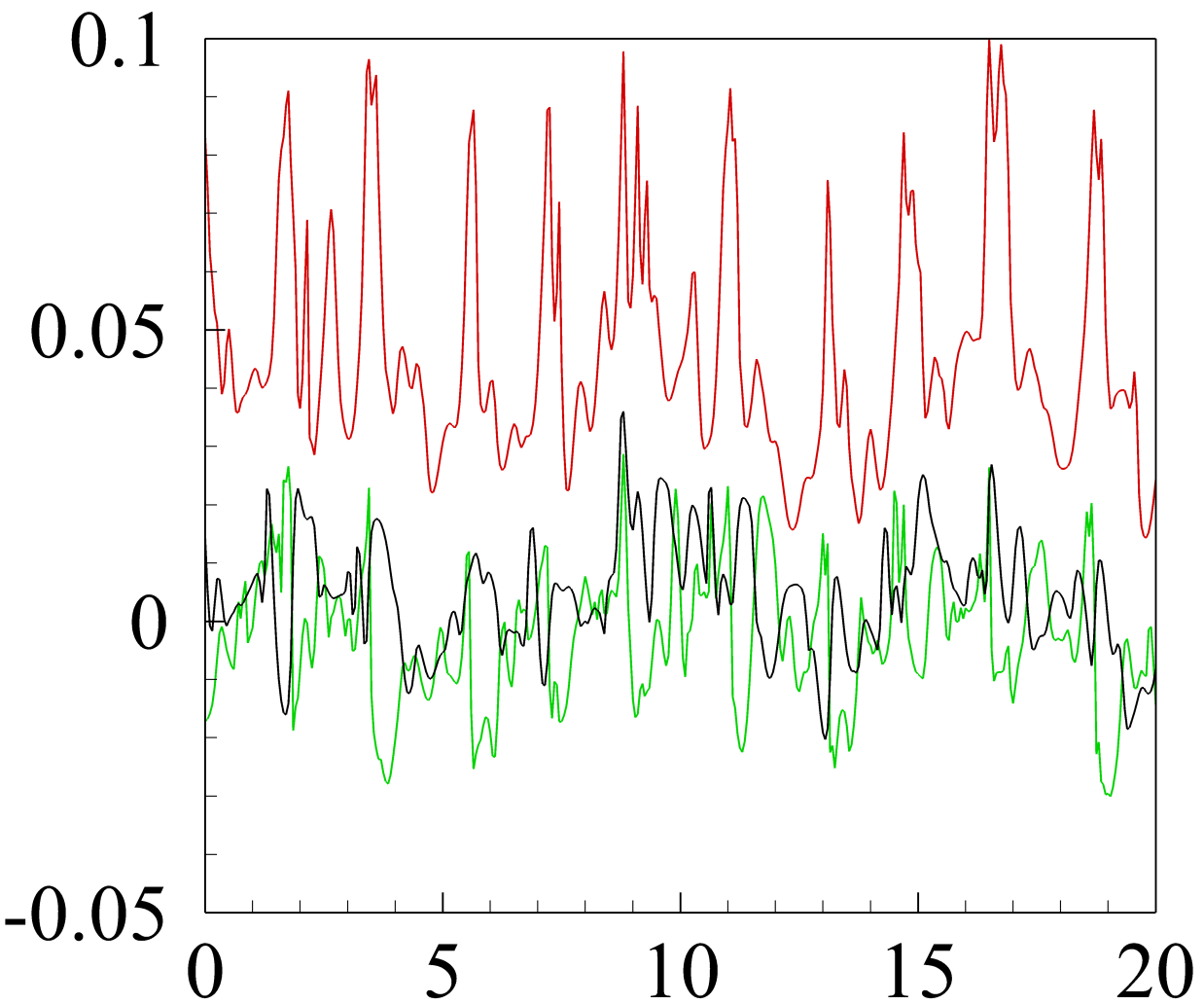}}; 
  	\node[left=of fig15_6, xshift=5.75cm ,yshift=2.5cm,rotate=0,font=\color{black}] {$({\it f})$ $Ha = 10^4, Gr = 10^{10}$};

	\node[left=of fig15_6, xshift=1.15cm ,yshift=0.15cm,rotate=0,scale=1.25,font=\color{black}] {$\theta$};s
	\node[below=of fig15_6, xshift=0.25cm ,yshift=1.00cm,rotate=0,scale=1.10,font=\color{black}] {$t$};

	
\node [below=of fig15_5, xshift=-1.00cm, yshift=0.65cm]  (fig15_l)  {\includegraphics[scale=0.25]{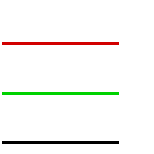}}; 

	\node[right=of fig15_l, xshift=-1.25cm ,yshift=0.30cm,rotate=0,scale=1.0,font=\color{black}]  {${\it z} = -0.2857$};
	\node[right=of fig15_l, xshift=-1.25cm ,yshift=-0.10cm,rotate=0,scale=1.0,font=\color{black}]  {${\it z} = 0$};
	\node[right=of fig15_l, xshift=-1.25cm ,yshift=-0.50cm,rotate=0,scale=1.0,font=\color{black}]  {${\it z} = 0.2857$};

\end{tikzpicture}

  \caption{Time signals of temperature measured at top and bottom walls and in the middle of the duct in fully developed flows at $Gr = 10^8, 10^9, 10^{10}$ are shown for $Ha = 5 \times 10^3$ in ($a, c, e$) and for $Ha = 10^4$ in ($b, d, f$). Results of 2D SM82 approximation are shown for $Gr = 10^8$ and $10^9$ ($a-d$). Results of 3D computational analysis are shown for $Gr = 10^{10}$ ($e-f$).}
  
\label{fig15}
\end{figure}

  The results concerning the nonlinear flow regimes are illustrated in figures \ref{fig8}($d,e,f$), \ref{fig13}, \ref{fig14}, and \ref{fig15}. DNS approach based on direct solution of the nonlinear governing equations is utilized. The 2D SM82 model (\ref{eqsm1})$-$(\ref{eqsm3}) and the computational domain of length $L_x = 4\pi$ are used for flows at $Gr=10^8$ and $10^9$. Full 3D equations (\ref{eq1})$-$(\ref{eq3}) are solved and the domain is reduced to $L_x = 2\pi$ at $Gr=10^{10}$. Other parameters of computational model are described in section \ref{gridstudy}. Each simulation starts with the streamwise-independent base flow (see section \ref{results2}) computed at the same $Gr$ and $Ha$, to which random small-amplitude ($\sim 10^{-3}$) random perturbations of velocity and temperature are added.
  

The typical flow evolution is illustrated by the curves of average kinetic energy shown in figure \ref{fig13}. The flow reaches a fully developed state after the instability and initial development. The evolution of the fully developed flow is computed for at least $500$ time units for the 2D model at $Gr = 10^8, 10^9$ and at least $100$ time units for the 3D model at $Gr = 10^{10}$. At this stage, the integral parameters fluctuate around steady means (at $Gr = 10^9$ and $10^{10}$) or remain steady (at $Gr = 10^8$). The amplitudes of the fluctuations are small at $Gr = 10^9$ and large, but still moderate at $Gr = 10^{10}$.

Structure of fully developed flows is illustrated in figures \ref{fig8} and \ref{fig14}. The velocity field shows finite-amplitude roll-like structures (hereafter referred to as rolls) resulting from the instability, which are superimposed on a streamwise-independent mean flow (see figures \ref{fig8}$d$ and \ref{fig14}$a,d$). The rolls cause variations of temperature (see figures \ref{fig8}$f$, \ref{fig14}$c$ and \ref{fig14}$f$). Transport of the rolls by the mean flow is a known reason of MCFs in horizontal channels \citep{Zikanov13, Zhang14, Zikanov21}.

Comparison of the flow structures in figures \ref{fig8}$d$-$f$ and \ref{fig14} reveal the effect of the value of $Gr$ on convection rolls. As anticipated, increase of $Gr$ leads to higher non-dimensional amplitude of the velocity fluctuations. This results in stronger vertical mixing as illustrated by the streamwise- and time-averaged profiles in figure \ref{fig14}$g, h$. In particular, a nearly uniform vertical distribution of average temperature with a thin (but still much thicker than the Shercliff layer) boundary layer at the bottom is observed at $Gr = 10^{10}$.

As we discussed earlier, MCFs caused by the instability have potentially critical implications for design and operation of liquid-metal components of nuclear fusion reactors. The DNS results allow us, for the first time, to evaluate the properties of the MCFs at the high values of $Gr$ and $Ha$ corresponding to the actual reactor conditions.

In addition to the instantaneous temperature distributions in figures \ref{fig8}$f$, \ref{fig14}$c$ and \ref{fig14}$f$, the discussion will be based on the point-signals of temperature measured at the top and bottom walls and in the middle of the duct (see figure \ref{fig15}). As discussed, e.g., by \citet{Zikanov21}, measuring such signals is the most reliable and commonly used tool for studying MCFs in experiments.

The evident conclusion from the DNS data is that MCFs are fully present in flows with $Gr = 10^8, 10^9, 10^{10}$ and the highest values $Ha = 5 \times 10^3$ and $10^4$ considered in this study. The fluctuations are observed in the entire duct. The temperature signals are regular and dominated by one or several low frequencies (the typical period is $2$-$3$ non-dimensional time units at $Gr = 10^8$ and $10^9$). The signal is less regular and characterized by higher typical frequencies at $Gr = 10^{10}$.

Interestingly, the non-dimensional amplitude of the temperature fluctuations decreases noticeably with growing $Gr$. Comparison of the signals in the two columns of figure \ref{fig15} demonstrates that value of $Ha$ has practically no effect on the MCFs. This can be attributed to the effect of nonlinearity, which distributes energy of the fluctuations to a range of streamwise modes.

Considering the practical implications, it is interesting to evaluate the parameters of the MCFs in dimensional units. We will do that for the temperature signals at the bottom of the duct ($z = -0.2857$) assuming the duct half-width $d = 5$ $cm$ and using the physical properties of PbLi at $573$ $K$ \citep{Zikanov21}. The wall heat rate is $q = 10.56$ $kW m^{-2}$ at $Gr = 10^8$, $q = 105.6$ $kW m^{-2}$ at $Gr = 10^9$ and $q = 1056$ $kW m^{-2}$ at $Gr = 10^{10}$. We find, by applying the temperature scale $qd/\kappa$, that the largest amplitude of fluctuations of wall temperature is in the range $5-6$ $K$ at $Gr = 10^8$, $44-62$ $K$ at $Gr = 10^9$, and somewhat unrealistic $180-300$ $K$ at $Gr = 10^{10}$. The typical time period of the fluctuations is $5.64$ s at $Gr = 10^8$, $6.45$ $s$ at $Gr = 10^9$, and $4.51$ $s$ at $Gr = 10^{10}$ for $Ha = 10^4$.

Similar evaluations have been done for the future experiments on the recently built experimental facility \citep[see, e.g.,][]{Belyaev17}, in which liquid mercury flows in the duct with the half-width of $d = 2.8$ $cm$. The physical properties of Hg are taken at $303$ $K$ \citep{Zikanov21}. The wall heat rate is $q = 9.59$ $kW m^{-2}$ at $Gr = 10^8$ and $q = 95.9$ $kW m^{-2}$ at $Gr = 10^9$. The results of nonlinear simulations allow us to predict the largest amplitude of fluctuations of temperature in the middle of the duct. The amplitudes are in the range of $2-3$ $K$ at $Gr = 10^8$ and $12-18$ $K$ at $Gr = 10^9$. The typical time period of the fluctuations is $1.1$ s at $Gr = 10^8$ and $1.58$ s at $Gr = 10^9$ for $Ha = 1000$.


 \section{\textbf{Concluding remarks}}
 
We have analysed mixed convection in a liquid metal flow in a duct with bottom heating and transverse magnetic field. The analysis is extended to much higher values of $Ha$ and $Gr$ than the previous analysis of similar effects by \citet{Zhang14}. 

The main conclusion of our work is that magnetoconvective fluctuations appear at the parameters anticipated for operational regimes of blankets and divertors of future fusion rectors. The fluctuations are not suppressed or even significantly reduced in amplitude by the very strong magnetic field. The amplitude remains high, reaching tens or hundreds degrees $K$ (depending on the value of $Gr$) in a typical duct geometry. This has significant far-reaching implications for mixing, heat and mass transfer, and structural integrity of reactor components. The most dangerous modes of the instability have the form of rolls localized in the lower half of the duct and having the streamwise wavelength measured in horizontal half-widths of the duct, approximately, between $0.8$ and $1.4$ at $Gr = 10^8$, $0.6$ at $Gr = 10^9$, and $0.4$ at $Gr = 10^{10}$.

Another conclusion concerns applicability of the two-dimensional approximation by \citet{Sommeria82} to flows with thermal convection. We have found that the approximation may become inaccurate at high values of $Gr$ even though the flow remains quasi-two-dimensional. Full reasons of this phenomenon remain to be understood. One of the reasons is, clearly, the geometry of the flow. The 2D model tends to be less accurate if applied to flows in ducts with larger aspect ratios and the magnetic field parallel to the long side. In general, the conclusion is important as a warning against application of the model without a proper verification.

It is pertinent to stress that the conclusions must be considered as preliminary because they are obtained for a single configuration of a horizontal duct flow with bottom heating and transverse magnetic field. At the same time, there are multiple indications that similar behaviours can be observed in other configurations related to the existing designs of liquid metal blankets of fusion reactors. This will need to be explored in future studies.

Further study of MCFs at high Hartmann and Grashof numbers is warranted by their practical importance and theoretical significance. Many interesting possible directions of future work can be suggested. We mention two of them. One is the exploration of the phenomenon for other geometries, where strong MCFs are known to exist, for example for downward flow in a vertical duct. Another particularly interesting direction is the analysis of the effects of finite thermal and electrical conductivities of the walls.

 {\textbf{Acknowledgements.}} Authors are thankful to D. Krasnov for continuing assistance with the numerical model and S. Molokov for interesting and useful discussions. 
 
 {\textbf{Funding.}} Work of R. Akhmedagaev and O. Zikanov is supported by the US NSF (Grant CBET 1803730 "Extreme magnetoconvection"). Work of Y. Listratov is supported by the Ministry of Science and Education of the Russian Federation (Grant 14.Z50.31.0042) and by the Russian Foundation for Basic Research (Grant NNIO  18-508-12005).

 {\textbf{Declaration of Interests.}} The authors report no conflict of interest.

\bibliographystyle{jfm}
\bibliography{duct_jfm}

\end{document}